\documentclass[11pt]{article}
\usepackage{latexformat}
\ifdefined\pdfoutput
  \pdfoutput=1
\fi
\usepackage{amsmath,bbm,array,amsfonts,graphicx,wrapfig,lscape,float,mathtools,multirow,longtable,amsthm,upgreek,mathrsfs}
\usepackage[dvipsnames]{xcolor}
\usepackage{stackrel}
\usepackage[all]{xy}
\usepackage[bottom]{footmisc}
\usepackage{tikz}
\usepackage{physics}
\usepackage{bm}
\usepackage[capitalize]{cleveref}
\AddToHook{cmd/appendix/before}{\crefalias{section}{appendix}}
\usepackage{color}
\usepackage{enumerate}
\usetikzlibrary{decorations.pathreplacing}
\usepackage{diagbox}
\numberwithin{equation}{section}
\numberwithin{figure}{section}
\numberwithin{table}{section}
\usepackage{pgfplots}
\pgfplotsset{compat=1.14}
\usepackage{makecell}
\usepackage[utf8]{inputenc}
\allowdisplaybreaks

\newtheorem{definition}{Definition}[section]
\newtheorem{theorem}{Theorem}[section]
\newtheorem{corollary}{Corollary}[theorem]
\newtheorem{lemma}[theorem]{Lemma}
\newtheorem{proposition}[theorem]{Proposition}
\newtheorem{conjecture}[theorem]{Conjecture} 
\newtheorem{remark}{Remark}

\newcommand{\U}{\mathrm{U}}
\newcommand{\SU}{\mathrm{SU}}

\newcommand{\SO}{\mathrm{SO}}

\newcommand{\USp}{\mathrm{USp}}

\newcommand{\ii}{\mathrm{i}}

\usepackage[toc,page]{appendix}

\usepackage{tocloft}

	\title{On \texorpdfstring{$\iota$Quiver}{iQuiver} (BPS) Algebras and (Strict) Partitions}
	
	\author[a,b,c]{Jiakang Bao}
	
	\affiliation[a]{
		Department of Physics, Yonsei University, Seoul 03722, Republic of Korea}
	\affiliation[b]{
		School of Physics, Korea Institute for Advanced Study, 85 Hoegi-ro, Dongdaemun-gu, Seoul 02455, Korea}
	\affiliation[c]{
		Department of Physics, Institute of Science Tokyo, Ookayama 2-12-1, Meguro-ku, Tokyo, 152-8551, Japan}
	
	\emailAdd{jiakang.bao@yonsei.ac.kr}
	
	\preprint{
		\begin{flushright}
			
		\end{flushright}
	}

	\abstract{We consider the $\iota$quiver algebras of split type associated with the quiver BPS algebras of affine type A. They are supposed to give twisted affine Yangians and (elliptic) $\iota$quantum toroidal algebras that could possibly arise from D-branes on $\mathbb{C}\times\mathbb{C}^2/\mathbb{Z}_n$ with orientifolds. We mention some properties of these algebras, such as the minimalistic presentations in the rational cases and the coideal structures. We also investigate how one might identify these algebras from the quiver theories with orientifold actions. Moreover, we discuss some representations with combinatorial bases of these algebras. In particular, the algebras with $n=1,2$ in certain limits of the parameters have representations labelled by (coloured) strict/symmetric partitions, which in the case of $n=1$ recovers the known BPS counting of certain theories.
	}

\begin{document}
	\maketitle

\section{Introduction and Summary}\label{intro}
The quiver BPS algebras \cite{Li:2020rij,Galakhov:2020vyb,Galakhov:2021xum,Noshita:2021ldl,Galakhov:2021vbo,Noshita:2021dgj} are constructed in recent years as algebras that encode the counting of $1/2$-BPS bound states for Type IIA on toric Calabi--Yau (CY) threefolds. In particular, they have representations with combinatorial bases given by the crystal melting models \cite{Okounkov:2003sp,Szendroi:2007nu,MR2836398,Ooguri:2009ijd,Dimofte:2009bv}.

Considering the supersymmetric quantum mechanics on the D-branes that wrap the compact cycles in the CY threefold, it can be described by a quiver, and the BPS counting problem becomes the Witten index computations for the quiver quantum mechanics. In particular, this quiver coincides with the one from the brane tiling on the CY threefold \cite{Feng:2000mi,Feng:2004uq,Hanany:2005ve,Franco:2005rj,Yamazaki:2008bt}, and the supersymmetric quantum mechanics is actually the dimensional reduction of the 4d $\mathcal{N}=1$ theory.

With the extensive study on these quantum algebras, including earlier mathematical works on affine Yangians, quantum toroidal algebras, and elliptic quantum toroidal algebras\footnote{Due to the vast literature on these topics, it is impossible to give an exhaustive list of references. We will mention a few when necessary. More works can be found in the references in the quiver BPS algebra papers mentioned here.} before the name quiver BPS algebras was introduced, many properties and their connections to other relevant fields in mathematics and physics have been discovered.

A next question one might wonder is whether we can have a similar story when we include orientifolds. The study of orientifolds on brane tilings was pioneered in \cite{Franco:2007ii}, and was later also discussed in, for example, \cite{Argurio:2020npm,Argurio:2020dko,Garcia-Valdecasas:2021znu}. In particular, the quiver would now not only have unitary gauge nodes, but also orthogonal and symplectic ones.

It turns out that in such cases, the BPS counting via localisation would in general lead to fractional numbers. Some of the theories were discussed, for example, in \cite{Yi:1997eg,Lee:2016dbm,Lee:2017lfw}. In fact, these rational invariants would coincide with the rational Donaldson--Thomas invariants \cite{Joyce:2008pc,Young:2014eqa,Young:2016vjz,Franzen:2016czs,bu2023enumerativeI,bu2023enumerativeII,Bu:2025nmp}. Although they are well-defined invariants, the physical counting of states should still be integers. The rational invariants would usually be caused by the non-compact directions in the moduli spaces. It could be possible to uplift these directions in the Higgs branch by introducing more flavours, but it would need some further regularisations when non-compact directions arise from the Coulomb branch, whose systematic prescriptions are still not clear.

Although it would be desirable to know the BPS counting and what kind of combinatorial modules one might expect\footnote{See also \cite{Krefl:2009mw,Krefl:2010sz} for the discussions from the perspective of real topological strings.}, there would still be some expectations of the algebras from other aspects. In particular, we should get some twisted versions of the quiver BPS algebras. For example, for affine Yangians, we should look for some twisted affine Yangians. Given a Yangian associated to a Lie algebra of finite type, the twisted Yangians, as coideal subalgebras of the Yangian, were introduced in \cite{molev2002representations,olshanskii2006twisted,guay2016twisted} in the $R$-matrix presentation. The study on their representations can be found, for example, in \cite{Molev:1994rs,nazarov2004representations,MacKay:2002at,guay2017representations,lu2025twisted}. The $J$-presentation of the twisted Yangias were discussed in \cite{Belliard:2014uja}. For twisted affine Yangians, some examples were constructed via the $J$-presentation in \cite{ueda2025twisted,harako2026new}.

There are various perspectives and approaches that might help us obtain the BPS algebras for orientifolded theories as we will mention shortly. Based on the study of quiver BPS algebras, the most direct analogue of understanding the twisted versions of the algebras would be considering the Drinfeld presentations. Recently, the Drinfeld presentations of the twisted Yangians were found in \cite{lu2025affine,Lu_2025,2024arXiv240806981L,Lu_2026,lu2026twisted,huang2026quasi,lu2026classification} for various cases (see also \cite{lu2021drinfeld,zhang2022drinfeld,lu2024isomorphism,lu2025affine,lu2026shifted,li2026gklo,li2026shifted,przezdziecki2025drinfeld,Li_2025,li2025boundary,li2026compatibility} on the study of affine $\iota$quantum groups and \cite{neguct2026folding,appel2026quantum} for recent discussions on twisted quantum toroidal algebras).

In particular, it should be possible to extend the Drinfeld presentations obtained there to the affine cases. In analogy with the quiver BPS algebras, one could then consider the trigonometric and the elliptic versions as well. In this paper, we shall discuss the $\iota$quiver (BPS) algebras based on the Drinfeld presentations. In particular, we shall mainly focus on the cases for $\mathbb{C}\times\mathbb{C}^2/\mathbb{Z}_n$ of split type. They should be some twisted version/$\iota$ version of the quiver BPS algebras, including affine Yangians, quantum toroidal algebras, and elliptic quantum toroidal algebras, associated to $\widehat{\mathfrak{gl}}_n$ (See also, for example, \cite{Gavshin:2025akn,Galakhov:2026cwv,Galakhov:2026hdx} on the discussions on those associated to the finite types).

Of course, whether this really serves as some BPS algebras would require further investigations, and this explains the use of the brackets in the name $\iota$quiver (BPS) algebras. In fact, for the $n=1$ case, we shall discuss some representations whose basis would recover the combinatorial structures of certain counting in \cite{Kac:1999av}. Nevertheless, we shall refer to such algebras as $\iota$quiver algebras in what follows.

For the quiver BPS algebras, there are various properties that indicate their role as BPS algebras with connections to various fields, such as those discussed in \cite{Litvinov:2020zeq,Chistyakova:2021yyd,Kolyaskin:2022tqi,Bao:2022fpk,Galakhov:2022uyu,ueda2022affine,kodera2022coproduct,Bao:2022jhy,ueda2025example,ueda2024guay,Negut:2023iia,Bao:2023ece,ueda2026coproduct,ueda2025affine1,ueda2025affine2,Negut:2026lzn,Negut:2026ugx,Li:2023zub,Galakhov:2023aev,Galakhov:2023mak,Galakhov:2023mpc,Galakhov:2024mbz,Galakhov:2024foa,Galakhov:2024bzs,Galakhov:2024cry,Galakhov:2024zqn,Galakhov:2025phf,Matsuo:2023lky,Kimura:2023bxy,Kimura:2024xpr,Kimura:2024osv,Noshita:2025bzg,Kimura:2025lfo}. For the twisted versions, the connections between twisted Yangians and $\mathcal{W}$-algebras (of finite type) were discussed in \cite{brown2009twisted,lu2025shiftedtwistedyangiansfinite,lu2026representationsshiftedsuperyangians}. For the twisted affine Yangians constructed in \cite{ueda2025twisted}, the connection to the rectangular $\mathcal{W}$-algebra of type D was also studied there. There are also other approaches to studying  the BPS algebras from various geometric and physical perspectives. For instance, the Coulomb branch approach constructs associative algebras through the equivariant Borel--Moore homology of spaces associated with 3d $\mathcal{N}=4$ gauge theories, leading to quantised Coulomb branches and their connections to Yangians and shifted Yangians \cite{Braverman:2016pwk,Braverman:2016wma}. See also \cite{Chen:2025xoe} for relevant discussions. In particular, those with symmetrisers involving non-simply laced cases were considered in \cite{Nakajima:2019olw}. Besides, cohomological Hall algebras give the BPS algebras from the cohomology of moduli stacks of quiver representations on CY threefolds \cite{Kontsevich:2010px,Davison:2013nza,Davison:2016bjk,Rapcak:2020ueh,davison2025bpsalgebrasgeneralisedkacmoody,Butson:2023eid,botta2025okounkovsconjecturebpslie,schiffmann2024cohomologicalhallalgebrasquivers}. Moreover, the stable envelopes provide geometric ways to construct $R$-matrices and quantum groups acting on their cohomologies \cite{maulik2018quantumgroupsquantumcohomology,Aganagic:2016jmx,Okounkov_2021,dinkins2021ellipticstableenvelopesaffine,Botta_2023,Ishtiaque:2023acr,Cao:2025yng,Cao:2026pir,Cao:2026vwf,Dedushenko:2021mds,Dedushenko:2023qjq}. More physically, in the twisted M-theory setting, the BPS algebras are realised as operator algebras on M2- and M5-branes in the $\Omega$-background \cite{Costello:2013zra,Costello:2016nkh,Costello:2017fbo,Gaiotto:2019wcc,Oh:2020hph,Oh:2021wes,Gaiotto:2020dsq,Gaiotto:2023ynn,Ashwinkumar:2024vys,Ishtiaque:2024orn}. These perspectives could also provide various insights and shed light on the study of the BPS algebra structures for more general physical theories.

\paragraph{Organisation of the paper} The paper is organised as follows. In \cref{iquiveralg}, we introduce the $\iota$quiver algebras of split type for $\mathbb{C}\times\mathbb{C}^2/\mathbb{Z}_n$. Then we discuss the minimalistic presentations (for the rational cases) and the coideal structures. In \cref{involutions}, we mention how the involutions on the algebra side could be related to the orientifold actions on the quivers. Although many details are still not clear, it would be important to have a dictionary on this when identifying which algebras would arise for different orientifolded quivers. In particular, we shall argue that the generators of the $\iota$quiver algebras would still be labelled by the nodes of the parent quiver which do not have a one-to-one correspondence with the nodes of the orientifolded quiver. In \cref{representations}, we consider some representations of the $\iota$quiver algebras. This includes certain representations labelled by strict/symmetric partitions although one needs to take certain limits of the parameters of the algebras. This recovers the counting for the $\mathcal{N}=16$ super Yang--Mills quantum mechanics with $\SO$ and $\USp$ gauge groups. In \cref{quiverbpsalg}, we recall the definition of the quiver BPS algebras for the $\mathbb{C}\times\mathbb{C}^2/\mathbb{Z}_n$ cases.

\section{The \texorpdfstring{$\iota$Quiver}{iQuiver} Algebras}\label{iquiveralg}
In this section, we shall introduce the $\iota$quiver algebras associated to the quiver BPS algebras of $\widehat{\mathfrak{gl}}_n$. This could potentially be related to the brane tilings on $\mathbb{C}\times\mathbb{C}^2/\mathbb{Z}_n$ with certain orientifolds. The $\iota$quiver algebras considered here may be viewed as the affine version of the $\iota$quantum algebras of split type in the finite Lie algebra case. We will then discuss their various properties.

To see how the quiver data would enter the definition of the $\iota$quiver algebras, let us first write down some expressions for a general quiver. These do not serve as any definition of the possible ($\iota$)quiver (BPS) algebras for a general quiver.

Given a quiver $Q$ with the set of vertices $Q_0$ and the set of arrows $Q_1$, we shall denote the set of arrows from $a$ to $b$ as $\{a\rightarrow b\}$ (and the number of such arrows as $|a\rightarrow b|$) for any $a,b\in Q_0$. We also say that a node $a$ is bosonic/even (resp.~fermionic/odd) if $|a\rightarrow a|$ is odd (resp.~even). Such a $\mathbb{Z}_2$-grading would be denoted as $|a|=0$ (resp.~$|a|=1$). Define the bond factor\footnote{In the context of quiver BPS algebras, the bond factor would need some extra sign in particular for the chiral quivers. A possible way is to include $(-1)^{|b\rightarrow a|}$. Then the factor $N_{ab}(u)$ introduced below should also have this factor $(-1)^{|b\rightarrow a|}$. Here, we shall focus on the symmetric quivers, so we do not have to consider the extra sign.}
\begin{equation}
	\phi^{ab}(u)=\frac{\prod\limits_{I\in{\{a\rightarrow b\}}}\zeta(u+\epsilon_I)}{\prod\limits_{I\in{\{b\rightarrow a\}}}\zeta(u-\epsilon_I)}\;,
\end{equation}
where we associate some weights $\epsilon_I$ to the arrows $I$, and we have introduced the odd function\footnote{The elliptic functions are given by $\eta(\tau)=q^{1/24}\prod\limits_{k=1}^\infty(1-q^k)$ and $\theta_1(\tau,u)=-\ii q^{1/8}U^{1/2}\prod\limits_{k=1}^\infty\left(1-q^k\right)\left(1-Uq^k\right)\left(1-U^{-1}q^{k-1}\right)$ with $q=\mathrm{e}^{2\pi\ii\tau}$ and $U=\mathrm{e}^{2\pi\ii u}$.}
\begin{equation}
	\zeta(u)=\begin{cases}
		\frac{\ii\theta_1(\tau,u)}{q^{1/12}\eta(\tau)}\;,&\text{elliptic}\;,\\
		2\ii\sin(\pi u)\;,&\text{trigonometric}\;,\\
		u\;,&\text{rational}\;.
	\end{cases}
\end{equation}
Notice that $\phi^{ab}(u)\phi^{ba}(-u)=1$ (for symmetric quivers or with the extra sign included for general quivers). We shall also denote the numerator and the denominator of $\phi^{ab}(u)$ as $N^{ab}(u)$ and $M^{ab}(u)$ respectively.

Consider the currents $h_\pm^{(a)}(u)$ and $b^{(a)}(u)$ labelled by the nodes $a\in Q_0$, where $h_\pm^{(a)}(u)$ are always even while $b^{(a)}(u)$ have the same $\mathbb{Z}_2$-grading as the nodes $a$. Write the following relations\footnote{In the trigonometric and elliptic quiver BPS algebras, there can be an extra central extension. Here, we shall not consider this.}
\begin{subequations}
	\begin{align}
		&h^{(a)}_\varsigma(u)h^{(b)}_{\varsigma'}(v)=h^{(b)}_{\varsigma'}(v)h^{(a)}_\varsigma(u)\;,\quad\varsigma,\varsigma'\in\{\pm\}\;,\label{hh}\\
		&h^{(\tau(a))}_+(u)=h^{(a)}_-(-u)\;,\label{h}\\
		&h^{(a)}_\pm(u)b^{(b)}(v)=\left[\frac{\phi^{ab}(u-v)}{\phi^{a\tau(b)}(u+v)}\right]_\pm b^{(b)}(v)h^{(a)}_\pm(u)\;,\label{hb}\\
		&M^{ab}(u-v)b^{(a)}(u)b^{(b)}(v)-(-1)^{|a||b|}N^{ab}(u-v)b^{(b)}(v)b^{(a)}(u)\nonumber\\
		=&-\delta_{b\tau(a)}c_a\left(M^{ab}(u-v)+N^{ab}(u-v)\right)\delta(u+v)\left(h^{(a)}_+(u)-h^{(a)}_-(-v)\right)\;\label{bb}.
	\end{align}
\end{subequations}
Here, $c_a$ is some constant for each node $a$, and $[f(u)]_+$ (resp.~$[f(u)]_-$) denotes the Laurent expansion around $U=+\infty$ (resp.~0) with $U=\mathrm{e}^{2\pi\ii u}$ (which are only needed for the trigonometric and the elliptic cases). In the rational case, we would only have $h^{(a)}(u)$ without extra $\pm$ in the subscript. The $\delta$-function is defined as
\begin{equation}
	\delta(u)=\begin{cases}
		\frac{1}{u}\;,&\text{rational}\;,\\
		\sum\limits_{k\in\mathbb{Z}}U^k\;,&\text{trigonometric}\;,\\
		\frac{2\pi\ii}{\zeta'(0)}\sum\limits_{k\in\mathbb{Z}}U^k\;,&\text{elliptic}\;.
	\end{cases}
\end{equation}
In the above relations, there is also another $\tau$ (as explained below) which is different from the one in the elliptic functions. Both are standard notations, and it should be clear what $\tau$ refers to from the context. Therefore, we shall not distinguish the two notations here. The same comment applies to the letter $b$ (for currents/modes and nodes) as well.

Again, we emphasise that these relations are only heuristic. This is also reflected in the following explanations that are needed for these relations:
\begin{itemize}
	\item There is some involution $\tau$ (i.e., $\tau^2=\mathrm{id}$) that acts (at least) on $Q_0$. This originates from the involutions on the Dynkin diagrams. We will discuss such involutions in more detail shortly. Nevertheless, this already indicates that it is not clear how this could be applied to a general quiver (although we shall mainly focus on the cases with $\tau=\mathrm{id}$ in what follows)\footnote{The orientifold also introduces some involution on the quiver. However, this does not necessarily coincide with $\tau$ in the Dynkin cases although they should be related as we will discuss in \cref{involutions}.}.
	\item The equal signs are understood as the equality of the Taylor (resp.~Laurent) expansions of the currents in the rational\footnote{In particular, we have not included the boundary terms for the rational cases in the current relations. They will be restored when we write the relations in terms of the modes.} (resp.~trigonometric/elliptic\footnote{For an elliptic algebra $\mathcal{A}$, the infinite expansions should be well-defined in the $q$-adic topology by considering the completion $\lim\limits_{\leftarrow}\left(\mathcal{A}/q^N\mathcal{A}\right)(\!(U)\!)$ \cite{Foda:1994xm,Konno:2016fmh,Konno:2021zvl}.}) cases. Notice that we have not specified the expansions of the currents yet. In this paper, we will concentrate on the cases for symmetric (i.e., non-chiral) quivers\footnote{In the definition of quiver BPS algebras, certain balanced bond factors (rather than just the bond factor with the extra sign above) are needed to avoid fractional powers in the current expansions for chiral quivers as well \cite{Galakhov:2021vbo}.}.
	\item The $bb$ relations lead to some consistency conditions. By considering the same relations but with $b^{(a)}(u)$ and $b^{(b)}(v)$ exchanged (which should be consistent with the original relations multiplied by $(-1)^{|a||b|}$ on both sides), we find that
	\begin{equation}
		c_a=(-1)^{|a||\tau(a)|}c_{\tau(a)}\;.
	\end{equation}
	If we have some involution $\tau$ on the quiver BPS algebras\footnote{The quiver BPS algebras are reviewed in \cref{quiverbpsalg}.}, it is natural to expect this condition. Moreover, using \eqref{h}, we may rewrite the term $\delta(u+v)\left(h^{(a)}_+(u)-h^{(a)}_-(u)\right)=\delta(u+v)\left(h^{(a)}_+(u)-h^{(\tau(a))}_+(v)\right)$ in the trigonometric and the elliptic cases.
	\item Another natural expectation of $\tau$ is that it should also act on the arrows. In other words, $\{a\rightarrow\tau(b)\}$ and $\{\tau(a)\rightarrow b\}$ should not only have the same number of arrows but also the same weights $\epsilon_I$ for the corresponding arrows. Then $\phi^{a\tau(b)}(u)\phi^{b\tau(a)}(-u)=1$. As a result, the coefficients in the $hb$ relations would satisfy
	\begin{equation}
		\frac{\phi^{ab}(u-v)}{\phi^{a\tau(b)}(u+v)}\frac{\phi^{ba}(v-u)}{\phi^{b\tau(a)}(-u-v)}=1\;.
	\end{equation}
\end{itemize}
Of course, all these comments on the involution $\tau$ require a well-defined notion of $\tau$. For the cases of affine type A discussed below, the above conditions are all satisfied.

\subsection{Dramatis Personae}\label{def}
Consider the quiver gauge theory on D3 probing $\mathbb{C}\times\mathbb{C}^2/\mathbb{Z}_n$ (written in the $\mathcal{N}=1$ language in what follows). The quiver can be viewed as the tripled quiver of the affine Dynkin diagram of $A^{(1)}_{n-1}$, and the superpotential is the standard one for the tripled quiver. More concretely, the quiver has $n$ nodes labelled by $a\in\mathbb{Z}_n$. The arrows are given by $I_{a,a+1}:a\rightarrow a+1$, $I_{a+1,a}:a+1\rightarrow a$, and $I_{aa}:a\rightarrow a$ with weights $\epsilon_1$, $\epsilon_2$, and $\epsilon_3$ respectively\footnote{For $n=1$, we have a single-node quiver with three adjoints followed from $a\equiv a+1$. For $n=2$, the quiver has two pairs of bidirectional arrows between the two nodes, together with an adjoint on each node, which can be obtained by considering $a$ and $a-1\equiv a+1$.} for any $a$. They satisfy the Calabi--Yau condition $\epsilon_1+\epsilon_2+\epsilon_3=0$. We shall also use $q_i$ to denote the exponentiated weights $q_i=\mathrm{e}^{2\pi\ii\epsilon_i}$ such that $q_1q_2q_3=1$. The corresponding quiver BPS algebra $\mathtt{A}$ can then be written following \cref{quiverbpsalg}. Here, let us introduce an $\iota$quiver algebra associated to the quiver.
\begin{definition}
	Given the quiver $Q$ as above, we say that the algebra generated by $h^{(a)}_i$ and $b^{(a)}_i$ (with an extra subscript $\alpha$ in the elliptic case) subject to the relations \eqref{hh}$\sim$\eqref{bb} is an $\iota$quiver algebra. In particular, $|a|=0$ for all the nodes. Moreover, we set
	\begin{equation}
		\begin{cases}
			\epsilon_1=\epsilon_2=-\frac{1}{2}\epsilon_3\;,&\text{rational}\;,\\
			\epsilon_1\equiv\epsilon_2\pmod{\mathbb{Z}}\;,&\text{trigonometric}\;,\\
			\epsilon_1\equiv\epsilon_2\pmod{\mathbb{Z}+\tau\mathbb{Z}}\;,&\text{elliptic}\;,
		\end{cases}
	\end{equation}
	for $n\geq3$ (whose reason will be clear shortly). The currents are defined as
	\begin{subequations}
		\begin{align}
		&h^{(a)}_\pm(u)=\begin{cases}
		1+\sum\limits_{i\in\mathbb{Z}_{\geq0}}\frac{h^{(a)}_i}{u^{i+1}}\;,&\text{rational}\;,\\
		\sum\limits_{i\in\mathbb{Z}_{\geq0}}\frac{h^{(a)}_{\pm,i}}{U^{\pm i}}\;,&\text{trigonometric}\;,\\
		\sum\limits_{i\in\mathbb{Z}_{\geq0}}\frac{h^{(a)}_{\pm,i,0}}{U^{\pm i}}+\sum\limits_{i\in\mathbb{Z}}\sum\limits_{\alpha\in\mathbb{Z}_{>0}}\frac{h^{(a)}_{\pm,i,\alpha}}{U^{\pm i}}q^\alpha\;,&\text{elliptic}\;,
		\end{cases}\\
		&b^{(a)}(u)=\begin{cases}
		\sum\limits_{i\in\mathbb{Z}_{\geq0}}\frac{b^{(a)}_i}{u^{i+1}}\;,&\text{rational}\;,\\
		\sum\limits_{i\in\mathbb{Z}}\frac{b^{(a)}_i}{U^i}\;,&\text{trigonometric}\;,\\
		\sum\limits_{i\in\mathbb{Z}}\sum\limits_{\alpha\in\mathbb{Z}_{\geq0}}\frac{b^{(a)}_{i,\alpha}}{U^i}q^\alpha\;,&\text{elliptic}\;.
		\end{cases}
		\end{align}
	\end{subequations}
	For later convenience, we shall also write $h^{(a)}_{-1}=1$ in the rational case. This $\iota$quiver algebra with $\tau=\mathrm{id}$ will be denoted as ${}^\iota A_n$, and we would often omit the subscript $n$ when it would not cause any confusion. For the rational (trigonometric, resp.~elliptic) cases, we shall also denote them as ${}^\iota Y_n$ (${}^\iota T_n$, resp.~${}^\iota E_n$).
\end{definition}
This can be viewed as an affine version of the twisted Yangians (in the rational case) whose Drinfeld presentations were found in \cite{Lu_2025,2024arXiv240806981L,lu2026twisted}. In particular, it is said to be of split type when $\tau=\mathrm{id}$. In fact, it is already pointed out in \cite{Lu_2025} that the Drinfeld presentation therein remains valid for the affine ADE types. We will discuss the involutions in more detail in \cref{involutions}.

\paragraph{Mode relations in the rational cases} With the current expansions, we may also write down the relations in terms of the modes. For the trigonometric and the elliptic cases, it suffices to work with the currents here. For the rational cases, although we may take the rational limit of the trigonometric cases, it would be helpful to spell out their mode relations. For the quivers associated to $\mathbb{C}\times\mathbb{C}^2/\mathbb{Z}_n$ with $\tau=\mathrm{id}$, we have
\begin{subequations}
	\begin{align}
		&\left[h^{(a)}_i,h^{(b)}_j\right]=0\;,\label{hhcomm}\\
		&h^{(a)}_{2i}=0\;,\label{h2i}\\
		&[h_{i+6},b_j]-3[h_{i+4},b_{j+2}]+3[h_{i+2},b_{j+4}]-[h_i,b_{j+6}]+2\sigma_2([h_{i+4},b_j]-[h_i,b_{j+4}])\nonumber\\
		&+\sigma_2^2([h_{i+2},b_j]-[h_i,b_{j+2}])-\sigma_3^2[h_i,b_j]\nonumber\\
		=&2\sigma_3(3\{h_{i+2},b_{j+1}\}+\{h_i,b_{j+3}\}+\sigma_2\{h_i,b_{j+1}\})\;,\quad i\geq-6\;,~n=1\;,\label{hbn1}\\
		&\left[h^{(a)}_{i+2},b^{(a)}_j\right]-\left[h^{(a)}_i,b^{(a)}_{j+2}\right]-\epsilon_3^2\left[h^{(a)}_i,b^{(a)}_j\right]=2\epsilon_3\left\{h^{(a)}_i,b^{(a)}_{j+1}\right\}\;,\quad i\geq-2\;,~n=2\;,\label{hbn2aa}\\
		&\left[h^{(a)}_{i+4},b^{(a+1)}_j\right]-2\left[h^{(a)}_{i+2},b^{(a+1)}_{j+2}\right]+\left[h^{(a)}_i,b^{(a+1)}_{j+4}\right]-\left(\epsilon_1^2+\epsilon_2^2\right)\left[h^{(a)}_{i+2},b^{(a+1)}_j\right]\nonumber\\
		&+\left(2\epsilon_1\epsilon_2+\epsilon_3^2\right)\left[h^{(a)}_i,b^{(a+1)}_{j+2}\right]+\epsilon_1^2\epsilon_2^2\left[h^{(a)}_i,b^{(a+1)}_j\right]\nonumber\\
		=&2\epsilon_3\left(\left\{h^{(a)}_i,b^{(a+1)}_{j+3}\right\}-\left\{h^{(a)}_{i+2},b^{(a+1)}_{j+1}\right\}+\epsilon_1\epsilon_2\left\{h^{(a)}_i,b^{(a+1)}_{j+1}\right\}\right)\;,\quad i\geq-4\;,~n=2\;,\label{hbn2aa+1}\\
		&\left[h^{(a)}_{i+2},b^{(a)}_j\right]-\left[h^{(a)}_i,b^{(a)}_{j+2}\right]-\epsilon_3^2\left[h^{(a)}_i,b^{(a)}_j\right]=2\epsilon_3\left\{h^{(a)}_i,b^{(a)}_{j+1}\right\}\;,\quad i\geq-2\;,~n\geq3\;,\label{hbn3aa}\\
		&\left[h^{(a)}_{i+2},b^{(a\pm1)}_j\right]-\left[h^{(a)}_i,b^{(a\pm1)}_{j+2}\right]-\frac{1}{4}\epsilon_3^2\left[h^{(a)}_i,b^{(a\pm1)}_j\right]=-\epsilon_3\left\{h^{(a)}_i,b^{(a\pm1)}_{j+1}\right\}\;,\quad i\geq-2\;,~n\geq3\;,\label{hbn3aa+-1}\\
		&\left[h^{(a)}_i,b^{(b)}_j\right]=0\;,\quad b\notin \{a,a\pm1\}\;,~n\geq3\;,\label{hbn3ab}\\
		&\left[b_{i+3},b_j\right]-3\left[b_{i+2},b_{j+1}\right]+3\left[b_{i+1},b_{j+2}\right]-\left[b_i,b_{j+3}\right]+\sigma_2(\left[b_{i+1},b_j\right]-\left[b_i,b_{j+1}\right])-\sigma_3\left\{b_i,b_j\right\}\nonumber\\
		=&-4c(-1)^j(4h_{i+j+3}+\sigma_2h_{i+j+1})\;,\quad n=1\;,\label{bbn1}\\
		&\left[b^{(a)}_{i+1},b^{(a)}_j\right]-\left[b^{(a)}_i,b^{(a)}_{j+1}\right]-\epsilon_3\left\{b^{(a)}_i,b^{(a)}_j\right\}=-4c_a(-1)^jh^{(a)}_{i+j+1}\;,\quad n=2\;,\label{bbn2aa}\\
		&\left[b^{(a)}_{i+2},b^{(a+1)}_j\right]-2\left[b^{(a)}_{i+1},b^{(a+1)}_{j+1}\right]+\left[b^{(a)}_i,b^{(a+1)}_{j+2}\right]+\epsilon_3\left(\left\{b^{(a)}_{i+1},b^{(a+1)}_j\right\}-\left\{b^{(a)}_i,b^{(a+1)}_{j+1}\right\}\right)\nonumber\\
		&+\epsilon_1\epsilon_2\left[b^{(a)}_i,b^{(a+1)}_j\right]=0\;,\quad n=2\label{bbn2aa+1}\;,\\
		&\left[b^{(a)}_{i+1},b^{(a)}_j\right]-\left[b^{(a)}_i,b^{(a)}_{j+1}\right]-\epsilon_3\left\{b^{(a)}_i,b^{(a)}_j\right\}=-4c_a(-1)^jh^{(a)}_{i+j+1}\;,\quad n\geq3\;,\label{bbn3aa}\\
		&\left[b^{(a)}_{i+1},b^{(a\pm1)}_j\right]-\left[b^{(a)}_i,b^{(a\pm1)}_{j+1}\right]+\frac{1}{2}\epsilon_3\left\{b^{(a)}_i,b^{(a\pm1)}_j\right\}=0\;,\quad n\geq3\;,\label{bbn3aa+-1}\\
		&\left[b^{(a)}_i,b^{(b)}_j\right]=0\;,\quad b\notin\{a,a\pm1\}\;,~n\geq3\label{n3bbab}\;.
	\end{align}
\end{subequations}
For the $\mathbb{C}^3$ case, we have omitted the label of the node since there is only one node in the quiver. Moreover, $\sigma_2=\epsilon_1\epsilon_2+\epsilon_1\epsilon_3+\epsilon_2\epsilon_3$ and $\sigma_3=\epsilon_1\epsilon_2\epsilon_3$.

Now, let us explain the choice $\epsilon_1=\epsilon_2$ in the $n\geq3$ cases. If $\epsilon_1$ and $\epsilon_2$ were to be different, then the $hb$ relations in terms of modes in the rational algebras would read
\begin{equation}
	\left[h^{(a)}_{i+2},b^{(a+1)}_j\right]+(\epsilon_1-\epsilon_2)\left[h^{(a)}_{i+1},b^{(a+1)}_j\right]-\left[h^{(a)}_i,b^{(a+1)}_{j+2}\right]-\epsilon_1\epsilon_2\left[h^{(a)}_i,b^{(a+1)}_j\right]=(\epsilon_1+\epsilon_2)\left\{h^{(a)}_i,b^{(a+1)}_{j+1}\right\}
\end{equation}
for neighbouring nodes. Since $h^{(a)}_0=h^{(a)}_2=0$, taking $i=0$ yields
\begin{equation}
	(\epsilon_1-\epsilon_2)\left[h^{(a)}_1,b^{(a\pm1)}_j\right]=0\;.
\end{equation}
Moreover, taking $i=-1$, we have
\begin{equation}
	\left[h^{(a)}_1,b^{(a\pm1)}_j\right]=2(\epsilon_1+\epsilon_2)b^{(a\pm1)}_{j+1}\;.
\end{equation}
Therefore,
\begin{equation}
	2(\epsilon_1-\epsilon_2)(\epsilon_1+\epsilon_2)b^{(a\pm1)}_{j+1}=0\;.
\end{equation}
If $\epsilon_1\neq\pm\epsilon_2$, all the generators $b^{(a)}_{i\geq1}$ would be zero, leaving us an almost trivial algebra. In order to keep the parameter $\epsilon_3$ generic and non-zero so that the algebra would not further degenerate (as can be seen from $\phi^{aa}(u)$), this forces $\epsilon_1=\epsilon_2$. We may also use the current relations to show this as discussed below.

\paragraph{Trigonometric and elliptic cases} There are also two important comments on the trigonometric and the elliptic cases:
\begin{itemize}
	\item In the rational cases, the even modes of $h^{(a)}(u)$ vanish. Likewise, in the trigonometric and elliptic cases, due to $h^{\tau(a)}_+(u)=h^{(a)}_-(-u)$ (with $\tau=\mathrm{id}$), or equivalently, $h^{(a)}_+(U)=h^{(a)}_-\left(U^{-1}\right)$, the modes in $h^{(a)}_+(u)$ and $h^{(a)}_-(u)$ would coincide. Henceforth, we would also write $h^{(a)}(u)=h^{(a)}_+(u)$ and $h^{(a)}(-u)=h^{(a)}_-(u)$ for simplicity.
	\item The parameters $\epsilon_1$ and $\epsilon_2$ are still not independent in the trigonometric and the elliptic cases for $n\geq3$. This can be seen from the current relations directly. Since $h^{(a)}_+(U)=h^{(a)}_-\left(U^{-1}\right)$, the coefficients should match when one looks at the relation of $h^{(a)}(-u)$ and $b^{(a\pm1)}(v)$:
	\begin{equation}
		\left[\frac{\phi^{ab}(1/UV)}{\phi^{ab}(V/U)}\right]_-=\left[\frac{\phi^{ab}(U/V)}{\phi^{ab}(UV)}\right]_-\;,
	\end{equation}
	where the left hand side may also be written as $\left[\frac{\phi^{ab}(U/V)}{\phi^{ab}(UV)}\right]_+\bigg|_{U\rightarrow U^{-1}}$. Taking the difference, we have
	\begin{align}
		\frac{\phi^{ab}(1/UV)}{\phi^{ab}(V/U)}-\frac{\phi^{ab}(U/V)}{\phi^{ab}(UV)}=&\frac{1}{\zeta(-u-v-\epsilon_2)\zeta(-u+v+\epsilon_1)\zeta(u-v-\epsilon_2)\zeta(u+v+\epsilon_1)}\nonumber\\
		&(\zeta(-u-v+\epsilon_1)\zeta(-u+v-\epsilon_2)\zeta(u-v-\epsilon_2)\zeta(u+v+\epsilon_1)\nonumber\\
		&-\zeta(u-v+\epsilon_1)\zeta(u+v-\epsilon_2)\zeta(-u-v-\epsilon_2)\zeta(-u+v+\epsilon_1))\nonumber\\
		=&\frac{\zeta(\epsilon_1-\epsilon_2)\zeta(\epsilon_1+\epsilon_2)\zeta(2u)\zeta(2v)}{\zeta(-u-v-\epsilon_2)\zeta(-u+v+\epsilon_1)\zeta(u-v-\epsilon_2)\zeta(u+v+\epsilon_1)}\;.
	\end{align}
	In the second equality, we have used the Riemann relation (see, for example, \cite[(29)]{kajiwara2004point}). Therefore, $\zeta(\epsilon_1-\epsilon_2)$ or $\zeta(\epsilon_1+\epsilon_2)$ should be zero. Again, since we do not want the algebras to further degenerate, this requires $\zeta(\epsilon_1-\epsilon_2)=0$. In the trigonometric cases\footnote{Alternatively, one may take the expansions on the two sides. This gives $1+\left(q_2-q_1^{-1}\right)\left(V-V^{-1}\right)U+\dots=1+\left(q_1-q_2^{-1}\right)\left(V-V^{-1}\right)U+\dots$, where the ellipses only contain terms of higher orders of $U$. Therefore, $q_2-q_1^{-1}=q_1-q_2^{-1}$, leading to $q_1=q_2$ as $q_1\neq q_2^{-1}$.}, this gives $\epsilon_1\equiv\epsilon_2\pmod{\mathbb{Z}}$, i.e., $q_1=q_2$. Likewise, in the elliptic cases, we have $\epsilon_1\equiv\epsilon_2\pmod{\mathbb{Z}+\tau\mathbb{Z}}$. In what follows, we shall mainly take $\epsilon_1=\epsilon_2$ for simplicity (in particular for the elliptic case due to the quasi-periodicity of the elliptic functions).
\end{itemize}

\paragraph{Serre relations} For quiver BPS algebras, there are also relations whose patterns would vary depending on different quivers, as opposed to the quadratic relations. These are known as the Serre relations. Although it is more difficult to determine the Serre relations, the quiver BPS algebras $\mathtt{A}$ which are reduced by the Serre relations are supposed to have nicer behaviours compared to the non-reduced ones $A$ in the following sense. One would expect that the reduced algebras would act faithfully on their crystal representations, and in the trigonometric cases, this is also reflected by the actions of the non-reduced algebras on these representations factoring through the shuffle algebra realisations of the quantum toroidal algebras. Moreover, the non-reduced affine Yangians associated to the affine super Dynkin cases with non-isotropic nodes do not seem to be isomorphic under Seiberg duality as observed in \cite{Bao:2023ece}. Including the right Serre relations could possibly render them isomorphic. Some discussions on finding the Serre relations for general quivers can be found in \cite{Galakhov:2021vbo,Negut:2023iia}. See also \cite{Negut:2026lzn,Negut:2026ugx} for the recent study on the super cases.

As certain $\iota$quiver algebras built out of the quiver BPS algebras\footnote{The meaning of this will be explained shortly.}, we would also need the Serre relations.
\begin{definition}
	We define an $\iota$quiver algebra ${}^\iota\mathtt{A}_n$ reduced from the above algebra ${}^\iota A_n$ given by ${}^\iota\mathtt{A}_n={}^\iota A_n/\langle\text{Serre relations}\rangle$. For the rational (trigonometric, resp.~elliptic) cases, we shall also denote them as ${}^\iota\mathtt{Y}_n$ (${}^\iota\mathtt{T}_n$, resp.~${}^\iota\mathtt{E}_n$).
\end{definition}

Define $\kappa=1$ for the rational cases and $\kappa=q_3^{1/2}$ otherwise. The Serre relations are listed as follows:
\begin{subequations}
	\begin{align}
		&\mathrm{Sym}_{u_1,u_2,u_3}\zeta_0(2(u_2-u_3))\mathfrak{D}([b(u_1),[b(u_2),b(u_3)]])=0\;,\quad n=1\;,\\
		&\mathrm{Sym}_{u_1,u_2,u_3}\mathfrak{D}\left(\left[b^{(a)}(u_1),\left[b^{(a)}(u_2),\left[b^{(a)}(u_3),b^{(a+1)}(v)\right]_{\kappa^2}\right]\right]_{\kappa^{-2}}\right)=0\;,\quad n=2\;,\\
		&\mathrm{Sym}_{u_1,u_2}\mathfrak{D}\left(\left[b^{(a)}(u_1),\left[b^{(a)}(u_2),b^{(a\pm1)}(v)\right]_{\kappa^{-1}}\right]_\kappa\right)=0\;,\quad n\geq3\;,
	\end{align}
\end{subequations}
where $\zeta_0(u)=\zeta(u)|_{q=0}$ indicates the $q^0$ term in the elliptic case (i.e., its trigonometric limit), and $[X,Y]_\mathfrak{q}=XY-\mathfrak{q}YX$. The factor $\mathfrak{D}$ is linear and is given by\footnote{Notice that $\left(1+\phi^{a_sa_s}(-2u_s)\right)\prod\limits_{r<t<s}\frac{1}{\phi^{a_sa_t}(u_s-u_t)}$ is understood as the expansion at large $U_s$.}
\begin{align}
	\mathfrak{D}\left(b^{(a_1)}(u_1)\dots b^{(a_m)}(u_m)\right)=&\left(\prod_{1\leq r<s\leq m}\prod_{I\in\{a_r\rightarrow a_s\}}\frac{\zeta_0(u_r-u_s+\epsilon_I)}{\zeta(u_r-u_s+\epsilon_I)}\right)\nonumber\\
	&\left(b^{(a_1)}(u_1)\dots b^{(a_m)}(u_m)\right.\nonumber\\
	&-\sum_{\substack{1\leq r<s\leq m\\a_r=a_s}}c_{a_s}\delta(u_r+u_s)\left(\left(1+\phi^{a_sa_s}(-2u_s)\right)\prod_{r<t<s}\frac{1}{\phi^{a_sa_t}(u_s-u_t)}\right)\nonumber\\
	&\left.b^{(a_1)}(u_1)\dots b^{(a_{r-1})}(u_{r-1})h^{(a_s)}(u_s)b^{(a_{r+1})}(u_{r+1})\dots\widehat{b}^{(a_s)}(u_s)\dots b^{(a_m)}(u_m)\right)\;.
\end{align}
The product $x_1\dots\widehat{x}_s\dots x_m$ indicates that the term $x_s$ is skipped. In the rational cases, the Serre relations in terms of the modes can be written as
\begin{subequations}
	\begin{align}
		&\mathrm{Sym}_{k_1,k_2,k_3}[b_{k_1},[b_{k_2},b_{k_3+1}]]=-\frac{c}{2}\sum_s(-1)^{k_j}C_{k_i+k_j,k_s}\;,\quad n=1\;,\label{serren1}\\
		&\mathrm{Sym}_{k_1,k_2,k_3}\left[b^{(a)}_{k_1},\left[b^{(a)}_{k_2},\left[b^{(a)}_{k_3},b^{(a+1)}_r\right]\right]\right]\nonumber\\
		=&c_a\sum_s(-1)^{k_j}\left(-\left[H^{(a)}_{k_i+k_j},\left[b^{(a)}_{k_s+1},b^{(a+1)}_r\right]\right]-2\left[b^{(a)}_{k_s+1},\left[H^{(a)}_{k_i+k_j},b^{(a+1)}_r\right]\right]-2\epsilon_3\left\{H^{(a)}_{k_i+k_j},\left[b^{(a)}_{k_s},b^{(a+1)}_r\right]\right\}\right.\nonumber\\
		&+\epsilon_3\left\{b^{(a)}_{k_s},\left[H^{(a)}_{k_i+k_j},b^{(a+1)}_r\right]\right\}+\frac{3}{\epsilon_1\epsilon_2}\left(\left[H^{(a)}_{k_i+k_j+2},\left[b^{(a)}_{k_s+1},b^{(a+1)}_r\right]-\left[b^{(a)}_{k_s},b^{(a+1)}_{r+1}\right]+\epsilon_3\left\{b^{(a)}_{k_s},b^{(a+1)}_r\right\}\right]\right.\nonumber\\
		&\left.\left.-\left[H^{(a)}_{k_i+k_j},\left[b^{(a)}_{k_s+2},b^{(a+1)}_{r+1}\right]-\left[b^{(a)}_{k_s+1},b^{(a+1)}_{r+2}\right]+\epsilon_3\left\{b^{(a)}_{k_s+1},b^{(a+1)}_{r+1}\right\}\right]\right)\right)\;,\quad n=2\;,\label{serren2}\\
		&\mathrm{Sym}_{k_1,k_2}\left[b^{(a)}_{k_1},\left[b^{(a)}_{k_2},b^{(a\pm1)}_r\right]\right]=2c_a(-1)^{k_1}\left[h^{(a)}_{k_1+k_2+1},b^{(a\pm1)}_{r-1}\right]\;,\quad n\geq3\label{serren3}\;,
	\end{align}
\end{subequations}
where
\begin{subequations}
	\begin{align}
	&C_{m\leq0,r}=0\;,~C_{m>0,r}=\frac{8}{3}[4h_{m+1}+\sigma_2h_{m-1},b_r]-\frac{1}{3}C_{m-2,r+2}-\frac{\sigma_2}{3}C_{m-2,r}\;,\\
	&H^{(a)}_m=\sum_{l=0}^{\lfloor m/2\rfloor}\left(\frac{\epsilon_3}{2}\right)^{2l}h^{(a)}_{m-1-2l}\;,\\
	&\{i,j,s\}=\{1,2,3\}\;,~i<j\;.
\end{align}
\end{subequations}
Moreover, \eqref{serren3} is understood as applying to both $r\geq1$ and $r=0$. For $r=0$, we introduce the formal notation $b^{(a)}_{-1}$ (which is not a new generator) determined by
\begin{equation}
	\left[h_{k+2}^{(a)},b_{-1}^{(a\pm1)}\right]=\left[h_k^{(a)},b_1^{(a\pm1)}\right]-\epsilon_3\left\{h_k^{(a)},b_0^{(a\pm1)}\right\}+\frac{\epsilon_3^2}{4}\left[h_k^{(a)},b_{-1}^{(a\pm1)}\right]\;,k\geq-1\;,\text{ and }\left[h_{-1}^{(a)},b_{-1}^{(a\pm1)}\right]=0\;.
\end{equation}
Alternatively, we may write the Serre relations for $n\geq3$ as (see, for example, \cite[Remark 5.2]{Lu_2025})
\begin{equation}
	\mathrm{Sym}_{k_1,k_2}\left[b^{(a)}_{k_1},\left[b^{(a)}_{k_2},b^{(a\pm1)}_r\right]\right]=2c_a(-1)^{k_1}\sum_{j=0}^{\lfloor(k_1+k_2)/2\rfloor}\left(\frac{\epsilon_3}{2}\right)^{2j}\left(\left[h_{k_1+k_2-2j-1}^{(a)},b_{r+1}^{(a\pm1)}\right]-\epsilon_3\left\{h_{k_1+k_2-2j-1}^{(a)},b_r^{(a\pm1)}\right\}\right)\;.\label{serren3alternative}
\end{equation}

It is straightforward to see that the algebras are independent of generic $c_a$ by taking the rescaling $b^{(a)}_i\rightarrow c_a^{1/2}b^{(a)}_i$.
\begin{proposition}
	The $\iota$quiver algebra ${}^\iota\mathtt{A}_n$ (and hence ${}^\iota A_n$) is isomorphic for different $c_a\neq0$ for all $a$.
\end{proposition}

The $\iota$quiver algebras ${}^\iota\mathtt{A}_n$ may be viewed as some twisted affine Yangians, $\iota$quantum toroidal algebras, or elliptic $\iota$quantum toroidal algebras of split type. For example, in the rational cases for $n\geq3$, one may take the Dynkin diagram/quiver of finite type with $\epsilon_3=-1/2$ and $c_a=1/2$. This recovers the Drinfeld presentation of the twisted Yangians associated to the finite Lie algebras as given in \cite{Lu_2025}, and that this extends to the affine cases has already been pointed out therein.

\subsection{A Minimalistic Presentation}\label{minimalistic}
For the rational cases, when working with the infinitely many modes with infinitely many relations, it would be helpful to find a minimalistic presentation that is finite. For the quiver Yangians, some cases were discussed in \cite{guay2018coproduct,ueda2023construction,Bao:2022jhy,Bao:2023ece}. For the twisted Yangians associated to the finite Lie algebras, such presentations were obtained in \cite{Lu_2026,huang2026quasi,lu2026classification}. Here, we shall consider the minimalistic presentations of the twisted affine Yangians at generic parameters in this paper.

\paragraph{The \texorpdfstring{$n\geq3$}{n>=3} cases} Let us start with the cases with $n\geq3$. Using the $hb$ and the $bb$ relations of the same node (with $i=-1$ in \eqref{hbn3aa} and $i=j$ in \eqref{bbn3aa}), we notice that
\begin{subequations}
	\begin{align}
		&\left[h^{(a)}_1,b^{(a)}_j\right]=4\epsilon_3b^{(a)}_{j+1}\;,\\
		&2\left[b^{(a)}_{i+1},b^{(a)}_i\right]-2\epsilon_3\left(b^{(a)}_i\right)^2=-4c_a(-1)^ih^{(a)}_{2i+1}\;.
	\end{align}
\end{subequations}
Therefore, the higher modes can be inductively obtained by
\begin{subequations}
	\begin{align}
		&b^{(a)}_k=\left(\frac{1}{4\epsilon_3}\right)^k\mathrm{ad}_{h^{(a)}_1}^kb^{(a)}_0\;,\label{higherbn3}\\
		&h^{(a)}_{2k+1}=\frac{(-1)^k}{2c_a}\left(\epsilon_3\left(b^{(a)}_k\right)^2-\left[b^{(a)}_{k+1},b^{(a)}_k\right]\right)\label{higherhn3}\;.
	\end{align}
\end{subequations}
We also recall that $h^{(a)}_{2k}=0$. In other words, the $\iota$quiver algebras are finitely generated.

Now, let us show that they are finitely presented.
\begin{theorem}
	The $\iota$quiver algebra ${}^\iota\mathtt{Y}_{n\geq3}$ is generated by $h^{(a)}_1$, $b^{(a)}_0$, $b^{(a)}_1$ with the relations
	\begin{subequations}
		\begin{align}
			&\left[h^{(a)}_1,h^{(b)}_1\right]=0\;,\\
			&\left[h^{(a)}_1,b^{(b)}_0\right]=2\epsilon_3A_{ab}b^{(b)}_1\;,\\
			&\left[b^{(a)}_1,b^{(a)}_0\right]=\epsilon_3\left(b^{(a)}_0\right)^2-2c_ah^{(a)}_1\;,\\
			&\left[b^{(a)}_1,b^{(a\pm1)}_0\right]-\left[b^{(a)}_0,b^{(a\pm1)}_1\right]=-\frac{1}{2}\epsilon_3\left\{b^{(a)}_0,b^{(a\pm1)}_0\right\}\;,\\
			&\left[b^{(a)}_0,b^{(b)}_0\right]=0\;,\quad A_{ab}=0\;,\\
			&\left[b^{(a)}_0,\left[b^{(a)}_0,b^{(a\pm1)}_0\right]\right]=-2c_a\epsilon_3b^{(a\pm1)}_0\;,
		\end{align}
	\end{subequations}
	where $A_{ab}$ are the elements in the Cartan matrix, i.e., $A_{ab}$ is $2$ ($-1$, resp.~$0$) for $b=a$ ($b=a\pm1$, resp.~otherwise).
\end{theorem}
\begin{proof}
	To prove this, we need to show that all the relations involving the higher modes in \eqref{higherbn3} and \eqref{higherhn3} can be obtained from the relations among $h^{(a)}_1$, $b^{(a)}_0$, $b^{(a)}_1$. First, we notice that the higher modes of $b^{(a\pm1)}(u)$ can also be generated using $h^{(a)}_1$. This can be shown by induction. For $b^{(a\pm1)}_1$, it is one of the given relations. Suppose that $\left[h^{(a)}_1,b^{(a\pm1)}_r\right]=-2\epsilon_3b^{(a\pm1)}_{r+1}$. Then
\begin{align}
	\left[h^{(a)}_1,b^{(a\pm1)}_{r+1}\right]=&\frac{1}{4\epsilon_3}\left[h^{(a)}_1,\left[h^{(a\pm1)}_1,b^{(a\pm1)}_r\right]\right]=\frac{1}{4\epsilon_3}\left[h^{(a\pm1)}_1,\left[h^{(a)}_1,b^{(a\pm1)}_r\right]\right]\nonumber\\
	=&-\frac{1}{2}\left[h^{(a\pm1)}_1,b^{(a\pm1)}_{r+1}\right]=-2\epsilon_3b^{(a\pm1)}_{r+2}\;.\label{ha1ba+-1}
\end{align}
For $b\neq a,a\pm1$, it is straightforward to see that $b^{(b)}_r$ commutes with $h^{(a)}_1$ for any $r$.

Now, let us consider all the relations for $A_{ab}=0$. Again, we prove this by induction. Let us only write the proof of the $bb$ relation here, and the $hh$ and $hb$ relations can be proven in a similar manner. Suppose that $\left[b^{(a)}_r,b^{(b)}_s\right]=0$. Then
\begin{equation}
	\left[b^{(a)}_{r+1},b^{(b)}_s\right]=\frac{1}{4\epsilon_3}\left[\left[h^{(a)}_1,b^{(a)}_r\right],b^{(b)}_s\right]=\frac{1}{4\epsilon_3}\left(\left[h^{(a)}_1,\left[b^{(a)}_r,b^{(b)}_s\right]\right]-\left[b^{(a)}_r,\left[h^{(a)}_1,b^{(b)}_s\right]\right]\right)=0\;.
\end{equation}

Next, let us prove the $bb$ relation for $b=a\pm1$. Suppose that this is true at level $(r,s)$, that is,
\begin{equation}
	\left[b^{(a)}_{r+1},b^{(a\pm1)}_s\right]-\left[b^{(a)}_r,b^{(a\pm1)}_{s+1}\right]+\frac{1}{2}\epsilon_3\left\{b^{(a)}_r,b^{(a\pm1)}_s\right\}=0\;.
\end{equation}
Then apply $2\mathrm{ad}_{h^{(a)}_1}+\mathrm{ad}_{h^{(a\pm1)}_1}$ to this relation gives
\begin{equation}
	6\epsilon_3\left(\left[b^{(a)}_{r+2},b^{(a\pm1)}_s\right]-\left[b^{(a)}_{r+1},b^{(a\pm1)}_{s+1}\right]+\frac{1}{2}\epsilon_3\left\{b^{(a)}_{r+1},b^{(a\pm1)}_s\right\}\right)=0\;,
\end{equation}
which is the relation at level $(r+1,s)$. The one at level $(r,s+1)$ can be proven in the same manner by applying $2\mathrm{ad}_{h^{(a\pm1)}_1}+\mathrm{ad}_{h^{(a)}_1}$.

For $b=a$, suppose that the $bb$ relation is true at levels $(r,s)$ with $r+s\leq2k-1$. At total level $r+s=2k-1$, the relation reads
\begin{equation}
	\left[b^{(a)}_{r+1},b^{(a)}_s\right]-\left[b^{(a)}_r,b^{(a)}_{s+1}\right]-\epsilon_3\left\{b^{(a)}_r,b^{(a)}_s\right\}=0\;\label{supposebb2k-1}.
\end{equation}
Then apply $\mathrm{ad}_{h^{(a)}_1}$ yields
\begin{equation}
	\left[b^{(a)}_{r+2},b^{(a)}_s\right]-\left[b^{(a)}_{r+1},b^{(a)}_{s+1}\right]-\epsilon_3\left\{b^{(a)}_{r+1},b^{(a)}_s\right\}=-\left[b^{(a)}_{r+1},b^{(a)}_{s+1}\right]+\left[b^{(a)}_r,b^{(a)}_{s+2}\right]+\epsilon_3\left\{b^{(a)}_r,b^{(a)}_{s+1}\right\}\;.
\end{equation}
In particular, the left hand side of this equation would pick up a minus sign every time the level changes as $(i,j)\rightarrow(i-1,j+1)$ at total level $2k-1$. Keep doing this sign change until one reaches the level $(k,k)$. We get
\begin{equation}
	\left[b^{(a)}_{r+2},b^{(a)}_s\right]-\left[b^{(a)}_{r+1},b^{(a)}_{s+1}\right]-\epsilon_3\left\{b^{(a)}_{r+1},b^{(a)}_s\right\}=(-1)^{r+1-k}\left(2\left[b^{(a)}_{k+1},b^{(a)}_k\right]-2\epsilon_3\left(b^{(a)}_k\right)^2\right)=-4c_a(-1)^{r+1}h^{(a)}_{2k+1}\;.\label{bbaa2k}
\end{equation}
This proves the $bb$ relation at even total level $2k$.

Before proving the $bb$ relation for $b=a$ at odd total level, let us prove the $hb$ relation for $b=a\pm1$. With the same assumption in \eqref{supposebb2k-1}, we have
\begin{align}
	&\left[\left[b^{(a)}_{k+1},b^{(a)}_k\right]-\epsilon_3\left(b^{(a)}_k\right)^2,b^{(a\pm1)}_s\right]\nonumber\\
	=&\left[b^{(a)}_{k+1},\left[b^{(a)}_k,b^{(a\pm1)}_s\right]\right]-\left[b^{(a)}_k,\left[b^{(a)}_{k+1},b^{(a\pm1)}_s\right]\right]-\epsilon_3\left\{b^{(a)}_k,\left[b^{(a)}_k,b^{(a\pm1)}_s\right]\right\}\nonumber\\
	=&-\left[\left[b^{(a)}_k,b^{(a)}_{k-1}\right],b^{(a\pm1)}_{s+2}\right]-\frac{1}{4}\epsilon_3^2\left[\left[b^{(a)}_k,b^{(a)}_{k-1}\right],b^{(a\pm1)}_s\right]+\epsilon_3\left\{\left[b^{(a)}_k,b^{(a)}_{k-1}\right],b^{(a\pm1)}_{s+1}\right\}\nonumber\\
	&+\epsilon_3\left[\left(b^{(a)}_{k-1}\right)^2,b^{(a\pm1)}_{s+2}\right]+\frac{1}{4}\epsilon_3^3\left[\left(b^{(a)}_{k-1}\right)^2,b^{(a\pm1)}_s\right]-\epsilon_3^2\left\{\left(b^{(a)}_{k-1}\right)^2,b^{(a\pm1)}_{s+1}\right\}\;.
\end{align}
Using \eqref{higherhn3}, we get the $hb$ relation for $b=a\pm1$:
\begin{equation}
	\left[h^{(a)}_{2k+1},b^{(a\pm1)}_s\right]-\left[h^{(a)}_{2k-1},b^{(a\pm1)}_{s+2}\right]-\frac{1}{4}\epsilon_3^2\left[h^{(a)}_{2k-1},b^{(a\pm1)}_s\right]=-\epsilon_3\left\{h^{(a)}_{2k-1},b^{(a\pm1)}_{s+1}\right\}\;.
\end{equation}

This relation can be rewritten as
\begin{equation}
	\left[h^{(a)}_{2k+1},b^{(a\pm1)}_s\right]=\left[h^{(a)}_{2k-1},b^{(a\pm1)}_{s+2}\right]+\frac{1}{4}\epsilon_3^2\left[h^{(a)}_{2k-1},b^{(a\pm1)}_s\right]-\epsilon_3\left[h^{(a)}_{2k-1},b^{(a\pm1)}_{s+1}\right]-2\epsilon_3b^{(a\pm1)}_{s+1}h^{(a)}_{2k-1}\;.
\end{equation}
Using  the $h^{(a)}_1b^{(a\pm1)}_r$ relation \eqref{ha1ba+-1} inductively, we have
\begin{equation}
	\left[h^{(a)}_{2k+1},b^{(b)}_s\right]=-2\epsilon_3\sum_{l=0}^k\sum_{j=0}^{2l}\binom{2l}{j}\left(-\frac{\epsilon_3}{2}\right)^{2l-j}b^{(b)}_{s+j+1}h^{(a)}_{2k-2l-1}\label{habbsum}
\end{equation}
(where $b=a\pm1$). There is another equation involving some sum over quadratic products of modes. It is obtained from the $bb$ relation \eqref{bbaa2k}:
\begin{equation}
	\left[b^{(b)}_{j+1},b^{(b)}_{r+s-1-j}\right]-\left[b^{(b)}_j,b^{(b)}_{r+s-j}\right]=\epsilon_3\left\{b^{(b)}_j,b^{(b)}_{r+s-1-j}\right\}-4c_b(-1)^jh^{(b)}_{r+s}\;,
\end{equation}
where $r+s=2k+1$. Summing all such equations for $j=r,\dots,s-1$, we get
\begin{equation}
	\left[b^{(b)}_r,b^{(b)}_s\right]=-\epsilon_3\sum_{j=r}^{s-1}b^{(b)}_jb^{(b)}_{r+s-1-j}+2c_b(-1)^rh^{(b)}_{r+s}
\end{equation}
after some simplifications. Applying $\mathrm{ad}_{h^{(b)}_1}$, together with the Jacobi identity and the $h^{(b)}_1b^{(b)}_i$ relation, yields
\begin{equation}
	\left[b^{(b)}_{r+1},b^{(b)}_{s'-1}\right]+\left[b^{(b)}_r,b^{(b)}_{s'}\right]=-\epsilon_3\sum_{j=r}^{s'-2}\left(b^{(b)}_{j+1}b^{(b)}_{r+s'-2-j}+b^{(b)}_jb^{(b)}_{r+s'-1-j}\right)+\frac{1}{2\epsilon_3}c_b(-1)^r\left[h^{(b)}_1,h^{(b)}_{2k+1}\right]\;,
\end{equation}
where $s'=s+1$ so that $r+s'=2k+2$. This leads to
\begin{equation}
	\left[b^{(b)}_r,b^{(b)}_{s'}\right]=-\epsilon_3\sum_{j=r}^{s'-1}b^{(b)}_jb^{(b)}_{r+s'-1-j}+\frac{1}{4\epsilon_3}c_b(-1)^r(s'-r)\left[h^{(b)}_1,h^{(b)}_{2k+1}\right]\label{bbrbbssum}
\end{equation}
by induction following the order $(r,s')=(k+1,k+1)\rightarrow(k,k+2)\rightarrow\dots$. Now, take $(r,s')=(0,2k+2)$ and $(1,2k+1)$ respectively in \eqref{bbrbbssum}. We have the following identity:
\begin{align}
	&\epsilon_3\left\{b^{(b)}_{2k+1},b^{(b)}_0\right\}-\left[b^{(b)}_{2k+2},b^{(b)}_0\right]-\left[b^{(b)}_1,b^{(b)}_{2k+1}\right]\nonumber\\
	=&\epsilon_3b^{(b)}_{2k+1}b^{(b)}_0+\epsilon_3b^{(b)}_0b^{(b)}_{2k+1}-\epsilon_3\sum_{l=0}^{2k+1}b^{(b)}_lb^{(b)}_{2k+1-l}+\epsilon_3\sum_{l=1}^{2k}b^{(b)}_lb^{(b)}_{2k+1-l}+\left(\frac{(2k+2)c_b}{4\epsilon_3}+\frac{2kc_b}{4\epsilon_3}\right)\left[h^{(b)}_1,h^{(b)}_{2k+1}\right]\nonumber\\
	=&\frac{(2k+1)c_b}{2\epsilon_3}\left[h^{(b)}_1,h^{(b)}_{2k+1}\right]\;.\label{bbbbj2k}
\end{align}
Applying $\mathrm{ad}_{h^{(a)}_{2k+1}}$ to the $\left[b^{(b)}_1,b^{(b)}_0\right]$ relation in the minimalistic presentation, we get an identity for $\left[h^{(a)}_{2k+1},h^{(b)}_1\right]$. Replacing the $hb$ commutators in the identity using \eqref{habbsum}, we have
\begin{align}
	c_b\left[h^{(a)}_{2k+1},h^{(b)}_1\right]=&-\epsilon_3\sum_{l=0}^k\sum_{j=0}^{2l}\binom{2l}{j}\left(-\frac{\epsilon_3}{2}\right)^{2l-j}\nonumber\\
	&\left(\epsilon_3\left\{b^{(b)}_{j+1}h^{(a)}_{2k-2l-1},b^{(b)}_0\right\}-\left[b^{(b)}_{j+2}h^{(a)}_{2k-2l-1},b^{(b)}_0\right]-\left[b^{(b)}_1,b^{(b)}_{j+1}h^{(a)}_{2k-2l-1}\right]\right)\nonumber\\
	=&-\epsilon_3\sum_{l=0}^k\sum_{j=0}^{2l}\binom{2l}{j}\left(-\frac{\epsilon_3}{2}\right)^{2l-j}\left(\epsilon_3\left\{b^{(b)}_{j+1},b^{(b)}_0\right\}-\left[b^{(b)}_{j+2},b^{(b)}_0\right]-\left[b^{(b)}_1,b^{(b)}_{j+1}\right]\right)h^{(a)}_{2k-2l-1}\nonumber\\
	&-\epsilon_3\sum_{l=0}^k\sum_{j=0}^{2l}\binom{2l}{j}\left(-\frac{\epsilon_3}{2}\right)^{2l-j}\left(\left(\epsilon_3b^{(b)}_{j+1}-b^{(b)}_{j+2}\right)\left[h^{(a)}_{2k-2l-1},b^{(b)}_0\right]+b^{(b)}_{j+1}\left[h^{(a)}_{2k-2l-1},b^{(b)}_1\right]\right)\;.
\end{align}
For brevity, let us denote the penultimate line and the last line in the last equality as $(*)$ and $(**)$ respectively.

In $(*)$, the combination of the (anti-)commutators coincides with the one in the $bb$ relation. In particular, the assumption for the induction gives $j=0,\dots,2k-1$, and \eqref{bbbbj2k} applies to $j=2k$. Then
\begin{equation}
	(*)=-\frac{(2k+1)}{2}c_b\left[h^{(b)}_1,h^{(b)}_{2k+1}\right]+2c_b\epsilon_3^2\sum_{l=1}^k\sum_{m=1}^l\binom{2l}{2m-1}\left(-\frac{\epsilon_3}{2}\right)^{2l-2m}h^{(b)}_{2m+1}h^{(a)}_{2k-2l-1}\;.\label{star}
\end{equation}
Likewise, using \eqref{habbsum}, we get
\begin{equation}
	(**)=2\epsilon_3^2\sum_{\substack{l,m\geq0\\l+m\leq k-1}}\sum_{j=0}^{2l}\sum_{i=0}^{2m}\binom{2l}{j}\binom{2m}{i}\left(-\frac{\epsilon_3}{2}\right)^{2l+2m-j-i}\left(\epsilon_3b^{(b)}_{j+1}b^{(b)}_{i+1}-b^{(b)}_{j+2}b^{(b)}_{i+1}+b^{(b)}_{j+1}b^{(b)}_{i+2}\right)h^{(a)}_{2k-2l-2m-3}\;.
\end{equation}
Add another copy of $(**)$ but with $(l,j)\leftrightarrow(m,i)$ to this identity. Using the $bb$ relation, we have
\begin{equation}
	(**)=-4c_b\epsilon_3^2\sum_{\substack{l,m\geq0\\l+m\leq k-1}}\sum_{j=0}^{2l}\sum_{i=0}^{2m}\binom{2l}{j}\binom{2m}{i}(-1)^i\left(-\frac{\epsilon_3}{2}\right)^{2l+2m-j-i}h^{(b)}_{j+i+3}h^{(a)}_{2k-2l-2m-3}\;.
\end{equation}
Notice that $h^{(b)}_{j+i+3}$ vanishes for odd $j+i$. Using the identity of the binomial coefficients,
\begin{equation}
	\sum_{l+m=L}\sum_{p=0}^{j+i}(-1)^p\binom{2l}{p}\binom{2m}{j+i-p}=\frac{1}{2}\binom{2L+2}{j+i+1}
\end{equation}
and defining $l'=l+m+1$, $m'=\frac{1}{2}(j+i)+1$ (since only even $j+i$ terms survive), we get
\begin{equation}
	(**)=-2c_b\epsilon_3^2\sum_{l'=1}^k\sum_{m'=1}^{l'}\binom{2l'}{2m'-1}\left(-\frac{\epsilon_3}{2}\right)^{2l'-2m'}h^{(b)}_{2m'+1}h^{(a)}_{2k-2l'-1}\;.
\end{equation}
This is exactly the negative of the second term in \eqref{star}. Therefore,
\begin{equation}
	c_b\left[h^{(a)}_{2k+1},h^{(b)}_1\right]=(*)+(**)=-(2k+1)c_b\left[h^{(b)}_1,h^{(b)}_{2k+1}\right]\;.
\end{equation}

Now, apply $\mathrm{ad}_{h^{(b)}_1}$ and $\mathrm{ad}_{h^{(a)}_1}$ to \eqref{higherhn3}. Using \eqref{ha1ba+-1}, we get
\begin{subequations}
	\begin{align}
		&\left[h^{(b)}_1,h^{(a)}_{2k+1}\right]=\frac{(-1)^k}{c_a}\epsilon_3\left(\left[b^{(a)}_{k+2},b^{(a)}_k\right]-\epsilon_3\left\{b^{(a)}_{k+1},b^{(a)}_k\right\}\right)\;,\\
		&\left[h^{(a)}_1,h^{(a)}_{2k+1}\right]=\frac{(-1)^k}{c_a}2\epsilon_3\left(-\left[b^{(a)}_{k+2},b^{(a)}_k\right]+\epsilon_3\left\{b^{(a)}_{k+1},b^{(a)}_k\right\}\right)\;.
	\end{align}
\end{subequations}
Then
\begin{equation}
	\left[h^{(b)}_1,h^{(a)}_{2k+1}\right]=-\frac{1}{2}\left[h^{(a)}_1,h^{(a)}_{2k+1}\right]\;.
\end{equation}
Together with $\left[h^{(a)}_{2k+1},h^{(b)}_1\right]=-(2k+1)\left[h^{(b)}_1,h^{(b)}_{2k+1}\right]$ from $(*)+(**)$, it is not hard to see that
\begin{equation}
	\left[h^{(a)}_1,h^{(a)}_{2k+1}\right]=\left[h^{(b)}_1,h^{(a)}_{2k+1}\right]=0\;.
\end{equation}

We can now prove the $bb$ relation for $b=a$. Recall that we have shown that
\begin{equation}
	\left[b^{(a)}_{r+1},b^{(a)}_s\right]-\left[b^{(a)}_r,b^{(a)}_{s+1}\right]-\epsilon_3\left\{b^{(a)}_r,b^{(a)}_s\right\}=-4c_a(-1)^rh^{(a)}_{2k+1}\;,
\end{equation}
where we have redefined $r\rightarrow r-1$ in \eqref{bbaa2k}. Apply $\mathrm{ad}_{h^{(a)}_1}$ to this identity. The right hand side is zero. The left hand side can be rewritten as
\begin{equation}
	\left[b^{(a)}_{r+2},b^{(a)}_s\right]-\left[b^{(a)}_{r+1},b^{(a)}_{s+1}\right]+\left[b^{(a)}_{r+1},b^{(a)}_{s+1}\right]-\left[b^{(a)}_r,b^{(a)}_{s+2}\right]-\epsilon_3\left\{b^{(a)}_{r+1},b^{(a)}_s\right\}-\epsilon_3\left\{b^{(a)}_r,b^{(a)}_{s+1}\right\}
\end{equation}
using $\left[h^{(a)}_1,b^{(a)}_j\right]=4\epsilon_3b^{(a)}_{j+1}$. Therefore,
\begin{equation}
	\left(\left[b^{(a)}_{r+2},b^{(a)}_s\right]-\left[b^{(a)}_{r+1},b^{(a)}_{s+1}\right]-\epsilon_3\left\{b^{(a)}_{r+1},b^{(a)}_s\right\}\right)+\left(\left[b^{(a)}_{r+1},b^{(a)}_{s+1}\right]-\left[b^{(a)}_r,b^{(a)}_{s+2}\right]-\epsilon_3\left\{b^{(a)}_r,b^{(a)}_{s+1}\right\}\right)=0\;.
\end{equation}
Equivalently,
\begin{equation}
	\left[b^{(a)}_{r+1},b^{(a)}_s\right]-\left[b^{(a)}_r,b^{(a)}_{s+1}\right]-\epsilon_3\left\{b^{(a)}_r,b^{(a)}_s\right\}=-\left(\left[b^{(a)}_{r+2},b^{(a)}_{s-1}\right]-\left[b^{(a)}_{r+1},b^{(a)}_s\right]-\epsilon_3\left\{b^{(a)}_{r+1},b^{(a)}_{s-1}\right\}\right)\;.
\end{equation}
In other words, there is an extra minus sign under $(r,s)\rightarrow(r+1,s-1)$. Apply this $(s-r)$ times, and we get
\begin{equation}
	\left[b^{(a)}_{r+1},b^{(a)}_s\right]-\left[b^{(a)}_r,b^{(a)}_{s+1}\right]-\epsilon_3\left\{b^{(a)}_r,b^{(a)}_s\right\}=(-1)^{s-r}\left(\left[b^{(a)}_{s+1},b^{(a)}_r\right]-\left[b^{(a)}_s,b^{(a)}_{r+1}\right]-\epsilon_3\left\{b^{(a)}_s,b^{(a)}_r\right\}\right)\;.
\end{equation}
In particular, $s-r$ is odd since $r+s$ is odd. However,
\begin{equation}
	\left[b^{(a)}_{r+1},b^{(a)}_s\right]-\left[b^{(a)}_r,b^{(a)}_{s+1}\right]-\epsilon_3\left\{b^{(a)}_r,b^{(a)}_s\right\}=-\left[b^{(a)}_s,b^{(a)}_{r+1}\right]+\left[b^{(a)}_{s+1},b^{(a)}_r\right]-\epsilon_3\left\{b^{(a)}_s,b^{(a)}_r\right\}\;.
\end{equation}
Therefore,
\begin{equation}
	\left[b^{(a)}_{r+1},b^{(a)}_s\right]-\left[b^{(a)}_r,b^{(a)}_{s+1}\right]-\epsilon_3\left\{b^{(a)}_r,b^{(a)}_s\right\}=0\;,
\end{equation}
and this proves the $bb$ relation.

Besides the Serre relations, the remaining relations to be proven are the $hb$ relation for $b=a$ and the $hh$ relation for $b=a,a\pm1$. Recall that the $hb$ relation reads
\begin{equation}
	\left[h^{(a)}_{2k+1},b^{(a)}_s\right]-\left[h^{(a)}_{2k-1},b^{(a)}_{s+2}\right]-\epsilon_3^2\left[h^{(a)}_{2k-1},b^{(a)}_s\right]=2\epsilon_3\left\{h^{(a)}_{2k-1},b^{(a)}_{s+1}\right\}\;.
\end{equation}
Let us only write the proof for $s=0$ here, and the relation with general $s$ can then be proven by induction. Using the $bb$ relation
\begin{equation}
	4c_ah^{(a)}_{2k+1}=\left[b^{(a)}_2,b^{(a)}_{2k-1}\right]-\left[b^{(a)}_1,b^{(a)}_{2k}\right]-\epsilon_3\left\{b^{(a)}_1,b^{(a)}_{2k-1}\right\}\;,
\end{equation}
we have
\begin{align}
    4c_a\left[h^{(a)}_{2k+1},b^{(a)}_0\right]=&\left[b^{(a)}_2,\left[b^{(a)}_{2k-1},b^{(a)}_0\right]\right]-2\epsilon_3\left\{b^{(a)}_1,\left[b^{(a)}_{2k-1},b^{(a)}_0\right]\right\}\nonumber\\
	&+\epsilon_3^2\left[b^{(a)}_0,\left[b^{(a)}_{2k-1},b^{(a)}_0\right]\right]+\left[b^{(a)}_1,\left[b^{(a)}_1,b^{(a)}_{2k-1}\right]\right]\nonumber\\
	&-\epsilon_3\left\{b^{(a)}_0,\left[b^{(a)}_1,b^{(a)}_{2k-1}\right]\right\}+8c_a\epsilon_3b^{(a)}_{2k+1}+4c_a\epsilon_3\left\{h^{(a)}_1,b^{(a)}_{2k-1}\right\}\;.
\end{align}
Further using the $bb$ relations
\begin{subequations}
	\begin{align}
		&\left[b^{(a)}_{2k-1},b^{(a)}_0\right]=\left[b^{(a)}_{2k-2},b^{(a)}_1\right]+\epsilon_3\left\{b^{(a)}_{2k-2},b^{(a)}_0\right\}-4c_ah^{(a)}_{2k-1}\;,\nonumber\\
		&\left[b^{(a)}_1,b^{(a)}_{2k-1}\right]=\left[b^{(a)}_2,b^{(a)}_{2k-2}\right]-\epsilon_3\left\{b^{(a)}_1,b^{(a)}_{2k-2}\right\}\;,\label{hbhb}
	\end{align}
\end{subequations}
we get
\begin{align}
	&4c_a\left(\left[h^{(a)}_{2k+1},b^{(a)}_0\right]-\left[h^{(a)}_{2k-1},b^{(a)}_2\right]-\epsilon_3^2\left[h^{(a)}_{2k-1},b^{(a)}_0\right]-2\epsilon_3\left\{h^{(a)}_{2k-1},b^{(a)}_1\right\}\right)\nonumber\\
	=&-2c_a\left(\left[h^{(a)}_3,b^{(a)}_{2k-2}\right]-4\epsilon_3b^{(a)}_{2k+1}-2\epsilon_3\left\{h^{(a)}_1,b^{(a)}_{2k-1}\right\}-4\epsilon_3^3b^{(a)}_{2k-1}\right)\;.
\end{align}
Due to \eqref{higherbn3}, the expressions in the parentheses coincide when $k=1$, and hence,
\begin{equation}
	\left[h^{(a)}_3,b^{(a)}_0\right]=4\epsilon_3b^{(a)}_3+2\epsilon_3\left\{h^{(a)}_1,b^{(a)}_1\right\}+4\epsilon_3^3b^{(a)}_1\;.
\end{equation}
As we have already shown that $h^{(a)}_1$ commutes with any $h^{(a)}_{2k+1}$, applying $\mathrm{ad}_{h^{(a)}_1}$ repeatedly to this gives the commutators of $h^{(a)}_3$ and $b^{(a)}_s$, which makes the right hand side of \eqref{hbhb} vanish. Therefore,
\begin{equation}
	\left[h^{(a)}_{2k+1},b^{(a)}_0\right]-\left[h^{(a)}_{2k-1},b^{(a)}_2\right]-\epsilon_3^2\left[h^{(a)}_{2k-1},b^{(a)}_0\right]-2\epsilon_3\left\{h^{(a)}_{2k-1},b^{(a)}_1\right\}=0\;.
\end{equation}

For the $hh$ relation, we shall only write the case with $b=a\pm1$ here, and the case with $b=a$ can be proven in a similar manner. Consider the commutator of $h^{(a)}_{2k+1}$ and $h^{(b)}_{2l+1}$. Using \eqref{higherhn3} on $h^{(b)}_{2l+1}$, we have
\begin{equation}
	\left[h^{(a)}_{2k+1},h^{(b)}_{2l+1}\right]=\frac{1}{2c_b}(-1)^l\left(\epsilon_3\left\{\left[h^{(a)}_{2k+1},b^{(b)}_l\right],b^{(b)}_l\right\}-\left[\left[h^{(a)}_{2k+1},b^{(b)}_{l+1}\right],b^{(b)}_l\right]-\left[b^{(b)}_{l+1},\left[h^{(a)}_{2k+1},b^{(b)}_l\right]\right]\right)\;.
\end{equation}
Plugging \eqref{habbsum} into this shows that the right hand side vanishes.

Now, let us prove the Serre relations. We start with $k_1=k_2=0$ and arbitrary $r$ in \eqref{serren3}. Suppose that this holds for $r=s$:
\begin{equation}
	\left[b^{(a)}_0,\left[b^{(a)}_0,b^{(a\pm1)}_s\right]\right]=-2c_a\epsilon_3b^{(a\pm1)}_s\;.
\end{equation}
Applying $\mathrm{ad}_{h^{(a)}_1}+2\mathrm{ad}_{h^{(a\pm1)}_1}$ yields
\begin{equation}
	6\epsilon_3\left[b^{(a)}_0,\left[b^{(a)}_0,b^{(a\pm1)}_{s+1}\right]\right]=-12c_a\epsilon_3^2b^{(a\pm1)}_{s+1}\;.
\end{equation}

Next, consider $k_1=k_2$, and take the form in \eqref{serren3alternative}. Suppose this holds for $k=k_1=k_2$. Using the $bb$ relation, we have
\begin{align}
	\left[b^{(a)}_{k+1},\left[b^{(a)}_{k+1},b^{(a\pm1)}_s\right]\right]=&\left[b^{(a)}_{k+1},\left[b^{(a)}_k,b^{(a\pm1)}_{s+1}\right]\right]-\frac{\epsilon_3}{2}\left[b^{(a)}_{k+1},\left\{b^{(a)}_k,b^{(a\pm1)}_s\right\}\right]\nonumber\\
	=&\left[b^{(a)}_k,\left[b^{(a)}_{k+1},b^{(a\pm1)}_{s+1}\right]\right]+\left[\left[b^{(a)}_{k+1},b^{(a)}_k\right],b^{(a\pm1)}_{s+1}\right]\nonumber\\
	&-\frac{\epsilon_3}{2}\left\{\left[b^{(a)}_{k+1},b^{(a)}_k\right],b^{(a\pm1)}_s\right\}-\frac{\epsilon_3}{2}\left\{b^{(a)}_k,\left[b^{(a)}_{k+1},b^{(a\pm1)}_s\right]\right\}\;.
\end{align}
Plugging
\begin{equation}
	\left[b^{(a)}_{k+1},b^{(a\pm1)}_s\right]=\left[b^{(a)}_k,b^{(a\pm1)}_{s+1}\right]-\frac{\epsilon_3}{2}\left\{b^{(a)}_k,b^{(a\pm1)}_s\right\}\;,\quad\left[b^{(a)}_{k+1},b^{(a)}_k\right]=\epsilon_3\left(b^{(a)}_k\right)^2-2c_a(-1)^kh^{(a)}_{2k+1}
\end{equation}
into this, the terms involving $\left\{b^{(a)}_k,\left[b^{(a)}_k,b^{(a\pm1)}_{s+1}\right]\right\}$ get cancelled. We get
\begin{align}
	\left[b^{(a)}_{k+1},\left[b^{(a)}_{k+1},b^{(a\pm1)}_s\right]\right]=&\left[b^{(a)}_k,\left[b^{(a)}_k,b^{(a\pm1)}_{s+2}\right]\right]-\frac{\epsilon_3^2}{4}\left[b^{(a)}_k,\left[b^{(a)}_k,b^{(a\pm1)}_s\right]\right]\nonumber\\
	&-2c_a(-1)^k\left(\left[h^{(a)}_{2k+1},b^{(a\pm1)}_{s+1}\right]-\frac{\epsilon_3}{2}\left\{h^{(a)}_{2k+1},b^{(a\pm1)}_s\right\}\right)\;.\label{bbbk+1}
\end{align}
Now, the assumption at $k=k_1=k_2$ with the $hb$ relation gives
\begin{align}
	\left[b^{(a)}_k,\left[b^{(a)}_k,b^{(a\pm1)}_{s+2}\right]\right]=&c_a(-1)^k\sum_{j=0}^k\left(\frac{\epsilon_3}{2}\right)^{2j}\left(\left[h^{(a)}_{2k-2j-1},b^{(a\pm1)}_{s+3}\right]-\epsilon_3\left\{h^{(a)}_{2k-2j-1},b^{(a\pm1)}_{s+2}\right\}\right)\nonumber\\
	=&c_a(-1)^k\left[h^{(a)}_{2k+1},b^{(a\pm1)}_{s+1}\right]\;.
\end{align}
Plugging this into \eqref{bbbk+1}, we have
\begin{align}
	\left[b^{(a)}_{k+1},\left[b^{(a)}_{k+1},b^{(a\pm1)}_s\right]\right]=&-c_a(-1)^k\left(\left[h^{(a)}_{2k+1},b^{(a\pm1)}_{s+1}\right]-\epsilon_3\left\{h^{(a)}_{2k+1},b^{(a\pm1)}_s\right\}\right)\nonumber\\
	&-\frac{\epsilon_3^2}{4}c_a(-1)^k\sum_{j=0}^k\left(\frac{\epsilon_3}{2}\right)^{2j}\left(\left[h^{(a)}_{2k-2j-1},b^{(a\pm1)}_{s+1}\right]-\epsilon_3\left\{h^{(a)}_{2k-2j-1},b^{(a\pm1)}_s\right\}\right)\nonumber\\
	=&c_a(-1)^{k+1}\sum_{j=0}^{k+1}\left(\frac{\epsilon_3}{2}\right)^{2j}\left(\left[h^{(a)}_{2k-2j+1},b^{(a\pm1)}_{s+1}\right]-\epsilon_3\left\{h^{(a)}_{2k-2j+1},b^{(a\pm1)}_s\right\}\right)\;.
\end{align}

For $k_1\neq k_2$, suppose that the Serre relations hold for all indices $k_1$ and $k_2$ summed up to $k+l$. Apply $2\mathrm{ad}_{h^{(a)}_1}+\mathrm{ad}_{h^{(a\pm1)}_1}$. We have
\begin{equation}
	\left[b^{(a)}_{k+1},\left[b^{(a)}_l,b^{(a\pm1)}_s\right]\right]+\left[b^{(a)}_l,\left[b^{(a)}_{k+1},b^{(a\pm1)}_s\right]\right]=-\left[b^{(a)}_k,\left[b^{(a)}_{l+1},b^{(a\pm1)}_s\right]\right]-\left[b^{(a)}_{l+1},\left[b^{(a)}_k,b^{(a\pm1)}_s\right]\right]\;.
\end{equation}
If $k+l+1=2r+1$ is odd, keep balancing the two indices until $k$ and $l$ both reach $r$. The expression then vanishes due to the symmetry of the expression. If $k+l+1=2r$ is even, keep balancing the two indices until both the subscripts of the $b^{(a)}$ modes reach $r$. We get
\begin{align}
	&\left[b^{(a)}_{k+1},\left[b^{(a)}_l,b^{(a\pm1)}_s\right]\right]+\left[b^{(a)}_l,\left[b^{(a)}_{k+1},b^{(a\pm1)}_s\right]\right]\nonumber\\
	=&2(-1)^{k+1-r}\left[b^{(a)}_r,\left[b^{(a)}_r,b^{(a\pm1)}_s\right]\right]\nonumber\\
	=&2c_a(-1)^{k+1}\sum_{j=0}^r\left(\frac{\epsilon_3}{2}\right)^{2j}\left(\left[h^{(a)}_{2r-2j-1},b^{(a\pm1)}_{s+1}\right]-\epsilon_3\left\{h^{(a)}_{2r-2j-1},b^{(a\pm1)}_s\right\}\right)\;.
\end{align}
This completes the proof.
\end{proof}

\paragraph{The \texorpdfstring{$n=2$}{n=2} case} Let us now consider the case with $n=2$. Again, the higher modes can be generated from the lower modes as in \eqref{higherbn3} and \eqref{higherhn3}, which are reproduced here:
\begin{subequations}
	\begin{align}
		&b^{(a)}_k=\left(\frac{1}{4\epsilon_3}\right)^k\mathrm{ad}_{h^{(a)}_1}^kb^{(a)}_0\;,\label{higherbn2}\\
		&h^{(a)}_{2k+1}=\frac{(-1)^k}{2c_a}\left(\epsilon_3\left(b^{(a)}_k\right)^2-\left[b^{(a)}_{k+1},b^{(a)}_k\right]\right)\label{higherhn2}\;.
	\end{align}
\end{subequations}

We shall only give the finite presentation for the algebra ${}^\iota Y_2$ without the Serre relations. For the Serre relations, we will keep a sub-family of the infinite relations.
\begin{theorem}
	The $\iota$quiver algebra ${}^\iota Y_2$ is generated by $h^{(a)}_1$, $b^{(a)}_0$, $b^{(a)}_1$ with the relations
	\begin{subequations}
		\begin{align}
			&\left[h^{(a)}_1,h^{(b)}_1\right]=0\;,\\
			&\left[h^{(a)}_1,h^{(a)}_3\right]=0\;,\\
			&\left[h^{(a)}_1,b^{(b)}_0\right]=2\epsilon_3A_{ab}b^{(b)}_1\;,\\
			&\left[\widetilde{h}^{(a)}_3,b^{(a+1)}_0\right]=-4\epsilon_3\left(b^{(a+1)}_3+\left(\epsilon_3^2-3\epsilon_1\epsilon_2\right)b^{(a+1)}_1\right)\;,\\
			&\left[b^{(a)}_1,b^{(a)}_0\right]=\epsilon_3\left(b^{(a)}_0\right)^2-2c_ah^{(a)}_1\;,\\
			&\left[b^{(a)}_2,b^{(a+1)}_0\right]-2\left[b^{(a)}_1,b^{(a+1)}_1\right]+\left[b^{(a)}_0,b^{(a+1)}_2\right]+\epsilon_1\epsilon_2\left[b^{(a)}_0,b^{(a+1)}_0\right]\nonumber\\
			=&-\epsilon_3\left(\left\{b^{(a)}_1,b^{(a+1)}_0\right\}-\left\{b^{(a)}_0,b^{(a+1)}_1\right\}\right)\;,
		\end{align}
	\end{subequations}
	where $\widetilde{h}^{(a)}_3=h^{(a)}_3-\frac{1}{2}\left(h^{(a)}_1\right)^2$, and $A_{ab}$ are the elements in the Cartan matrix, i.e., $A_{ab}$ is $2$ (resp.~$-2$) for $b=a$ (resp.~$b=a+1$). The higher modes are given by \eqref{higherbn2} and \eqref{higherhn2}.
\end{theorem}
\begin{proof}
	Similar to the $n\geq3$ cases, the higher modes of $b^{(b)}(u)$ may be obtained from $h^{(a)}_1$ with $b\neq a$ as well:
	\begin{equation}
		\left[h^{(a)}_1,b^{(a+1)}_j\right]=-4\epsilon_3b^{(a+1)}_{j+1}\;.
	\end{equation}
	This gives the negative coefficient compared to $\left[h^{(a)}_1,b^{(a)}_j\right]$. In particular,
	\begin{equation}
		\left[h^{(a)}_1,h^{(a+1)}_{2j+1}\right]=-\left[h^{(a+1)}_1,h^{(a+1)}_{2j+1}\right]\;.
	\end{equation}

	Now, let us first consider the relations with $b=a$ (i.e., all the relations involving only $a$). The proof can be done by induction in the same manner as in the $n\geq3$ cases. The only difference would be the $hh$ relation. Therefore, we shall only write the $hh$ relation here. This is also indicated by the extra relation $\left[h^{(a)}_1,h^{(a)}_3\right]=0$. With this extra relation, we have
	\begin{equation}
		\left[h^{(a)}_3,b^{(a)}_s\right]=4\epsilon_3b^{(a)}_{s+3}+2\epsilon_3\left\{h^{(a)}_1,b^{(a)}_{s+1}\right\}+4\epsilon_3^3b^{(a)}_{s+1}\;.
	\end{equation}
	Equivalently,
	\begin{equation}
		\left[\widetilde{h}^{(a)}_3,b^{(a)}_r\right]=4\epsilon_3\left(b^{(a)}_{r+3}+\epsilon_3^2b^{(a)}_{r+1}\right)\;.\label{htildeaban2}
	\end{equation}
	This identity can be obtained as follows. Suppose that this holds for $s=k$ (the $s=0$ case is similar to the calculations in the $n\geq3$ cases). For $s=k+1$, we have
	\begin{align}
		4\epsilon_3\left[h^{(a)}_3,b^{(a)}_{k+1}\right]=&\left[h^{(a)}_3,\left[h^{(a)}_1,b^{(a)}_k\right]\right]=\left[h^{(a)}_1,\left[h^{(a)}_3,b^{(a)}_k\right]\right]\nonumber\\
		=&4\epsilon_3\left[h^{(a)}_1,b^{(a)}_{k+3}\right]+2\epsilon_3\left[h^{(a)}_1,\left\{h^{(a)}_1,b^{(a)}_{k+1}\right\}\right]+4\epsilon_3^3\left[h^{(a)}_1,b^{(a)}_{k+1}\right]\nonumber\\
		=&16\epsilon_3^2b^{(a)}_{k+4}+8\epsilon_3^2\left\{h^{(a)}_1,b^{(a)}_{k+2}\right\}+16\epsilon_3^4b^{(a)}_{k+2}\;,
	\end{align}
	where the extra commutation relation of $h^{(a)}_1$ and $h^{(a)}_3$ enters the second equality in the first line. Then following the same procedure as in the $n\geq3$ cases, we can get $\left[h^{(a)}_{2k+1},h^{(a)}_{2l+1}\right]=\left[h^{(a)}_1,h^{(a)}_{2k+1}\right]=0$.

	Next, let us prove the $bb$ relation for $b=a+1$. As $h^{(a+1)}_1$ commutes with both $h^{(a)}_3$ and $\left(h^{(a)}_1\right)^2$, we have
	\begin{equation}
		\left[\widetilde{h}^{(a)}_3,b^{(a+1)}_r\right]=-4\epsilon_3\left(b^{(a+1)}_{r+3}+\left(\epsilon_3^2-3\epsilon_1\epsilon_2\right)b^{(a+1)}_{r+1}\right)\;.
	\end{equation}
	Together with \eqref{htildeaban2}, we get
	\begin{equation}
		\left[\widetilde{h}^{(a)}_3+\widetilde{h}^{(a+1)}_3,b^{(b)}_r\right]=\left(4\epsilon_3^3-4\epsilon_3(\epsilon_3^2-3\epsilon_1\epsilon_2)\right)b^{(b)}_{r+1}=12\epsilon_1\epsilon_2\epsilon_3b^{(b)}_{r+1}
	\end{equation}
	for either $b=a$ or $b=a+1$. Combining this with the commutator of $h^{(a)}_1$ and $b^{(b)}_r$ gives
	\begin{subequations}
		\begin{align}
		&\left[\widetilde{h}^{(a)}_3+\widetilde{h}^{(a+1)}_3+3\epsilon_1\epsilon_2h^{(a)}_1,b^{(a)}_r\right]=24\epsilon_1\epsilon_2\epsilon_3b^{(a)}_{r+1}\;,\\
		&\left[\widetilde{h}^{(a)}_3+\widetilde{h}^{(a+1)}_3+3\epsilon_1\epsilon_2h^{(a)}_1,b^{(a+1)}_r\right]=0\;.
	\end{align}
	\end{subequations}
	Suppose that the $bb$ relation holds at level $(r,s)$. Then applying $\mathrm{ad}_{\widetilde{h}^{(a)}_3+\widetilde{h}^{(a+1)}_3+3\epsilon_1\epsilon_2h^{(a)}_1}$ gives the relation at the next levels. This proves the $bb$ relation for $b=a+1$.

	The remaining relations to be proven are the $hh$ relation and the $hb$ relation for $b=a+1$. The $hh$ relation can be proven similarly as in the $n\geq3$ cases. Therefore, we shall only write the $hb$ relation for $b=a+1$. For brevity, define
	\begin{equation}
		B_{r,s}=\left[b^{(a)}_{r+1},b^{(a+1)}_s\right]-\left[b^{(a)}_r,b^{(a+1)}_{s+1}\right]+\epsilon_3\left\{b^{(a)}_r,b^{(a+1)}_s\right\}\;.
	\end{equation}
	In particular, the $bb$ relation for $b=a+1$ can be written as
	\begin{equation}
		B_{r+1,s}-B_{r,s+1}+\epsilon_1\epsilon_2\left[b^{(a)}_r,b^{(a+1)}_s\right]=0\;.\label{Bidentity}
	\end{equation}
	We want to show that
	\begin{align}
		&\left[h^{(a)}_{i+4},b^{(a+1)}_j\right]-2\left[h^{(a)}_{i+2},b^{(a+1)}_{j+2}\right]+\left[h^{(a)}_i,b^{(a+1)}_{j+4}\right]-\left(\epsilon_1^2+\epsilon_2^2\right)\left[h^{(a)}_{i+2},b^{(a+1)}_j\right]\nonumber\\
		&+\left(2\epsilon_1\epsilon_2+\epsilon_3^2\right)\left[h^{(a)}_i,b^{(a+1)}_{j+2}\right]+\epsilon_1^2\epsilon_2^2\left[h^{(a)}_i,b^{(a+1)}_j\right]\nonumber\\
		=&2\epsilon_3\left(\left\{h^{(a)}_i,b^{(a+1)}_{j+3}\right\}-\left\{h^{(a)}_{i+2},b^{(a+1)}_{j+1}\right\}+\epsilon_1\epsilon_2\left\{h^{(a)}_i,b^{(a+1)}_{j+1}\right\}\right)\;.
	\end{align}
	Take $i=2m+1$. Replacing all the $h$ modes with the $b$ modes using \eqref{higherhn2}, we have
	\begin{align}
		\text{LHS}-\text{RHS}=&\frac{(-1)^m}{2c_a}\left(\left[b^{(a)}_{m},B_{m+2,s+2}-B_{m+1,s+3}+\epsilon_1\epsilon_2\left[b^{(a)}_{m+1},b^{(a+1)}_{s+2}\right]\right]\right.\nonumber\\
		&-\left[b^{(a)}_{m+1},B_{m+1,s+2}-B_{m,s+3}+\epsilon_1\epsilon_2\left[b^{(a)}_m,b^{(a+1)}_{s+2}\right]\right]\nonumber\\
		&-2\left[b^{(a)}_{m+1},B_{m+2,s+1}-B_{m+1,s+2}+\epsilon_1\epsilon_2\left[b^{(a)}_{m+1},b^{(a+1)}_{s+1}\right]\right]\nonumber\\
		&+2\left[b^{(a)}_{m+2},B_{m+1,s+1}-B_{m,s+2}+\epsilon_1\epsilon_2\left[b^{(a)}_m,b^{(a+1)}_{s+1}\right]\right]\nonumber\\
		&+\left[b^{(a)}_{m+2},B_{m+2,s}-B_{m+1,s+1}+\epsilon_1\epsilon_2\left[b^{(a)}_{m+1},b^{(a+1)}_s\right]\right]\nonumber\\
		&-\left[b^{(a)}_{m+3},B_{m+1,s}-B_{m,s+1}+\epsilon_1\epsilon_2\left[b^{(a)}_m,b^{(a+1)}_s\right]\right]\nonumber\\
		&+\epsilon_3\left\{b^{(a)}_{m},B_{m+1,s+2}-B_{m,s+3}+\epsilon_1\epsilon_2\left[b^{(a)}_m,b^{(a+1)}_{s+2}\right]\right\}\nonumber\\
		&-\epsilon_3\left\{b^{(a)}_{m},B_{m+2,s+1}-B_{m+1,s+2}+\epsilon_1\epsilon_2\left[b^{(a)}_{m+1},b^{(a+1)}_{s+1}\right]\right\}\nonumber\\
		&-\epsilon_3\left\{b^{(a)}_{m+1},B_{m+1,s+1}-B_{m,s+2}+\epsilon_1\epsilon_2\left[b^{(a)}_m,b^{(a+1)}_{s+1}\right]\right\}\nonumber\\
		&+\epsilon_3\left\{b^{(a)}_{m+1},B_{m+2,s}-B_{m+1,s+1}+\epsilon_1\epsilon_2\left[b^{(a)}_{m+1},b^{(a+1)}_s\right]\right\}\nonumber\\
		&-\epsilon_3^2\left[b^{(a)}_m,B_{m+1,s+1}-B_{m,s+2}+\epsilon_1\epsilon_2\left[b^{(a)}_m,b^{(a+1)}_{s+1}\right]\right]\nonumber\\
		&+\epsilon_1\epsilon_2\left[b^{(a)}_m,B_{m+2,s}-B_{m+1,s+1}+\epsilon_1\epsilon_2\left[b^{(a)}_{m+1},b^{(a+1)}_s\right]\right]\nonumber\\
		&+\left(\epsilon_3^2-\epsilon_1\epsilon_2\right)\left[b^{(a)}_{m+1},B_{m+1,s}-B_{m,s+1}+\epsilon_1\epsilon_2\left[b^{(a)}_m,b^{(a+1)}_s\right]\right]\nonumber\\
		&\left.+\epsilon_1\epsilon_2\epsilon_3\left\{b^{(a)}_m,B_{m+1,s}-B_{m,s+1}+\epsilon_1\epsilon_2\left[b^{(a)}_m,b^{(a+1)}_s\right]\right\}\right)\;.
	\end{align}
	The whole expression vanishes due to \eqref{Bidentity}. This completes the proof.
\end{proof}

For the reduced algebra ${}^\iota\mathtt{Y}_2$, there are also Serre relations. We find that there is a subset of infinitely many relations that would recover the full set of the Serre relations.
\begin{theorem}
	The rational $\iota$quiver algebra ${}^\iota\mathtt{Y}_2$ is generated by $h^{(a)}_1$, $b^{(a)}_0$, $b^{(a)}_1$ with the relations in the above minimalistic presentation of ${}^\iota Y_2$ and
	\begin{align}
		&\mathrm{Sym}_{k_1,k_2,k_3}\left[b^{(a)}_{k_1},\left[b^{(a)}_{k_2},\left[b^{(a)}_{k_3},b^{(a+1)}_r\right]\right]\right]\nonumber\\
		=&c_a\sum_s(-1)^{k_j}\left(-\left[H^{(a)}_{k_i+k_j},\left[b^{(a)}_{k_s+1},b^{(a+1)}_r\right]\right]-2\left[b^{(a)}_{k_s+1},\left[H^{(a)}_{k_i+k_j},b^{(a+1)}_r\right]\right]-2\epsilon_3\left\{H^{(a)}_{k_i+k_j},\left[b^{(a)}_{k_s},b^{(a+1)}_r\right]\right\}\right.\nonumber\\
		&+\epsilon_3\left\{b^{(a)}_{k_s},\left[H^{(a)}_{k_i+k_j},b^{(a+1)}_r\right]\right\}+\frac{3}{\epsilon_1\epsilon_2}\left(\left[H^{(a)}_{k_i+k_j+2},\left[b^{(a)}_{k_s+1},b^{(a+1)}_r\right]-\left[b^{(a)}_{k_s},b^{(a+1)}_{r+1}\right]+\epsilon_3\left\{b^{(a)}_{k_s},b^{(a+1)}_r\right\}\right]\right.\nonumber\\
		&\left.\left.-\left[H^{(a)}_{k_i+k_j},\left[b^{(a)}_{k_s+2},b^{(a+1)}_{r+1}\right]-\left[b^{(a)}_{k_s+1},b^{(a+1)}_{r+2}\right]+\epsilon_3\left\{b^{(a)}_{k_s+1},b^{(a+1)}_{r+1}\right\}\right]\right)\right)
	\end{align}
	for all $(k_1,k_2,k_3,r)=(2m,0,0,0)$ with $m\geq0$, where
	\begin{equation}
		\{i,j,s\}=\{1,2,3\}\;,~i<j\;.
	\end{equation}
	Recall that
	\begin{equation}
		H^{(a)}_m=\sum_{l=0}^{\lfloor m/2\rfloor}\left(\frac{\epsilon_3}{2}\right)^{2l}h^{(a)}_{m-1-2l}\;.
	\end{equation}
\end{theorem}
\begin{proof}
	Again, we prove this by induction. First, let us show that they recover $(2m,0,0,r)$ for arbitrary $r$. Suppose that this is true for $r=l$. Recall that
	\begin{equation}
		B_{r,s}=\left[b^{(a)}_{r+1},b^{(a+1)}_s\right]-\left[b^{(a)}_r,b^{(a+1)}_{s+1}\right]+\epsilon_3\left\{b^{(a)}_r,b^{(a+1)}_s\right\}\;.
	\end{equation}
	We have
	\begin{subequations}
		\begin{align}
		&\left[\widetilde{h}^{(a)}_3+\widetilde{h}^{(a+1)}_3-3\epsilon_1\epsilon_2h^{(a)}_1,b^{(a)}_r\right]=0\;,\\
		&\left[\widetilde{h}^{(a)}_3+\widetilde{h}^{(a+1)}_3-3\epsilon_1\epsilon_2h^{(a)}_1,b^{(a+1)}_r\right]=24\epsilon_1\epsilon_2\epsilon_3b^{(a+1)}_{r+1}\;,\\
		&\left[\widetilde{h}^{(a)}_3+\widetilde{h}^{(a+1)}_3-3\epsilon_1\epsilon_2h^{(a)}_1,B_{s,r}\right]=24\epsilon_1\epsilon_2\epsilon_3B_{s,r+1}\;.
	\end{align}
	\end{subequations}
	Applying this to the relation at $r=l$ gives the relation at $r=l+1$.

	Next, let us consider arbitrary $k_i\neq0$. The cases with one or two $k_i$ being zero can be proven in a similar manner, and hence we shall only write the case for all non-vanishing $k_i$ below. Suppose that all the relations hold for $k_1+k_2+k_3\leq K$. For brevity, write the difference between the left hand side and the right hand side of the Serre relation as $\mathcal{D}_{k_1,k_2,k_3}$ (with $r$ fixed). The goal is to show that this vanishes for any possible combinations with $K+1$.

	Define
	\begin{equation}
		\widetilde{h}^{(a)}_5=h^{(a)}_5-h^{(a)}_3h^{(a)}_1+\frac{1}{3}\left(h^{(a)}_1\right)^3\;.
	\end{equation}
	Its commutator with $b^{(a)}_r$ can be computed as
	\begin{subequations}
		\begin{align}
			&\left[\widetilde{h}^{(a)}_5,b^{(a)}_r\right]=4\epsilon_3b^{(a)}_{r+5}+\frac{40}{3}\epsilon_3^3b^{(a)}_{r+3}+4\epsilon_3^5b^{(a)}_{r+1}\;,\\
			&\left[\widetilde{h}^{(a)}_5,b^{(a+1)}_r\right]=-4\epsilon_3b^{(a+1)}_{r+5}-\frac{40}{3}\epsilon_3\left(\epsilon_3^2-3\epsilon_1\epsilon_2\right)b^{(a+1)}_{r+3}-4\epsilon_3\left(\epsilon_3^4-5\epsilon_1\epsilon_2\epsilon_3^2+5\epsilon_1^2\epsilon_2^2\right)b^{(a+1)}_{r+1}\;.
		\end{align}
	\end{subequations}
	Therefore,
	\begin{equation}
		\left[\widetilde{h}^{(a)}_5+\widetilde{h}^{(a+1)}_5,b^{(b)}_r\right]=40\epsilon_1\epsilon_2\epsilon_3b^{(b)}_{r+3}+20\epsilon_1\epsilon_2\epsilon_3\left(\epsilon_3^2-\epsilon_1\epsilon_2\right)b^{(b)}_{r+1}
	\end{equation}
	for either $b=a$ or $b=a+1$. Take $(k_1,k_2,k_3)$ such that $k_1+k_2+k_3=K+1$. Applying $\mathrm{ad}_{\widetilde{h}^{(a)}_5+\widetilde{h}^{(a+1)}_5}$ to $\mathcal{D}_{k_1-1,k_2-1,k_3-1}$ yields
	\begin{equation}
		\mathcal{D}_{k_1+2,k_2-1,k_3-1}+\mathcal{D}_{k_1-1,k_2+2,k_3-1}+\mathcal{D}_{k_1-1,k_2-1,k_3+2}=0\;.
	\end{equation}
	Likewise, applying $\mathrm{ad}_{\widetilde{h}^{(a)}_3+\widetilde{h}^{(a+1)}_3+3\epsilon_1\epsilon_2h^{(a)}_1}$ to $\mathcal{D}_{k_1+1,k_2-1,k_3-1}$, $\mathcal{D}_{k_1-1,k_2+1,k_3-1}$, and $\mathcal{D}_{k_1-1,k_2-1,k_3+1}$ yields
	\begin{subequations}
		\begin{align}
			&\mathcal{D}_{k_1+2,k_2-1,k_3-1}+\mathcal{D}_{k_1+1,k_2,k_3-1}+\mathcal{D}_{k_1+1,k_2-1,k_3}=0\;,\\
			&\mathcal{D}_{k_1,k_2+1,k_3-1}+\mathcal{D}_{k_1-1,k_2+2,k_3-1}+\mathcal{D}_{k_1-1,k_2+1,k_3}=0\;,\\
			&\mathcal{D}_{k_1,k_2-1,k_3+1}+\mathcal{D}_{k_1-1,k_2,k_3+1}+\mathcal{D}_{k_1-1,k_2-1,k_3+2}=0\;.
		\end{align}
	\end{subequations}
	Adding the latter three equations and subtract the first equation involving $\mathcal{D}$, we can see that
	\begin{equation}
		\mathcal{D}_{k_1+1,k_2,k_3-1}+\mathcal{D}_{k_1+1,k_2-1,k_3}+\mathcal{D}_{k_1,k_2+1,k_3-1}+\mathcal{D}_{k_1-1,k_2+1,k_3}+\mathcal{D}_{k_1,k_2-1,k_3+1}+\mathcal{D}_{k_1-1,k_2,k_3+1}=0\;.
	\end{equation}
	Now, apply $\mathrm{ad}_{\widetilde{h}^{(a)}_3+\widetilde{h}^{(a+1)}_3+3\epsilon_1\epsilon_2h^{(a)}_1}^3$ to $\mathcal{D}_{k_1-1,k_2-1,k_3-1}$. Due to the above vanishing sums of $\mathcal{D}$, the remaining term gives
	\begin{equation}
		6\mathcal{D}_{k_1,k_2,k_3}=0\;.
	\end{equation}
	This completes the proof.
\end{proof}

\begin{remark}
	At level $0$, the Serre relations read
	\begin{subequations}
		\begin{align}
			&\left[b^{(0)}_0,\left[b^{(0)}_0,\left[b^{(0)}_0,b^{(1)}_0\right]\right]\right]=-8c_0\epsilon_3\left[b^{(0)}_0,b^{(1)}_0\right]\;,\\
			&\left[b^{(1)}_0,\left[b^{(1)}_0,\left[b^{(1)}_0,b^{(0)}_0\right]\right]\right]=-8c_1\epsilon_3\left[b^{(1)}_0,b^{(0)}_0\right]\;.
		\end{align}
	\end{subequations}
	This is the Dolan--Grady presentation of the Onsager algebra \cite{dolan1982conserved}. This is not surprising since the Onsager algebra is isomorphic to the subalgebra of $\widehat{\mathfrak{sl}}_2$ invariant under the Chevalley involution.
\end{remark}

\paragraph{The \texorpdfstring{$n=1$}{n=1} case} Let us now consider the case with $n=1$. The higher modes can be generated from the lower modes via
\begin{subequations}
	\begin{align}
		&b_k=\left(\frac{1}{12\sigma_3}\right)^k\mathrm{ad}_{h_3}^kb_0\;,\label{higherbn1}\\
		&h_{2k+3}=-\frac{\sigma_2}{4}h_{2k+1}-\frac{(-1)^k}{8c}\left([b_{k+3},b_k]-3[b_{k+2},b_{k+1}]+\sigma_2[b_{k+1},b_k]-\sigma_3b_k^2\right)\label{higherhn1}\;.
	\end{align}
\end{subequations}
The expressions are different from the above cases as one may expect.

Here, we shall give a presentation with a subset of relations as given in the definition of ${}^\iota\mathtt{Y}_1$ although there are still infinitely many of them.
\begin{theorem}
	The $\iota$quiver algebra ${}^\iota\mathtt{Y}_1$ is generated by $h_1$, $h_3$, $b_0$ with the relations
	\begin{subequations}
		\begin{align}
			&[h_{2k+1},h_{2l+1}]=0\;,\quad 0\leq k<l\;,\\
			&[h_1,b_0]=0\;,\\
			&[b_3,b_0]-3[b_2,b_1]+\sigma_2[b_1,b_0]-\sigma_3b_0^2=-8ch_3-2c\sigma_2h_1\;,\\
			&[h_{r+6},b_0]-3[h_{r+4},b_2]+3[h_{r+2},b_4]-[h_r,b_6]+2\sigma_2\left([h_{r+4},b_0]-[h_r,b_4]\right)\nonumber\\
			&+\sigma_2^2\left([h_{r+2},b_0]-[h_r,b_2]\right)-\sigma_3^2[h_r,b_0]\nonumber\\
			=&2\sigma_3\left(3\{h_{r+2},b_1\}+\{h_r,b_3\}+\sigma_2\{h_r,b_1\}\right)\;,\quad r=2m-1,~m\geq0\;,\\
			&\mathrm{Sym}_{k_1,k_2,k_3}[b_{k_1},[b_{k_2},b_{k_3+1}]]=-\frac{c}{2}\sum_s(-1)^{k_j}C_{k_i+k_j,k_s}\;,\quad(k_1,k_2,k_3)=(2m,0,0)\;,~m\geq0\;.
		\end{align}
	\end{subequations}
	Recall that
	\begin{subequations}
	    \begin{align}
	        &C_{m\leq0,r}=0\;,~C_{m>0,r}=\frac{8}{3}[4h_{m+1}+\sigma_2h_{m-1},b_r]-\frac{1}{3}C_{m-2,r+2}-\frac{\sigma_2}{3}C_{m-2,r}\;,\\
	        &\{i,j,s\}=\{1,2,3\}\;,~i<j\;.
        \end{align}
    \end{subequations}
\end{theorem}
\begin{proof}
	Again, we prove this by induction. For the Serre relations, they are similar to those in the $n=2$ case. Therefore, we shall only write the other relations here.

	First, it would be useful to notice that $h_1$ is a central element. This can be seen as follows. Suppose that $h_1$ commutes with $b_k$. Then
	\begin{equation}
		[h_1,b_{k+1}]=\frac{1}{12\sigma_3}[h_1,[h_3,b_k]]=\frac{1}{12\sigma_3}\left([[h_1,h_3],b_k]+[h_3,[h_1,b_k]]\right)=0\;.
	\end{equation}
	Thus, $h_1$ commutes with all the $b$ modes. Likewise, suppose that $h_1$ commutes with $h_{2k+1}$. Then
	\begin{align}
		[h_1,h_{2k+3}]=&-\frac{\sigma_2}{4}[h_1,h_{2k+1}]-\frac{(-1)^k}{8c}([[h_1,b_{k+3}],b_k]+[b_{k+3},[h_1,b_k]]-3[[h_1,b_{k+2}],b_{k+1}]\nonumber\\
		&-3[b_{k+2},[h_1,b_{k+1}]]+\sigma_2[[h_1,b_{k+1}],b_k]+\sigma_2[b_{k+1},[h_1,b_k]]-\sigma_3\{[h_1,b_k],b_k\})=0\;.
	\end{align}
	Thus, $h_1$ commutes with all the $h$ modes.

	Now, let us prove the $bb$ relation. We want to show that
	\begin{align}
		&[b_{r+3},b_s]-3[b_{r+2},b_{s+1}]+3[b_{r+1},b_{s+2}]-[b_r,b_{s+3}]+\sigma_2\left([b_{r+1},b_s]-[b_r,b_{s+1}]\right)-\sigma_3\{b_r,b_s\}\nonumber\\
		=&-4c(-1)^s\left(4h_{r+s+3}+\sigma_2h_{r+s+1}\right)\;.
	\end{align}
	For $r=s=k$, this holds following \eqref{higherhn1}. Next, suppose that this holds for all $r+s\leq K$. Take $r'+s'=K+1$. Apply $\mathrm{ad}_{h_3}$ to this relation. We get
	\begin{equation}
	    [b_{r+4},b_s]-2[b_{r+3},b_{s+1}]+2[b_{r+1},b_{s+3}]-[b_r,b_{s+4}]+\sigma_2\left([b_{r+2},b_s]-[b_r,b_{s+2}]\right)-\sigma_3\left(\{b_{r+1},b_s\}+\{b_r,b_{s+1}\}\right)=0\;.\label{bb+bb}
	\end{equation}
	In particular, the left hand side is the sum of the left hand sides of the $bb$ relation at $(r+1,s)$ and $(r,s+1)$. The sum of the right hand sides of the $bb$ relation at $(r+1,s)$ and $(r,s+1)$ is also zero:
	\begin{equation}
		-4c(-1)^s\left(4h_{r+s+4}+\sigma_2h_{r+s+2}\right)-4c(-1)^{s+1}\left(4h_{r+s+4}+\sigma_2h_{r+s+2}\right)=0\;.
	\end{equation}

	For even $K+1$, the case with $k=r'=s'$ is already proven. Then taking \eqref{bb+bb} at $(k,k-1)$, $(k+1,k-2)$, $\dots$, $(2k-1,0)$ proves the cases with $(k+1,k-1)$, $(k+2,k-2)$, $\dots$, $(2k,0)$. The remaining half can be obtained by exchanging the two indices.

	For odd $K+1$, let $K=2k$. At $(k,k)$, \eqref{bb+bb} becomes
	\begin{equation}
		2\left([b_{k+4},b_k]-2[b_{k+3},b_{k+1}]+\sigma_2[b_{k+2},b_k]-\sigma_3\{b_{k+1},b_k\}\right)=0\;.
	\end{equation}
	This proves the relation at $(k+1,k)$. For the remaining relations at $K+1$, they can be proven in the same manner as the even $K+1$ case.

	Now, let us prove the $hb$ relation. We want to show that
	\begin{align}
		&[h_{r+6},b_s]-3[h_{r+4},b_{s+2}]+3[h_{r+2},b_{s+4}]-[h_r,b_{s+6}]+2\sigma_2([h_{r+4},b_s]-[h_r,b_{s+4}])\nonumber\\
		&+\sigma_2^2([h_{r+2},b_s]-[h_r,b_{s+2}])-\sigma_3^2[h_r,b_s]\nonumber\\
		=&2\sigma_3(3\{h_{r+2},b_{s+1}\}+\{h_r,b_{s+3}\}+\sigma_2\{h_r,b_{s+1}\})\;,\quad r=2m-5\;,~m\geq0\;.
	\end{align}
	The cases with $r=-5$ and $r=-3$ are already known. Therefore, we need to prove those for $r\geq-1$. Suppose that this holds for $s=k$. Notice that
	\begin{subequations}
		\begin{align}
			&[h_3,[h_j,b_{k+l}]]=[[h_3,h_j],b_{k+l}]+[h_j,[h_3,b_{k+l}]]=12\sigma_3[h_j,b_{k+l+1}]\;,\\
			&[h_3,\{h_j,b_{k+l}\}]=\{[h_3,h_j],b_{k+l}\}+\{h_j,[h_3,b_{k+l}]\}=12\sigma_3\{h_j,b_{k+l+1}\}\;.
		\end{align}
	\end{subequations}
	With these two identities, apply $\mathrm{ad}_{h_3}$ to the relation at $s=k$. We have
	\begin{align}
		&[h_{r+6},b_{k+1}]-3[h_{r+4},b_{k+3}]+3[h_{r+2},b_{k+5}]-[h_r,b_{k+7}]+2\sigma_2\left([h_{r+4},b_{k+1}]-[h_r,b_{k+5}]\right)\nonumber\\
		&+\sigma_2^2\left([h_{r+2},b_{k+1}]-[h_r,b_{k+3}]\right)-\sigma_3^2[h_r,b_{k+1}]=2\sigma_3\left(3\{h_{r+2},b_{k+2}\}+\{h_r,b_{k+4}\}+\sigma_2\{h_r,b_{k+2}\}\right)\;,
	\end{align}
	which is the relation at $s=k+1$. This completes the proof.
\end{proof}

Let us also give an example to see why the above discussions do not produce a finite presentation. For $h_5$, it can be obtained by
\begin{equation}
	h_5=-\frac{\sigma_2}{4}h_3+\frac{1}{8c}\left([b_4,b_1]-3[b_3,b_2]+\sigma_2[b_2,b_1]-\sigma_3b_1^2\right)\;.
\end{equation}
Apply $\mathrm{ad}_{h_3}$ to this equation. We get
\begin{align}
	[h_3,h_5]=&\frac{12\sigma_3}{8c}\left([b_5,b_1]+[b_4,b_2]-3[b_4,b_2]-3[b_3,b_3]+\sigma_2[b_3,b_1]+\sigma_2[b_2,b_2]-\sigma_3\{b_2,b_1\}\right)\nonumber\\
	=&\frac{3\sigma_3}{2c}\left([b_5,b_1]-2[b_4,b_2]+\sigma_2[b_3,b_1]-\sigma_3\{b_2,b_1\}\right)\;.
\end{align}
Equivalently,
\begin{equation}
	[b_5,b_1]-2[b_4,b_2]+\sigma_2[b_3,b_1]-\sigma_3\{b_2,b_1\}=\frac{2c}{3\sigma_3}[h_3,h_5]\;.
\end{equation}
The left hand side is the left hand side of the $bb$ relation at $(r,s)=(2,1)$. Without providing $[h_3,h_5]=0$, we cannot further say that this vanishes.

\subsection{Embeddings and Coideal Structures}\label{coideals}
In this subsection, we shall discuss some more properties of the $\iota$quiver algebras. In particular, they are right coideals of the quiver BPS algebras.

It would be helpful to recall some basic aspects and set the notations of the untwisted affine Lie algebras of type A. Let $R=R_+\cup R_-$ be the roots composed of positive and negative roots. The sets of real and imaginary roots are denoted by $R^{\text{re}}$ and $R^{\text{im}}$ respectively. Given a root $\alpha$, its root space is denoted by $\mathfrak{g}_\alpha$, and the simple roots are $\alpha_a$. Then $R^\text{re}_+=\mathring{R}_+\cup\left\{m\delta+\alpha\;|\;m\in\mathbb{Z}_{>0},~\alpha\in\mathring{R}\right\}$ and $R^\text{im}_+=\{m\delta\;|\;m\in\mathbb{Z}_{>0}\}$, where $\mathring{R}$ is the set of roots of the underlying finite Lie algebra and $\delta$ is the minimal positive imaginary root. For each positive root $\alpha$, we choose a basis denoted by $\left\{e^{(\alpha,k)}\right\}$ of $\mathfrak{g}_\alpha$ with a dual basis $\left\{f^{(\alpha,k)}\right\}$ of $\mathfrak{g}_{-\alpha}$ such that $\left(e^{(\alpha,k)},f^{(\alpha,l)}\right)=\delta_{kl}\epsilon_3$, where $(\text{-},\text{-})$ is an invariant inner product on the affine Lie algebra\footnote{Here, we choose the normalisation such that the convention would match the one in \cref{quiverbpsalg}. The modes here and the ones often used in Lie algebras and affine Yangians are related as $e^{(a)}_0=\epsilon_3^{1/2}\mathtt{x}_+^{(a)}$, $f^{(a)}_0=\epsilon_3^{1/2}\mathtt{x}_-^{(a)}$, and $\psi^{(a)}_0=\epsilon_3\mathtt{h}^{(a)}$. For $\alpha_a$, we still have $(\alpha_a,\alpha_a)=2$.}. Here, $k=1,\dots,\dim\mathfrak{g}_\alpha$. When $\alpha$ is a real root, $\dim\mathfrak{g}_\alpha=1$, and we shall simply write $e^{(\alpha)}=e^{(\alpha,1)}$, $f^{(\alpha)}=f^{(\alpha,1)}$, which form an $\mathfrak{sl}_2$ triple with $\psi^{(\alpha)}$. In particular, given a simple root $\alpha_a$, we have $e^{(a)}_0=e^{\left(\alpha_a\right)}$, $f^{(a)}_0=f^{\left(\alpha_a\right)}$, and $\psi^{(a)}_0=\psi^{\left(\alpha_a\right)}$. This is consistent with the fact that the zero modes of the affine Yangian of $\widehat{\mathfrak{gl}}_n$ give (the universal enveloping algebra of) the underlying affine Lie algebra. In what follows, we would also need to take the completions of the algebras (see, for example, \cite{guay2018coproduct,ueda2023construction}), which will be denoted as $\widehat{\mathtt{A}}_n$ and ${}^\iota\mathtt{A}_n\widehat{\otimes}\mathtt{A}_n$.

\paragraph{Embeddings} Let us first give their embeddings into the quiver BPS algebras. Such embeddings would often be denoted as $\iota$. This explains the name of the $\iota$quiver algebras.

We shall start with the rational cases. The affine Yangians were recalled in \cref{quiverbpsalg}.
\begin{theorem}
	For $n\geq3$, there is an injective homomorphism $\iota$ from the $\iota$quiver algebra ${}^\iota\mathtt{Y}_n$ to the affine Yangian $\widehat{\mathtt{Y}}_n$ with $\epsilon_1=\epsilon_2$ given by
	\begin{subequations}
		\begin{align}
			&\iota\left(h^{(a)}_1\right)=2\psi^{(a)}_1-\left(\psi^{(a)}_0\right)^2+\epsilon_3\sum_{\alpha\in R^\textup{re}_+}(\alpha_a,\alpha)\left(e^{(\alpha)}\right)^2\;,\\
			&\iota\left(b^{(a)}_0\right)=(2c_a)^{1/2}\left(e^{(a)}_0-f^{(a)}_0\right)\;.
		\end{align}
	\end{subequations}
	The images of the higher modes can be obtained from \eqref{higherbn3} and \eqref{higherhn3}.
\end{theorem}
\begin{proof}
	The proof is completely analogous to the proof for the twisted Yangians of (finite) type A as in \cite[Theorem 4.1]{Lu_2026}. Therefore, we shall omit it here.
\end{proof}

It would be useful to give the explicit embedding of $b^{(a)}_1$ here:
\begin{equation}
	\iota\left(b^{(a)}_1\right)=(2c_a)^{1/2}\left(e^{(a)}_1+f^{(a)}_1+\frac{1}{4}\sum_{\alpha\in R^\text{re}_+}(\alpha_a,\alpha)\left[\left(e^{(\alpha)}\right)^2,e^{(a)}_0-f^{(a)}_0\right]\right)\;.
\end{equation}
Equivalently,
\begin{equation}
	\iota\left(b^{(a)}_1\right)=(2c_a)^{1/2}\left(e^{(a)}_1+f^{(a)}_1+\frac{1}{2}\sum_{\alpha\in R_+}\sum_{k=1}^{\mathrm{mult}(\alpha)}\left\{\left[e^{(a)}_0,e^{(\alpha,k)}\right],e^{(\alpha,k)}\right\}-\frac{1}{2}\left\{e^{(a)}_0,\psi^{(a)}_0\right\}\right)\;.
\end{equation}

Next, let us consider the $n=1,2$ cases. Notice that we have two independent parameters $\epsilon_1$ and $\epsilon_2$ here. We have the following conjectured embeddings, and we have not checked all the relations.
\begin{conjecture}\label{embeddingconjY2}
	There is an injective homomorphism $\iota$ from the $\iota$quiver algebra ${}^\iota\mathtt{Y}_2$ to the affine Yangian $\widehat{\mathtt{Y}}_2$ given by
	\begin{subequations}
		\begin{align}
			&\iota\left(h^{(a)}_1\right)=2\psi^{(a)}_1-\left(\psi^{(a)}_0\right)^2+\epsilon_3\sum_{\alpha\in R^\text{re}_+}(\alpha_a,\alpha)\left(e^{(\alpha)}\right)^2\;,\\
			&\iota\left(b^{(a)}_0\right)=(2c_a)^{1/2}\left(e^{(a)}_0-f^{(a)}_0\right)\;.
		\end{align}
	\end{subequations}
	The images of the higher modes can be obtained from \eqref{higherbn2} and \eqref{higherhn2}.
\end{conjecture}
\begin{conjecture}\label{embeddingconjY1}
	There is an injective homomorphism $\iota$ from the $\iota$quiver algebra ${}^\iota\mathtt{Y}_1$ to the affine Yangian $\widehat{\mathtt{Y}}_1$ given by\footnote{For $\mathtt{Y}_1$, we are using a slightly different normalisation of the mode expansions from the other cases since this is used in most of the literature (including the physics ones). See \cref{quiverbpsalg}. This can also be seen from the extra $\sigma_3$ coefficients in some of the terms.}
	\begin{subequations}
		\begin{align}
			&\iota(h_1)=2\sigma_3\psi_1-\sigma_3^2\psi_0^2\;,\\
			&\iota(h_3)=2\sigma_3\psi_3-2\sigma_3^2\psi_0\psi_2+\sigma_3^2\psi_1^2+6\sigma_3^2\sum_{m\geq1}mJ_{-m}^2\;,\\
			&\iota(b_0)=(2c\sigma_3)^{1/2}(e_0-f_0)\;.
		\end{align}
	\end{subequations}
	Here,
	\begin{equation}
		J_{-m}=\frac{1}{(m-1)!}\left(\mathrm{ad}_{e_1}\right)^{m-1}e_0\;,\quad m\geq1\;.
	\end{equation}
	The images of the higher modes can be obtained from \eqref{higherbn1} and \eqref{higherhn1}.
\end{conjecture}
We shall also write
\begin{equation}
	J_m=\frac{1}{(m-1)!}\left(\mathrm{ad}_{(-f_1)}\right)^{m-1}(-f_0)\;,\quad m\geq1\;.
\end{equation}

Now, let us consider the trigonometric and the elliptic cases.
\begin{theorem}\label{embeddingtrigell}
	There is an injective homomorphism $\iota$ from the $\iota$quiver algebra ${}^\iota\mathtt{T}_n$ (resp.~${}^\iota\mathtt{E}_n$) to the quantum toroidal algebra $\widehat{\mathtt{T}}_n$ (resp.~the elliptic quantum toroidal algebra $\widehat{\mathtt{E}}_n$) given by
	\begin{subequations}
		\begin{align}
			&\iota\left(h_+^{(a)}(u)\right)=\psi_+^{(a)}(u)\psi_-^{(a)}(-u)\;,\\
			&\iota\left(h_-^{(a)}(u)\right)=\psi_-^{(a)}(u)\psi_+^{(a)}(-u)\;,\\
			&\iota\left(b^{(a)}(u)\right)=\psi_+^{(a)}(-u)e^{(a)}(u)+c_a\left(1+\phi^{aa}(-2u)\right)f^{(a)}(-u)\;.
		\end{align}
	\end{subequations}
	For $n\geq3$, we take $\epsilon_1=\epsilon_2$ in $\widehat{\mathtt{A}}_n$. For $n=1,2$, we have two independent parameters $\epsilon_1$ and $\epsilon_2$.
\end{theorem}
Before we prove this, let us mention that the product of the currents is understood as the product of the modes in the formal expansions of the currents. Moreover, recall that $h_+^{(a)}(u)$ and $h_-^{(a)}(u)$ contain the same modes, and the above embedding also gives the same modes (as the $\psi$ modes commute with each other).
\begin{proof}
	We need to check that the right hand sides of the above expressions satisfy the relations of the $\iota$quiver algebras. Here, we shall only write the proof for the $n\geq3$ cases. The cases with $n=1,2$ are similar. For the two sets of $hh$ relations, it is straightforward as everything commutes.

	For the $hb$ relations, the first term in the image of $b^{(a)}(u)$ satisfies
	\begin{align}
		\psi^{(a)}_\pm(u)\psi^{(a)}_{\mp}(-u)\psi^{(b)}_+(-v)e^{(b)}(v)=&\phi^{ab}(u-v)\phi^{ab}(-u-v)\psi^{(b)}_+(-v)e^{(b)}(v)\psi^{(a)}_\pm(u)\psi^{(a)}_{\mp}(-u)\nonumber\\
		=&\frac{\phi^{ab}(u-v)}{\phi^{ab}(u+v)}\psi^{(b)}_+(-v)e^{(b)}(v)\psi^{(a)}_\pm(u)\psi^{(a)}_{\mp}(-u)\;,
	\end{align}
	where we have used $\phi^{ab}(-u)=\phi^{ab}(u)^{-1}$ due to the identification of the parameters $\epsilon_1$, $\epsilon_2$. The second term is similar.

	For the $bb$ relations, notice that
	\begin{align}
		\psi^{(a)}_+(-u)e^{(a)}(u)\psi^{(b)}_+(-v)e^{(b)}(v)=&\phi^{ab}(u+v)\psi^{(a)}_+(-u)\psi^{(b)}_+(-v)e^{(a)}(u)e^{(b)}(v)\nonumber\\
		=&\phi^{ab}(u+v)\phi^{ab}(u-v)\psi^{(a)}_+(-u)\psi^{(b)}_+(-v)e^{(b)}(v)e^{(a)}(u)\nonumber\\
		=&\phi^{ab}(u-v)\psi^{(b)}_+(-v)e^{(b)}(v)\psi^{(a)}_+(-u)e^{(a)}(u)\;,
	\end{align}
	and likewise for the second term. Therefore, only the cross terms would contribute. For the cross terms, notice that
	\begin{align}
		&\psi^{(a)}_+(-u)e^{(a)}(u)f^{(b)}(-v)-\phi^{ab}(u-v)f^{(b)}(-v)\psi^{(a)}_+(-u)e^{(a)}(u)\nonumber\\
		=&\psi^{(a)}_+(-u)\left[e^{(a)}(u),f^{(b)}(-v)\right]\nonumber\\
		=&-\delta_{ab}\delta(u+v)\psi^{(a)}_+(-u)\left(\psi^{(a)}_+(u)-\psi^{(a)}_-(u)\right)\;,
	\end{align}
	and likewise,
	\begin{align}
		&f^{(a)}(-u)\psi^{(b)}_+(-v)e^{(b)}(v)-\phi^{ab}(u-v)\psi^{(b)}_+(-v)e^{(b)}(v)f^{(a)}(-u)\nonumber\\
		=&\delta_{ab}\delta(u+v)\phi^{aa}(u-v)\psi^{(a)}_+(-v)\left(\psi^{(a)}_+(v)-\psi^{(a)}_-(v)\right)\;.
	\end{align}
	If $b\neq a$, they both vanish, and the relations are satisfied. If $b=a$, multiply the first equation by $c_a\left(1+\phi^{aa}(-2v)\right)$ and the second equation by $c_a\left(1+\phi^{aa}(-2u)\right)$. Using the property of the $\delta$-function and the fact that $\phi^{aa}(2u)\phi^{aa}(-2u)=1$, their sum gives
	\begin{equation}
		c_a\delta(u+v)\left(1+\phi_{aa}(2u)\right)\left(\psi^{(a)}_+(-u)\psi^{(a)}_-(u)-\psi^{(a)}_+(u)\psi^{(a)}_-(-u)\right)
	\end{equation}
	as desired.

	For the Serre relations, after expanding the brackets and using the $bb$ relations, they read
	\begin{align}
		&\mathrm{Sym}_{u_1,u_2}\left\{\frac{\zeta_0(u_1-u_2+\epsilon_3)}
        {\zeta(u_1-u_2+\epsilon_3)}\left[\prod_{j=1}^2\frac{\zeta_0(u_j-v-\epsilon_3/2)}{\zeta(u_j-v-\epsilon_3/2)}b^{(a)}(u_1)b^{(a)}(u_2)b^{(b)}(v)\right.\right.\nonumber\\
	    &-\left(\kappa+\kappa^{-1}\right)\frac{\zeta_0(u_1-v-\epsilon_3/2)\zeta_0(v-u_2-\epsilon_3/2)}{\zeta(u_1-v-\epsilon_3/2)\zeta(v-u_2-\epsilon_3/2)}b^{(a)}(u_1)b^{(b)}(v)b^{(a)}(u_2)\nonumber\\
	    &\left.\left.+\prod_{j=1}^2\frac{\zeta_0(v-u_j-\epsilon_3/2)}{\zeta(v-u_j-\epsilon_3/2)}b^{(b)}(v)b^{(a)}(u_1)b^{(a)}(u_2)\right]\right\}\nonumber\\
	    =&c_a\delta(u_1+u_2)\frac{\zeta_0(\epsilon_3)\zeta_0(2u_1-\epsilon_3)\zeta_0(2u_1+\epsilon_3)}{\zeta(v-u_1+\epsilon_3/2)\zeta(v+u_1+\epsilon_3/2)}\nonumber\\
	    &\times\left(\frac{1}{\zeta(2u_1-\epsilon_3)}+\frac{1}{\zeta(2u_1+\epsilon_3)}\right)\left(h_+^{(a)}(u_1)-h_-^{(a)}(u_1)\right)b^{(b)}(v)\;,
	\end{align}
	where $b=a\pm1$. As $\zeta$ and $\zeta_0$ are odd functions, the factors in front of the three products on the left hand side are
	\begin{align}
        &\frac{\zeta_0(u_1-u_2+\epsilon_3)}{\zeta(u_1-u_2+\epsilon_3)}\frac{\zeta_0(v-u_1+\epsilon_3/2)\zeta_0(v-u_2+\epsilon_3/2)}{\zeta(v-u_1+\epsilon_3/2)\zeta(v-u_2+\epsilon_3/2)}\;,\nonumber\\
        &-\left(\kappa+\kappa^{-1}\right)\frac{\zeta_0(u_1-u_2+\epsilon_3)}{\zeta(u_1-u_2+\epsilon_3)}\frac{\zeta_0(v-u_1+\epsilon_3/2)\zeta_0(v-u_2-\epsilon_3/2)}{\zeta(v-u_1+\epsilon_3/2)\zeta(v-u_2-\epsilon_3/2)}\;,\nonumber\\
        &\frac{\zeta_0(u_1-u_2+\epsilon_3)}{\zeta(u_1-u_2+\epsilon_3)}\frac{\zeta_0(v-u_1-\epsilon_3/2)\zeta_0(v-u_2-\epsilon_3/2)}{\zeta(v-u_1-\epsilon_3/2)\zeta(v-u_2-\epsilon_3/2)}\;,\label{eq:Serre-factors}
    \end{align}
    respectively. For both $u=u_1$ and $u=u_2$, we have
	\begin{equation}
		\phi^{ab}(v-u)=\frac{\zeta(v-u-\epsilon_3/2)}{\zeta(v-u+\epsilon_3/2)}\;,\quad\phi^{ab}(u-v)=\frac{\zeta(v-u+\epsilon_3/2)}{\zeta(v-u-\epsilon_3/2)}\;.\label{eq:Serre-phi-ab}
	\end{equation}
	Therefore, multiplying the factors in \eqref{eq:Serre-factors} by such bond factors only changes the signs in front of $\epsilon_3/2$ in their denominators.
	
	Apart from the common factor $\zeta_0(u_1-u_2+\epsilon_3)/\zeta(u_1-u_2+\epsilon_3)$, the numerators of \eqref{eq:Serre-factors} satisfy
	\begin{align}
		&\zeta_0(v-u_1+\epsilon_3/2)\zeta_0(v-u_2+\epsilon_3/2)-\left(\kappa+\kappa^{-1}\right)\zeta_0(v-u_1+\epsilon_3/2)\zeta_0(v-u_2-\epsilon_3/2)\nonumber\\
		&+\zeta_0(v-u_1-\epsilon_3/2)\zeta_0(v-u_2-\epsilon_3/2)=-\zeta_0(\epsilon_3)\zeta_0(u_1-u_2-\epsilon_3)\;.\label{eq:Serre-key-a}
	\end{align}
    Exchanging $u_1$ and $u_2$ in \eqref{eq:Serre-key-a}, we also have
	\begin{align}
		&\zeta_0(v-u_2+\epsilon_3/2)\zeta_0(v-u_1+\epsilon_3/2)-\left(\kappa+\kappa^{-1}\right)\zeta_0(v-u_2+\epsilon_3/2)\zeta_0(v-u_1-\epsilon_3/2)\nonumber\\
		&+\zeta_0(v-u_2-\epsilon_3/2)\zeta_0(v-u_1-\epsilon_3/2)=\zeta_0(\epsilon_3)\zeta_0(u_1-u_2+\epsilon_3)\;.\label{eq:Serre-key-b}
	\end{align}
	In particular, if the left hand sides of \eqref{eq:Serre-key-a} and \eqref{eq:Serre-key-b} are multiplied by $\zeta_0(u_1-u_2+\epsilon_3)$ and $\zeta_0(u_1-u_2-\epsilon_3)$ respectively, their sum would vanish. We also notice that
	\begin{subequations}
		\begin{align}
			&\frac{\zeta_0(u_2-u_1+\epsilon_3)}{\zeta(u_2-u_1+\epsilon_3)}\phi^{aa}(u_2-u_1)=\frac{\zeta_0(u_1-u_2-\epsilon_3)}{\zeta(u_1-u_2+\epsilon_3)}\;,\label{eq:Serre-kappa}\\
			&\frac{1+\phi^{aa}(2u)}{\zeta(2u+\epsilon_3)}=\frac{1}{\zeta(2u-\epsilon_3)}+\frac{1}{\zeta(2u+\epsilon_3)}\;.\label{eq:Serre-g}
		\end{align}
	\end{subequations}

    Now, recall that $\iota\left(b^{(a)}(u)\right)$ is the sum of its $e$ part $\psi^{(a)}_+(-u)e^{(a)}(u)$ and its $f$ part $c_a\left(1+\phi^{aa}(-2u)\right)f^{(a)}(-u)$. Moreover,
	\begin{equation}
		\iota\left(h^{(a)}_+(u)-h^{(a)}_-(u)\right)=\psi^{(a)}_+(u)\psi^{(a)}_-(-u)-\psi^{(a)}_-(u)\psi^{(a)}_+(-u)\;.\label{eq:Serre-iota-h}
	\end{equation}
	Using the relations among $e$, $f$, and $\psi_\pm$, we have
	\begin{subequations}
		\begin{align}
			&f^{(b)}(-v)\psi^{(a)}_+(-u)e^{(a)}(u)=\phi^{ab}(v-u)\psi^{(a)}_+(-u)e^{(a)}(u)f^{(b)}(-v)\;,\nonumber\\
			&f^{(a)}(-u)\psi^{(b)}_+(-v)e^{(b)}(v)=\phi^{ab}(u-v)\psi^{(b)}_+(-v)e^{(b)}(v)f^{(a)}(-u)\;,\label{eq:Serre-ex1}\\
			&f^{(a)}(-u_2)\psi^{(a)}_+(-u_1)e^{(a)}(u_1)=\phi^{aa}(u_2-u_1)\Big(\psi^{(a)}_+(-u_1)e^{(a)}(u_1)f^{(a)}(-u_2)\nonumber\\
			&\qquad+\delta(u_1+u_2)\psi^{(a)}_+(-u_1)\left(\psi^{(a)}_+(u_1)-\psi^{(a)}_-(u_1)\right)\Big)\;,\label{eq:Serre-ex2}\\
			&f^{(a)}(-u_2)f^{(a)}(-u_1)=\phi^{aa}(u_2-u_1)f^{(a)}(-u_1)f^{(a)}(-u_2)\;,\label{eq:Serre-ex3a}\\
			&f^{(b)}(-v)f^{(a)}(-u)=\phi^{ab}(v-u)f^{(a)}(-u)f^{(b)}(-v)\;,\label{eq:Serre-ex3b}\\
			&\psi^{(a)}_+(-u)\psi^{(a)}_{\pm}(u)\psi^{(b)}_+(-v)e^{(b)}(v)=\frac{\phi^{ab}(u-v)}{\phi^{ab}(u+v)}\psi^{(b)}_+(-v)e^{(b)}(v)\psi^{(a)}_+(-u)\psi^{(a)}_{\pm}(u)\;,\nonumber\\
			&f^{(b)}(-v)\psi^{(a)}_+(-u)\psi^{(a)}_{\pm}(u)=\phi^{ab}(v-u)\phi^{ab}(v+u)\psi^{(a)}_+(-u)\psi^{(a)}_{\pm}(u)f^{(b)}(-v)\;.\label{eq:Serre-ex4}
		\end{align}
	\end{subequations}

	Consider the degrees of the currents in $\mathtt{A}_n$ given by their nodes. In other words, we may denote the degree of the current $e^{(a)}(u)$ ($f^{(a)}(u)$, resp.~$\psi_\pm^{(a)}(u)$) by (the corresponding simple root) $\alpha_a$ ($-\alpha_a$, resp.~$0$). Now, the left hand side of Serre relation is given either by its $e$ part or by its $f$ part. The resulting terms have degree $2\alpha_a+\alpha_b$ (resp.~$-2\alpha_a-\alpha_b$) if all the three currents are replaced by their $e$ parts (resp.~$f$ parts). They have degree $2\alpha_a-\alpha_b$ (resp.~$-2\alpha_a+\alpha_b$) if the two currents of colour $a$ are replaced by their $e$ parts (resp.~$f$ parts) and the current of colour $b$ by its $f$ part (resp.~$e$ part). They have degree $\alpha_b$ (resp. $-\alpha_b$) if the two currents of colour $a$ are replaced by one $e$ part and one $f$ part, and the current of colour $b$ by its $e$ part (resp.~$f$ part). By \eqref{eq:Serre-iota-h}, the right hand side only has components of degrees $\pm\alpha_b$. Since $\alpha_a$ and $\alpha_b$ are linearly independent, these six degrees are pairwise distinct, and the Serre relation would hold if and only if it holds in each degree separately.

	First, consider the terms of degree $2\alpha_a+\alpha_b$. Moving all the $\psi$ factors to the left by the $\psi e$ relations, each of the three products of $\psi^{(a)}_+(-u_1)e^{(a)}(u_1)$, $\psi^{(a)}_+(-u_2)e^{(a)}(u_2)$ and $\psi^{(b)}_+(-v)e^{(b)}(v)$ becomes
	\begin{equation}
		\phi^{aa}(u_1+u_2)\phi^{ab}(u_1+v)\phi^{ab}(u_2+v)\psi^{(a)}_+(-u_1)\psi^{(a)}_+(-u_2)\psi^{(b)}_+(-v)
	\end{equation}
	multiplied by the product of $e^{(a)}(u_1)$, $e^{(a)}(u_2)$ and $e^{(b)}(v)$ in the same order. Indeed, each pair of the currents contributes its bond factor with the sum of their spectral parameters, which does not depend on their order. As this prefactor is symmetric in $u_1$ and $u_2$, the terms of degree $2\alpha_a+\alpha_b$ vanish by the Serre relation of $\mathtt{A}_n$ for the $e$ currents. This is similar for the terms of degree $-2\alpha_a-\alpha_b$.

    Next, consider the terms of degree $2\alpha_a-\alpha_b$. Using \eqref{eq:Serre-ex1}, we move $f^{(b)}(-v)$ to the right. By \eqref{eq:Serre-factors} and \eqref{eq:Serre-phi-ab}, the sum of their coefficients is
	\begin{equation}
		\frac{\zeta_0(u_1-u_2+\epsilon_3)}{\zeta(u_1-u_2+\epsilon_3)\zeta(v-u_1+\epsilon_3/2)\zeta(v-u_2+\epsilon_3/2)}
	\end{equation}
	multiplied by the left hand side of \eqref{eq:Serre-key-a}. Hence, by \eqref{eq:Serre-key-a}, the terms of degree $2\alpha_a-\alpha_b$ are
	\begin{align}
		&-c_b\left(1+\phi^{bb}(-2v)\right)\zeta_0(\epsilon_3)\mathrm{Sym}_{u_1,u_2}\bigg(\frac{\zeta_0(u_1-u_2+\epsilon_3)\zeta_0(u_1-u_2-\epsilon_3)}{\zeta(u_1-u_2+\epsilon_3)\zeta(v-u_1+\epsilon_3/2)\zeta(v-u_2+\epsilon_3/2)}\nonumber\\
		&\psi^{(a)}_+(-u_1)e^{(a)}(u_1)\psi^{(a)}_+(-u_2)e^{(a)}(u_2)f^{(b)}(-v)\bigg)\;.
	\end{align}
	Together with the term with $u_1$ and $u_2$ exchanged, the two terms in the symmetrisation cancel. This is similar for the terms of degree $-2\alpha_a+\alpha_b$.

	Now, consider the terms of degree $\alpha_b$. The terms where the $f$ part is at $u_2$ and those where it is at $u_1$ are exchanged under $u_1\leftrightarrow u_2$. Therefore, let us collect all the terms involving the $f$ part with $u_2$. In these terms, we move $f^{(a)}(-u_2)$ to the right of the other two currents, and then $\psi^{(b)}_+(-v)e^{(b)}(v)$ to the right of $\psi^{(a)}_+(-u_1)e^{(a)}(u_1)$. Up to $\delta$-terms, every term then becomes a multiple of
	\begin{equation}
		\psi^{(a)}_+(-u_1)e^{(a)}(u_1)\psi^{(b)}_+(-v)e^{(b)}(v)f^{(a)}(-u_2)\;.\label{eq:Serre-normal-plus}
	\end{equation}
	As for the terms of degree $\pm(2\alpha_a-\alpha_b)$, the coefficients of \eqref{eq:Serre-normal-plus} coming from the two summands of the symmetrisation are given by the left hand sides of \eqref{eq:Serre-key-a} and \eqref{eq:Serre-key-b}, multiplied by $\zeta_0(u_1-u_2+\epsilon_3)$ and $\zeta_0(u_1-u_2-\epsilon_3)$ respectively, with a common denominator. By \eqref{eq:Serre-key-a} and \eqref{eq:Serre-key-b}, they cancel each other. The only remaining terms are the $\delta$-terms produced by \eqref{eq:Serre-ex2}, which arise in the summand where the current with $u_2$ is the first current of colour $a$.

	For the $\delta$-terms, we move $\psi_+^{(b)}(-v)e^{(b)}(v)$ to the right of $\psi_+^{(a)}(-u_1)\left(\psi_+^{(a)}(u_1)-\psi_-^{(a)}(u_1)\right)$ using \eqref{eq:Serre-ex4}. Including the factor $c_a\left(1+\phi^{aa}(-2u_2)\right)$ of the $f$ part and using the property of the formal $\delta$-function, together with \eqref{eq:Serre-key-b} and \eqref{eq:Serre-g}, these $\delta$-terms add up to
	\begin{align}
		&c_a\delta(u_1+u_2)\frac{\zeta_0(\epsilon_3)\zeta_0(2u_1-\epsilon_3)\zeta_0(2u_1+\epsilon_3)}{\zeta(v-u_1+\epsilon_3/2)\zeta(v+u_1+\epsilon_3/2)}\left(\frac{1}{\zeta(2u_1-\epsilon_3)}+\frac{1}{\zeta(2u_1+\epsilon_3)}\right)\nonumber\\
		&\psi^{(a)}_+(-u_1)\left(\psi^{(a)}_+(u_1)-\psi^{(a)}_-(u_1)\right)\psi^{(b)}_+(-v)e^{(b)}(v)\;.
	\end{align}
	The first line without $c_a\delta(u_1+u_2)$ is the factor on the right hand side of the Serre relation. It is odd in $u_1$. Therefore, adding the image under $u_1\leftrightarrow u_2$ replaces $\psi^{(a)}_+(-u_1)\left(\psi^{(a)}_+(u_1)-\psi^{(a)}_-(u_1)\right)$ by $\psi^{(a)}_+(u_1)\psi^{(a)}_-(-u_1)-\psi^{(a)}_-(u_1)\psi^{(a)}_+(-u_1)$. Therefore, the terms of degree $\alpha_b$ agree with the right hand side of the Serre relation. This is similar for the terms of degree $-\alpha_b$. Adding the components of all the degrees shows that the Serre relation is satisfied.

	Now, let us prove injectivity. Similar to the rational cases (see \cite[Section 4.3]{Lu_2026}), we pass it to the associated graded algebras. On $\mathtt{A}$, there is a grading such that all the $e$ ($f$, resp.~$\psi_\pm$) modes have degree $1$ ($-1$, resp.~$0$). On ${}^\iota\mathtt{A}$, there is a filtration such that all the $b$ (resp.~$h$) modes have degree $1$ (resp.~$0$). The embedding $\iota$ is a filtered homomorphism, and it induces a homomorphism $\mathrm{gr}(\iota):~\mathrm{gr}{}^\iota\mathtt{A}\rightarrow\mathrm{gr}\mathtt{A}$ by
	\begin{equation}
		b^{(a)}(u)\mapsto\psi_+^{(a)}(-u)e^{(a)}(u)\;,\quad h^{(a)}(u)\mapsto\psi_+^{(a)}(u)\psi_-^{(a)}(-u)\;.
	\end{equation}
	Also define the homogeneous algebra ${}^\iota\mathtt{A}_H$ by keeping only the parts with the highest degrees of the relations of ${}^\iota\mathtt{A}$ (in the sense of the above filtration). In other words, the relations of ${}^\iota\mathtt{A}_H$ are (i) \eqref{hh}, (ii) \eqref{h}, (iii) \eqref{hb}, (iv) \eqref{bb} with its right hand side replaced by zero, and (v) the Serre relations with the terms proportional to $c_{a_s}$ in $\mathfrak{D}$ removed. In particular, ${}^\iota\mathtt{A}_H$ surjects onto $\mathrm{gr}{}^\iota\mathtt{A}$.

	Moreover, consider the enlarged algebra ${}^\iota\mathtt{A}_E$ generated by ${}^\iota\mathtt{A}_H$ together with the modes of $\psi_+^{(a)}(-u)$ and $\left(\psi^{(a)}_{+,0}\right)^{-1}$ for all $a$. Besides the relations as stated in ${}^\iota\mathtt{A}_H$, the enlarged algebra ${}^\iota\mathtt{A}_E$ also has the relations
	\begin{subequations}\label{AErelations}
		\begin{align}
			&\left[\psi^{(a)}_+(-u),\psi^{(b)}_+(-v)\right]=\left[\psi^{(a)}_+(-u),h^{(b)}_\pm(v)\right]=0\;,\\
			&\psi^{(a)}_{+,0}\left(\psi^{(a)}_{+,0}\right)^{-1}=\left(\psi^{(a)}_{+,0}\right)^{-1}\psi^{(a)}_{+,0}=1\;,\\
			&\psi^{(a)}_+(-u)b^{(b)}(v)=\phi^{ab}(u+v)^{-1}b^{(b)}(v)\psi^{(a)}_+(-u)\;,\label{psibE}
		\end{align}
	\end{subequations}
	where $\phi^{ab}(u+v)^{-1}$ is expanded in non-negative powers of $UV$ since $\psi^{(a)}_+(-u)=\sum\limits_{i\geq0}\psi^{(a)}_{+,i}U^i$ in the expansion\footnote{In the elliptic cases, the mode relations involve infinite sums. In particular, the rescaling map $\sigma^{(a)}_0$ and the action $\psi_i\cdot(y\otimes p)$ below would be sums over $k\in\mathbb{Z}$. Therefore, we should consider the reduction modulo $q$ as in the $q$-adic topology. Alternatively, we need to assume a completion for the discussions.}. The first two lines are needed for the inverse map below. We now show that the natural map $\mathfrak{j}:~{}^\iota\mathtt{A}_H\rightarrow{}^\iota\mathtt{A}_E$ is injective by constructing an ${}^\iota\mathtt{A}_E$-module where ${}^\iota\mathtt{A}_H$ acts faithfully.
	
	For each node $a$, define $\sigma^{(a)}(u)$ on the generators of ${}^\iota\mathtt{A}_H$ by
	\begin{equation}
		\sigma^{(a)}(u)\left(h^{(b)}_\pm(v)\right)=h^{(b)}_\pm(v)\;,\quad\sigma^{(a)}(u)\left(b^{(b)}(v)\right)=\phi^{ab}(u+v)^{-1}b^{(b)}(v)\;.
	\end{equation}
	In terms of the modes, write $\phi^{ab}(u+v)^{-1}=\sum\limits_{k\geq0}\upphi^{ab}_k(UV)^k$. Then $\sigma^{(a)}(u)\left(b^{(b)}_j\right)=\sum\limits_{k\geq0}\upphi^{ab}_kU^kb^{(b)}_{j+k}$. Every relation of ${}^\iota\mathtt{A}_H$ is a linear combination, with scalar coefficient functions, of products of the same currents $b^{(a_1)}(u_1),\dots,b^{(a_m)}(u_m)$ (and possibly $h$ currents) in different orders. Therefore, $\sigma^{(a)}(u)$ maps each relation to itself multiplied by the common factor $\prod\limits_{r=1}^m\phi^{aa_r}(u+u_r)^{-1}$, and hence extends to an algebra homomorphism
	\begin{equation}
		\sigma^{(a)}(u)=\sum_{k\geq0}U^k\sigma^{(a)}_k:~{}^\iota\mathtt{A}_H\rightarrow{}^\iota\mathtt{A}_H[\![U]\!]\;.
	\end{equation}
	The map $\sigma^{(a)}_0$ rescales each $b^{(b)}_j$ by the non-zero constant $\upphi^{ab}_0$, and is therefore invertible. Moreover, all the maps $\sigma^{(a)}(u)$ commute with each other since they act on the generators by rescaling.
	
	Now, let $\mathcal{P}=\mathbb{C}\left[\psi^{(a)}_{+,i},\left(\psi^{(a)}_{+,0}\right)^{-1}\right]$ be the commutative algebra with $a\in Q_0$ and $i\in\mathbb{Z}_{\geq0}$. Consider the vector space $\mathcal{E}={}^\iota\mathtt{A}_H\otimes\mathcal{P}$. For any $x,y\in{}^\iota\mathtt{A}_H$ and $p\in\mathcal{P}$, define
	\begin{subequations}
		\begin{align}
			&x\cdot(y\otimes p)=xy\otimes p\;,\\
			&\psi^{(a)}_+(-u)\cdot(y\otimes p)=\sigma^{(a)}(u)(y)\otimes\psi^{(a)}_+(-u)p\;,\\
			&\left(\psi^{(a)}_{+,0}\right)^{-1}\cdot(y\otimes p)=\left(\sigma^{(a)}_0\right)^{-1}(y)\otimes\left(\psi^{(a)}_{+,0}\right)^{-1}p\;.
		\end{align}
	\end{subequations}
	In terms of the modes, the second line reads $\psi^{(a)}_{+,i}\cdot(y\otimes p)=\sum\limits_{k=0}^i\sigma^{(a)}_k(y)\otimes\psi^{(a)}_{+,i-k}p$, which is a finite sum. These operators satisfy all the relations of ${}^\iota\mathtt{A}_E$ as can be seen as follows:
	\begin{itemize}
		\item The relations of ${}^\iota\mathtt{A}_H$ hold since ${}^\iota\mathtt{A}_H$ acts by left multiplication.
		\item The $\psi$ modes commute with each other since the maps $\sigma^{(a)}(u)$ commute and $\mathcal{P}$ is commutative. They commute with the $h$ modes since $\sigma^{(a)}(u)(xy)=x\sigma^{(a)}(u)(y)$ for any $h$ mode $x$. As $\psi^{(a)}_{+,0}$ acts as $\sigma^{(a)}_0\otimes\psi^{(a)}_{+,0}$, the operator defined for $\left(\psi^{(a)}_{+,0}\right)^{-1}$ is indeed its inverse.
		\item For \eqref{psibE}, since $\sigma^{(a)}(u)$ is a homomorphism, we have
		\begin{align}
			\psi^{(a)}_+(-u)b^{(b)}(v)\cdot(y\otimes p)=&\sigma^{(a)}(u)\left(b^{(b)}(v)\right)\sigma^{(a)}(u)(y)\otimes\psi^{(a)}_+(-u)p\nonumber\\
			=&\phi^{ab}(u+v)^{-1}b^{(b)}(v)\psi^{(a)}_+(-u)\cdot(y\otimes p)\;.
		\end{align}
	\end{itemize}
	Therefore, $\mathcal{E}$ is an ${}^\iota\mathtt{A}_E$-module. For any $y\in{}^\iota\mathtt{A}_H$, we have $\mathfrak{j}(y)\cdot(1\otimes1)=y\otimes1$. If $\mathfrak{j}(y)=0$, then $y\otimes1=0$, i.e., $y=0$. Hence, $\mathfrak{j}$ is injective\footnote{In fact, since the products $yp$ span ${}^\iota\mathtt{A}_E$ by \eqref{AErelations} and $yp\cdot(1\otimes1)=y\otimes p$, the same module shows that ${}^\iota\mathtt{A}_E\cong{}^\iota\mathtt{A}_H\otimes\mathcal{P}$ as vector spaces.}. Therefore, there is an embedding ${}^\iota\mathtt{A}_H\hookrightarrow{}^\iota\mathtt{A}_E$.

	Following the map
	\begin{equation}
		b^{(a)}(u)\mapsto\psi_+^{(a)}(-u)e^{(a)}(u)\;,\quad h^{(a)}(u)\mapsto\psi_+^{(a)}(u)\psi_-^{(a)}(-u)\;,\quad\psi_+^{(a)}(-u)\mapsto\psi_+^{(a)}(-u)\;,
	\end{equation}
	with the inverse map
	\begin{equation}
		e^{(a)}(u)\mapsto\psi_+^{(a)}(-u)^{-1}b^{(a)}(u)\;,\quad\psi_-^{(a)}(-u)\mapsto h^{(a)}_+(u)\psi_+^{(a)}(u)^{-1}\;,\quad\psi_+^{(a)}(-u)\mapsto\psi_+^{(a)}(-u)\;,
	\end{equation}
	we have ${}^\iota\mathtt{A}_E\cong\mathtt{A}^{\geq0}$. Since the natural embedding $\mathtt{A}^{\geq0}\hookrightarrow\mathtt{A}$ is injective, the composition of the following maps
	\begin{equation}
		{}^\iota\mathtt{A}_H\hookrightarrow{}^\iota\mathtt{A}_E\xrightarrow{\cong}\mathtt{A}^{\geq0}\hookrightarrow\mathtt{A}
	\end{equation}
	is injective. Therefore, following the diagram
	\begin{equation}
		{}^\iota\mathtt{A}_H\twoheadrightarrow\mathrm{gr}{}^\iota\mathtt{A}\xhookrightarrow{\mathrm{gr}(\iota)}\mathrm{gr}\mathtt{A}\;,
	\end{equation}
	the map $\mathrm{gr}(\iota)$ is injective, and hence $\iota$ is injective.
\end{proof}

\paragraph{The \texorpdfstring{$J$-presentation}{J-presentation}} Let us make a comment on the $J$-presentations. The $J$-presentation of the affine Yangians is a useful presentation and has been used to find the coproducts of the affine Yangians \cite{guay2018coproduct,ueda2023construction} and prove the isomorphisms under Seiberg duality for the super cases \cite{Bao:2022jhy}. With the embedding $\iota$, it could be possible to write down the $J$-presentation of ${}^\iota\mathtt{A}_{n\geq3}$. It would then be interesting to compare this with the definition of some twisted affine Yangians in \cite{ueda2025twisted,harako2026new}.

\paragraph{Coideal structures} With the above embeddings, we may write down the right coactions for ${}^\iota\mathtt{A}$. In particular, we shall take the following coproducts $\Delta_A:~\mathtt{A}\rightarrow\mathtt{A}\widehat{\otimes}\mathtt{A}$. For $\mathtt{Y}_{n\geq3}$ \cite{guay2018coproduct},
\begin{subequations}
	\begin{align}
		&\Delta_A\left(\psi^{(a)}_0\right)=\psi^{(a)}_0\otimes1+1\otimes\psi^{(a)}_0\;,\\
		&\Delta_A\left(\psi^{(a)}_1\right)=\psi^{(a)}_1\otimes1+1\otimes\psi^{(a)}_1+\psi^{(a)}_0\otimes\psi^{(a)}_0-\epsilon_3\sum_{\alpha\in R^\text{re}_+}(\alpha_a,\alpha)f^{(\alpha)}\otimes e^{(\alpha)}\;,\\
		&\Delta_A\left(e^{(a)}_0\right)=e^{(a)}_0\otimes1+1\otimes e^{(a)}_0\;,\\
		&\Delta_A\left(f^{(a)}_0\right)=f^{(a)}_0\otimes1+1\otimes f^{(a)}_0\;.
	\end{align}
\end{subequations}
For $\mathtt{Y}_1$ \cite{schiffmann2013cherednik,Prochazka:2015deb},
\begin{subequations}
	\begin{align}
		&\Delta_A(\psi_0)=\psi_0\otimes1+1\otimes\psi_0\;,\\
		&\Delta_A(\psi_1)=\psi_1\otimes1+1\otimes\psi_1+\sigma_3\psi_0\otimes\psi_0\;,\\
		&\Delta_A(\psi_2)=\psi_2\otimes1+1\otimes\psi_2+\sigma_3\psi_1\otimes\psi_0+\sigma_3\psi_0\otimes\psi_1\;,\\
		&\Delta_A(\psi_3)=\psi_3\otimes1+1\otimes\psi_3+\sigma_3\psi_2\otimes\psi_0+\sigma_3\psi_1\otimes\psi_1+\sigma_3\psi_0\otimes\psi_2+6\sigma_3\sum_{m>0}mJ_m\otimes J_{-m}\;,\\
		&\Delta_A(e_0)=e_0\otimes1+1\otimes e_0\;,\\
		&\Delta_A(f_0)=f_0\otimes1+1\otimes f_0\;.
	\end{align}
\end{subequations}
For $\mathtt{Y}_2$, the explicit expression of a similar coproduct has not yet been written down to the best of our knowledge. For the trigonometric and the elliptic cases \cite{Noshita:2021ldl,Galakhov:2021vbo},
\begin{subequations}
	\begin{align}
		&\Delta_A\left(\psi_\pm^{(a)}(u)\right)=\psi_\pm^{(a)}(u)\otimes\psi_\pm^{(a)}(u)\;,\\
		&\Delta_A\left(e^{(a)}(u)\right)=e^{(a)}(u)\otimes1+\psi^{(a)}_-(u)\otimes e^{(a)}(u)\;,\\
		&\Delta_A\left(f^{(a)}(u)\right)=f^{(a)}(u)\otimes\psi_+^{(a)}(u)+1\otimes f^{(a)}(u)\;.
	\end{align}
\end{subequations}
\begin{corollary}
	Consider the homomorphisms $\Delta:~{}^\iota\mathtt{A}_n\rightarrow{}^\iota\mathtt{A}_n\widehat{\otimes}\mathtt{A}_n$ in the rational cases given by
	\begin{subequations}
		\begin{align}
			&\Delta\left(h^{(a)}_1\right)=h^{(a)}_1\otimes1+1\otimes h^{(a)}_1+2\epsilon_3\sum_{\alpha\in R^\text{re}_+}(\alpha_a,\alpha)\left(e^{(\alpha)}-f^{(\alpha)}\right)\otimes e^{(\alpha)}\;,\\
			&\Delta\left(b^{(a)}_0\right)=b^{(a)}_0\otimes1+1\otimes b^{(a)}_0\;,
		\end{align}
	\end{subequations}
	for $n\geq3$, and in the trigonometric and elliptic cases given by
	\begin{subequations}
		\begin{align}
			&\Delta\left(h_\pm^{(a)}(u)\right)=h_\pm^{(a)}(u)\otimes\psi_\pm^{(a)}(u)\psi_\mp^{(a)}(-u)\;,\\
			&\Delta\left(b^{(a)}(u)\right)=b^{(a)}(u)\otimes\psi_+^{(a)}(-u)+h_-^{(a)}(u)\otimes\psi_+^{(a)}(-u)e^{(a)}(u)+c_a(1+\phi^{aa}(-2u))(1\otimes f^{(a)}(-u))\;,
		\end{align}
	\end{subequations}
	for $n\geq1$. They satisfy the compatibility conditions
	\begin{subequations}
		\begin{align}
			&(\iota\otimes\mathrm{id})\Delta=\Delta_A\iota\;,\\
			&(\mathrm{id}\otimes\varepsilon_A)\Delta=\mathrm{id}\;,
		\end{align}
	\end{subequations}
	where $\varepsilon_A$ is the counit on $\mathtt{A}_n$ that sends the $e^{(a)}(u)$, $f^{(a)}(u)$ currents to $0$, and the $\psi_\pm^{(a)}(u)$ currents to $1$. If \cref{embeddingconjY2} is true and the above coproduct for $\mathtt{Y}_{n\geq3}$ extends to $\mathtt{Y}_2$, then the above $\Delta$ for ${}^\iota\mathtt{Y}_{n\geq3}$ also extends to ${}^\iota\mathtt{Y}_2$. If \cref{embeddingconjY1} is true, then $\Delta$ given by
	\begin{subequations}
		\begin{align}
			&\Delta(h_1)=h_1\otimes1+1\otimes h_1\;,\\
			&\Delta(h_3)=h_3\otimes1+1\otimes h_3+h_1\otimes h_1+12\sigma_3^2\sum_{m\geq1}m(J_m+J_{-m})\otimes J_{-m}\;,\\
			&\Delta(b_0)=b_0\otimes1+1\otimes b_0
		\end{align}
	\end{subequations}
	for ${}^\iota\mathtt{Y}_1$ satisfies the above compatibility conditions for $\Delta$.
\end{corollary}
We notice that
\begin{equation}
	(\Delta_A\otimes\mathrm{id})\Delta_A\iota=(\Delta_A\otimes\mathrm{id})(\iota\otimes\mathrm{id})\Delta=(\Delta_A\iota\otimes\mathrm{id})\Delta=(\iota\otimes\mathrm{id}\otimes\mathrm{id})(\Delta\otimes\mathrm{id})\Delta\;.
\end{equation}
Then
\begin{equation}
	(\iota\otimes\mathrm{id}\otimes\mathrm{id})(\Delta\otimes\mathrm{id})\Delta=(\Delta_A\otimes\mathrm{id})\Delta_A\iota=(\mathrm{id}\otimes\Delta_A)\Delta_A\iota=(\iota\otimes\mathrm{id}\otimes\mathrm{id})(\mathrm{id}\otimes\Delta_A)\Delta\;,
\end{equation}
where we have used the coassociativity of $\Delta_A$ in the second equality. Since $\iota$ is injective, the compatibility condition implies that
\begin{equation}
	(\Delta\otimes\mathrm{id})\Delta=(\mathrm{id}\otimes\Delta_A)\Delta\;.
\end{equation}

\section{Involutions and Orientifolds}\label{involutions}
As the $\iota$quiver algebras involve some involutions, it is natural to wonder if they can be related to the orientifold actions on the quivers and brane tilings. It is tempting to simply identify the involution $\tau$ above as the orientifold action $\sigma$ on the quiver. However, it does not seem to be the case.

\paragraph{Twisted Yangians and involutions} Let us start with the discussions of $\tau$, along with some other involutions, and see how they would arise on the algebra side. We briefly recall that given a Yangian associated to some (finite) Lie algebra $\mathfrak{g}$, a twisted Yangian may be constructed via the symmetric pair $\left(\mathfrak{g},\mathfrak{g}^\theta\right)$. Here, $\theta$ is some involution on $\mathfrak{g}$ such that it decomposes $\mathfrak{g}$ into the positive eigenspace $\mathfrak{l}=\mathfrak{g}^\theta=\{x\in\mathfrak{g}\;|\;\theta(x)=x\}$ and the negative eigenspace $\mathfrak{m}=\{x\in\mathfrak{g}\;|\;\theta(x)=-x\}$. The Casimir element of $\mathfrak{g}$ may be decomposed into $C_\mathfrak{l}+C_\mathfrak{m}$. Choose a basis $\{X_i\}$ (resp.~$\{Y_j\}$) of $\mathfrak{l}$ (resp.~$\mathfrak{m}$). The twisted Yangian is isomorphic to the subalgebra of the Yangian generated by $X_i$ and $B(Y_j)$, where $B(Y_j)=J(Y_j)+\frac{\hbar}{4}[Y_j,C_\mathfrak{l}]$ \cite{Belliard:2014uja}. Here, $\hbar$ is the parameter of the Yangian, and $J(Y_j)$ denote the level-one generators in the $J$-presentation.

Now, a twisted Yangian is said to be of split type if $\theta$ is conjugate to the Chevalley involution $\varpi$. The Chevalley involution $\varpi$ acts on the $\mathfrak{sl}_2$ triples as $\mathtt{x}_+^{(a)}\rightarrow-\mathtt{x}_-^{(a)}$, $\mathtt{x}_-^{(a)}\rightarrow-\mathtt{x}_+^{(a)}$, and $\mathtt{h}^{(a)}\rightarrow-\mathtt{h}^{(a)}$. It is said to be of quasi-split type if $\theta$ is conjugate to $\varpi\circ\tau$, where $\tau$ is an involution of the Dynkin diagram. As we are taking $\tau=\mathrm{id}$ here, we are considering the split type. Recently, the Drinfeld presentations of the twisted Yangians were found in \cite{Lu_2025,lu2025affine,Lu_2026} for the split type and in \cite{2024arXiv240806981L,huang2026quasi} for the quasi-split type. See also \cite{lu2026classification}.

\paragraph{Satake diagrams} For different $\tau$, it would be convenient to consider the Satake diagrams \cite{satake1960representations}. Given a symmetric pair with $\mathfrak{g}$ semisimple, the Satake diagrams were classified in \cite{araki1962root}. The Satake diagram is the Dynkin diagram with decorations determined by the construction as follows. Consider some involution $\varrho$ acting on the root system $R$. Denote the set of $\varrho$-fixed roots as $R_\bullet$. Let $\alpha_{1,\dots,m}$ be the simple roots of $R$ with $\alpha_{p+1,\dots,m}$ being the simple roots of $R_\bullet$. Then this defines $\tau$ by
\begin{equation}
	\varrho(\alpha_a)\in-\alpha_{\tau(a)}+\mathbb{Z}R_\bullet\;,\quad a=1,\dots,p\;.
\end{equation}
The Satake diagram is the Dynkin diagram with the nodes corresponding to $\alpha_{p+1,\dots,m}$ painted black, and we draw a bidirectional arrow between each pair of white nodes that are exchanged by $\tau$. On the black nodes, $\tau$ acts as the opposition involution, i.e., $\tau|_\bullet=-w_\bullet$, where $w_\bullet$ is the longest Weyl element generated by the root system of the black nodes. In particular, $\varrho=-w_\bullet\circ\tau$. When there are no black nodes and $\tau=\mathrm{id}$ (resp.~$\tau\neq\mathrm{id}$), we say that this is of split (resp.~quasi-split) type. In what follows, we shall mainly focus on the cases without black nodes.

\paragraph{Orientifolds} Now, consider the brane tiling and the quiver gauge theory with an orientifold action \cite{Franco:2007ii}. It acts on the D3-brane probing CY3 system by $\omega\xi(-1)^{F_L}$. Here, $\omega$ reverses the string worldsheet orientation, and $\xi$ is an involution of the CY3. Moreover, $F_L$ is the left-moving fermion number. To preserve supersymmetry, the holomorphic 3-form of the CY3 should transform as $\Omega\rightarrow-\Omega$. For simplicity, we shall collectively denote the orientifold action as $\sigma$.

As a result, there could be various orientifold actions on the brane tiling and the quiver. For instance, there are point and line orientifolds depending on their projections on the 2-torus of the brane tiling system as discussed in \cite{Franco:2007ii}. Besides, there are orientifolds that do not have fixed points on the brane tiling, giving rise to the Klein bottle \cite{Garcia-Valdecasas:2021znu}. In particular, a glide shift would still preserve supersymmetry while a shift of the fundamental cell would break supersymmetry. Readers are also referred to \cite{Argurio:2020npm,Argurio:2020dko} for more discussions on orientifolds and brane tilings.

To discuss the ($\iota$)quiver (BPS) algebras associated to these gauge theories, it would therefore be important to find out how the involutions on the algebra side would match with the orientifold actions on the gauge theory side. So far, we do not have a concrete dictionary on this. Here, we would only give some very naive discussions for a small family of quivers. Even within these cases, there are still many things unclear.

\paragraph{Finite types} Although they are not from toric CY, let us first consider the cases associated to finite Lie algebras. The gauge theories are described by the tripled quivers of the Dynkin diagrams. In \cite{li2018quiver}, an involution on the Nakajima quiver variety was considered. This involution should be the one induced by the orientifold action $\sigma$. The fixed-point subvariety is dubbed $\sigma$-quiver variety somewhat coincidentally (it is a quite common choice to use $\sigma$, so it may not be that coincidental). Here, we shall mainly follow the discussions in \cite[Section 3]{Nakajima:2025utt}.

Following the McKay correspondence, one can associate an irreducible representation $\rho_a$ of $\Gamma$, where $\Gamma$ is a finite $\SU(2)$ subgroup, to each node in the corresponding affine ADE Dynkin diagram. This gives the corresponding fractional brane in the Chan--Paton space \cite{Nakajima:1994nid,Douglas:1996sw,Diaconescu:1997br}:
\begin{equation}
	\mathcal{C}=\bigoplus_aV_a\otimes\rho_a\;,
\end{equation}
where $V_a$ is the multiplicity space with the number of fractional branes given by $\dim V_a$. Consider the involution $\star$ acting on the nodes such that $\rho_{a^\star}\cong\rho_a^*$, where $*$ denotes the dual representation.

Then this induces an involution on the underlying Dynkin diagram of finite type since it fixes the trivial representation. On the Dynkin diagram with the affine node removed, it can be written as $\alpha_{a^\star}=-w_0(\alpha_a)$ for any simple root $\alpha_a$, where $w_0$ is the longest Weyl element. An illustration can be found in \cite[Figure 1]{Nakajima:2025utt}.

Recall that we are mainly focusing on the split and quasi-split types. The orientifold action $\sigma$ is an automorphism on the quiver variety, and its fixed point subset is the $\sigma$-quiver variety. Overall, the action $\sigma$ acts on the quiver variety via a composition of $\tau$, the reflection functor associated to $w_0$, the involution $\star$, and some transpose involution. We shall not go into the details here. It suffices to know that the gauge node is determined by the action of $\tau\circ\star$ on the parent unitary quiver. We shall write it as
\begin{equation}
	\sigma_Q=\tau\circ\star
\end{equation}
when considering how the nodes and arrows would be mapped under the orientifold action.

Therefore, $\sigma$ does not coincide with $\tau$ when acting on the nodes. The currents in the $\iota$quiver algebras (regardless of the splitness) do not have a strict match with the nodes in the orientifolded quivers. In fact, from the Drinfeld presentations in \cite{Lu_2025} for the split type and in \cite{2024arXiv240806981L} for the quasi-split type, we can see that the generators are always labelled by the nodes of the parent quiver rather than the orientifolded one.

Let us illustrate this with an example:
\begin{equation}
	\includegraphics[width=10cm]{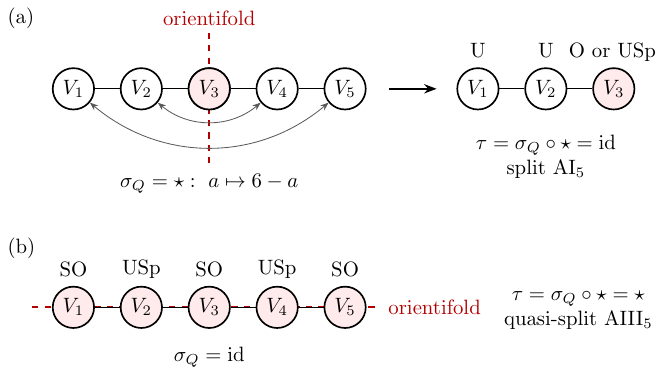}\;.
\end{equation}
In case (a), we have the split type called AI, where the arrows on the Dynkin diagram indicate the action of $\star$ (rather than $\tau$). The Drinfeld presentation of the corresponding twisted Yangian can be written down following \cite{Lu_2025}. In case (b), we have the quasi-split type called AIII. Notice that $\sigma_Q=\mathrm{id}$ does not mean that $\sigma$ is trivial. The Drinfeld presentation of the corresponding twisted Yangian can be written down following \cite{2024arXiv240806981L}.

\paragraph{Affine types} Let us now discuss the cases of affine type. In particular, the toric quivers and their brane tilings on $\mathbb{C}\times\mathbb{C}^2/\mathbb{Z}_n$ correspond to those of affine type A.

For symmetrisable Kac--Moody algebras, the symmetric pairs can be classified by the admissible pairs \cite{kolb2014quantum} (see also \cite{mudrov2026affine} for the untwisted affine Lie superalgebra cases). Given a Dynkin diagram, denote the set of nodes as $\mathtt{Q}$. The admissible pair $(\mathtt{Q}_\bullet,\tau)$ consists of a subset $\mathtt{Q}_\bullet$ of finite type and an involution $\tau$ such that (i) $\tau(\mathtt{Q}_\bullet)=\mathtt{Q}_\bullet$, (ii) $\tau|_{\mathtt{Q}_\bullet}=-w_\bullet$, (iii) $\alpha_a\left(\rho^\vee_\bullet\right)\in\mathbb{Z}$ if $a\in\mathtt{Q}\backslash\mathtt{Q}_\bullet$ and $\tau(a)=a$. Here, $w_\bullet$ is the longest Weyl element generated by the root system $R_\bullet$ of $\mathtt{Q}_\bullet$, and $\rho^\vee_\bullet$ is half of the sum of the positive coroots of $R_\bullet$. This admissible pair then gives rise to the involution $\varrho=-w_\bullet\circ\tau$ on the root system $R$ of $\mathtt{Q}$.

Again, we shall consider the cases without the black nodes. Although we may take some involution $\tau$ on the affine Dynkin diagram, unlike the finite types, the affine Weyl group does not have a longest element. Nevertheless, as the finite types discussed above come from removing the affine node in the affine Dynkin diagrams, it is natural to conjecture that the orientifolded quivers and the $\iota$quiver algebras would still have the correspondence $\sigma_Q=\tau\circ\star$. This is consistent with Theorem A.1 in \cite{Nakajima:2025utt} for $\tau=\mathrm{id}$.

If this is the case, then we can match the orientifolded quivers arising from $\mathbb{C}\times\mathbb{C}^2/\mathbb{Z}_n$ with the involutions $\tau$ and the potential $\iota$quiver algebras determined by $\tau$ (to some extent). For the split type with $\tau=\mathrm{id}$, the quiver should then be a quiver with one (resp.~two) orthogonal/symplectic gauge nodes (while the remaining $\left\lfloor\frac{n-1}{2}\right\rfloor$ gauge nodes are unitary) for odd (resp.~even) $n$. This would come from the point orientifold. If $\tau$ coincides with $\star$, then all the nodes would be orthogonal/symplectic. This would come from the line orientifold (or the point orientifold as well for $n=1,2$). However, this does not fully determine the quiver gauge theory as there is still information yet to be determined, such as how the matter fields would transform under the gauge groups.

It could be possible that this would still be sufficient to determine the $\iota$quiver (BPS) algebra which could be independent of the matter representations. This would depend on the BPS counting of these theories, which may or may not be realised by some combinatorial modules of the same algebra. Moreover, the cases with black nodes are not clear as well, and the use of $w_\bullet$, rather than just $w_0$, could be involved. For more general quivers, even fewer things are known, and there is no known construction of any twisted version or $\iota$ version of these quiver BPS algebras so far.

Of course, to answer these questions and find the BPS algebras, we need to get the correct BPS counting. There are also some other puzzles that already arise in the discussions in this paper.

In the next section, we are going to discuss some representations of the $\iota$quiver algebras. For the case with $n=1$, since there is only one node, we cannot tell whether this would correspond to the point or line orientifolds just from $\tau=\mathrm{id}$, let alone the matter contents, although it could be possible that they are all governed by the same algebra. Nevertheless, for a single $\SO$ or $\USp$ node with three adjoints, the BPS counting is known, and there is a representation whose basis would encode this counting.

However, this also brings another question. In this representation as constructed below, the parameters $\epsilon_1$ and $\epsilon_2$ are not independent. The gauge theory with a single $\SO$ or $\USp$ node and three adjoints comes from the point orientifold which preserves the two $\U(1)$ mesonic flavour symmetries. Nevertheless, taking a one-parameter limit would naturally lead to a representation with basis labelled by 2-dimensional diagrams. Moreover, preserving the full $\U(1)^3$ isometry does not necessarily determine the number of parameters after all, where more parameters could arise from the refinement of the counting, or this question could be resolved using some other tricks/facts that are not yet considered.

\section{Partitions as Representations}\label{representations}
In this section, we shall discuss some combinatorial representations of the $\iota$quiver algebras. For the $n=1$ case, we shall see that some of the BPS counting could be recovered.

\subsection{Restricted Representations}\label{restrictedrepresentations}
As we have the embeddings from the $\iota$quiver algebras to the corresponding quiver BPS algebras, we can write down the representations of ${}^\iota\mathtt{A}$ inherited from those of $\mathtt{A}$. As the $\iota$quiver algebras for $n\geq3$ have the parameter identification $\epsilon_1=\epsilon_2$, we shall first consider the cases with $n=1,2$.

Let us recall the vector, Fock, and MacMahon representations for general $\mathtt{A}_n$. In the vector representation, there is a basis labelled by a 1-dimensional chain of boxes (which can extend arbitrarily in both directions). In the Fock representation, there is a basis labelled by partitions. In the MacMahon representation, there is a basis labelled by plane partitions. All these boxes can have different colours labelled by $a\in Q_0$, and we shall use $\mathfrak{a}$ to denote a box. In other words, the bases can be labelled by 1-, 2-, and 3-dimensional Young diagrams $\Lambda$ respectively. We shall uniformly denote a vector/state in such a basis by $|\Lambda\rangle$. We also write $\mathrm{Add}(\Lambda)$ (resp.~$\mathrm{Rem}(\Lambda)$) as the set of addable (resp.~removable) boxes such that the resulting $k$-dimensional Young diagram is still a valid Young diagram after adding (resp.~removing) the box.
\begin{theorem}{(\cite{Feigin:2010qea,feigin2013representations,tsymbaliuk2017affine,Li:2020rij,Noshita:2021ldl,Galakhov:2021vbo,jimbo2024combinatorial,laurie2024a}, $\dots$)}\label{crystalrep}
	Let $V$ be a $\mathbb{C}$-vector space with the basis $\{|\Lambda\rangle\}$. Then there is an action of $\mathtt{A}$ with generic parameters on $V$ given by
	\begin{subequations}
		\begin{align}
			&\psi^{(a)}_{\pm}(u)|\Lambda\rangle=\left[\Psi^{(a)}_\Lambda(u)\right]_\pm|\Lambda\rangle\;,\\
			&e^{(a)}(u)|\Lambda\rangle=\sum_{\mathfrak{a}\in\mathrm{Add}(\Lambda)}\varsigma_{\Lambda\rightarrow\Lambda+\mathfrak{a}}\delta(u-\epsilon(\mathfrak{a}))\left(\lim_{x\rightarrow\epsilon(\mathfrak{a})}\zeta(x-\epsilon(\mathfrak{a}))\Psi^{(a)}_\Lambda(x)\right)^{1/2}|\Lambda+\mathfrak{a}\rangle\;,\\
			&f^{(a)}(u)|\Lambda\rangle=\sum_{\mathfrak{a}\in\mathrm{Rem}(\Lambda)}\varsigma_{\Lambda\rightarrow\Lambda-\mathfrak{a}}\delta(u-\epsilon(\mathfrak{a}))\left(\lim_{x\rightarrow\epsilon(\mathfrak{a})}\zeta(x-\epsilon(\mathfrak{a}))\Psi^{(a)}_\Lambda(x)\right)^{1/2}|\Lambda-\mathfrak{a}\rangle\;.
		\end{align}
	\end{subequations}
	Here, $\epsilon(\mathfrak{a})$ is the weight of the box $\mathfrak{a}$ of colour $a$, and $[\dots]_{\pm}$ indicates the Laurent expansion at large/small $U$. The function\footnote{Here, we need to take the weight $w$ of the box at the origin to be generic such that $\pm(w+\epsilon(\mathfrak{a}))$ would not coincide. This also includes the requirement that $w$ is non-zero.}
	\begin{equation}
		\Psi^{(a)}_\Lambda(u)={}^\#\phi^{(a)}(u)\prod_{\mathfrak{b}\in\Lambda}\phi^{ab}(u-\epsilon(\mathfrak{b}))
	\end{equation}
	is called the charge function in the physics literature and the $\ell$-weight in the quantum toroidal algebra literature. The signs $\varsigma_{\Lambda\rightarrow\Lambda\pm\mathfrak{a}}$ depend on different cases and are not important here. In what follows, we shall slightly abuse the notation and write $\mathfrak{a}=\epsilon(\mathfrak{a})$ if it would not cause any confusion.
\end{theorem}
By generic parameters, we mean that the parameters in the algebra (including the framing) would be generic so that it would not cause issues such as truncations or higher order poles (in the coefficients) in the actions.

The factors ${}^\#\phi^{(a)}(u)$ are determined by the framing of the quiver. Readers are referred to the above listed references for its explicit expression. With different framings, one may have different shapes of combinatorial modules. They could, for example, correspond to the counting of open BPS states. See \cite{Galakhov:2021xum,Noshita:2021dgj} for more details.

We can then write down the corresponding representation restricting to ${}^\iota\mathtt{A}_n$ for $n=1,2$.
\begin{proposition}\label{iotaquiveralgYoungdiagramrep}
	Let $V$ be a $\mathbb{C}$-vector space with the basis $\{|\Lambda\rangle\}$. Then there is an action of the trigonometric/elliptic ${}^\iota\mathtt{A}_n$ for $n=1,2$ with generic parameters on $V$ given by
	\begin{subequations}
		\begin{align}
			&h^{(a)}_{\pm}(u)|\Lambda\rangle=\left[\Psi^{(a)}_\Lambda(u)\right]_\pm\left[\Psi^{(a)}_\Lambda(-u)\right]_\mp|\Lambda\rangle\;,\\
			&b^{(a)}(u)|\Lambda\rangle=\sum_{\mathfrak{a}\in\mathrm{Add}(\Lambda)}\left[\Psi^{(a)}_{\Lambda+\mathfrak{a}}(-u)\right]_+\varsigma_{\Lambda\rightarrow\Lambda+\mathfrak{a}}\delta(u-\mathfrak{a})\left(\lim_{x\rightarrow\mathfrak{a}}\zeta(x-\mathfrak{a})\Psi^{(a)}_\Lambda(x)\right)^{1/2}|\Lambda+\mathfrak{a}\rangle\nonumber\\
			&\qquad\qquad\quad+c_a(1+\phi^{aa}(-2u))\sum_{\mathfrak{a}\in\mathrm{Rem}(\Lambda)}\varsigma_{\Lambda\rightarrow\Lambda-\mathfrak{a}}\delta(-u-\mathfrak{a})\left(\lim_{x\rightarrow\mathfrak{a}}\zeta(x-\mathfrak{a})\Psi^{(a)}_\Lambda(x)\right)^{1/2}|\Lambda-\mathfrak{a}\rangle\;.
		\end{align}
	\end{subequations}
\end{proposition}
For the rational case, the embeddings would be different where there would be extra correction terms in terms of the modes. Nevertheless, we expect that the above representations would still hold by taking the rational limit of the trigonometric case, i.e., $\frac{\zeta(\beta u)}{2\pi\ii\beta}$ with $\beta\rightarrow0$. See also \cite[Section 4]{tsymbaliuk2017affine}.

Call a module thin (aka tame) if the action of $\mathtt{A}^0$ is semisimple and all the eigenspaces are 1-dimensional. In particular, the charge functions/$\ell$-weights for different states are different. The representations labelled by $|\Lambda\rangle$ are thin and irreducible. In the $\iota$quiver algebra case, we may likewise call modules thin if $\mathtt{A}^0$ is replaced by the subalgebra generated by the $h$ generators.

For the Young diagrams, we can show that the irreducibility remains in ${}^\iota\mathtt{A}$.
\begin{proposition}\label{thinirredrep}
	The modules $V$ of ${}^\iota\mathtt{A}$ above are thin and irreducible.
\end{proposition}
\begin{proof}
	Let us first show the thinness. Since $h^{(a)}_\pm(u)$ act diagonally on the basis $\{|\Lambda\rangle\}$, we only need to show that their eigenvalues distinguish different states. Write the eigenvalues as $\left[H^{(a)}_\Lambda(u)\right]_\pm$, where $H^{(a)}_\Lambda(u)=\Psi^{(a)}_\Lambda(u)\Psi^{(a)}_\Lambda(-u)$. As a meromorphic function is determined by its expansion, it suffices to show that $H^{(a)}_\Lambda(u)=H^{(a)}_{\Lambda'}(u)$ for all $a\in Q_0$ implies that $\Lambda=\Lambda'$.

	Take $\Gamma$ to be $\{0\}$ ($\mathbb{Z}$, resp.~$\mathbb{Z}+\tau\mathbb{Z}$) in the rational (trigonometric, resp.~elliptic cases). Then $\zeta(u)$ has simple zeros (but no poles) exactly at $u\in\Gamma$. Let $\mathfrak{L}=\mathbb{Z}\epsilon_1+\mathbb{Z}\epsilon_2$ so that $\epsilon(\mathfrak{a})\in w+\mathfrak{L}$ for all $\mathfrak{a}$. By generic parameters, we mean that
	\begin{itemize}
		\item $m_1\epsilon_1+m_2\epsilon_2\notin\Gamma$ for any integer pairs $(m_1,m_2)\neq(0,0)$;
		\item $2w\notin\mathfrak{L}+\Gamma$.
	\end{itemize}
	Given a meromorphic function $f$, denote its order of vanishing at $x$ (which is negative at a pole) by $\mathrm{ord}_xf$. In particular, $\mathrm{ord}_x(fg)=\mathrm{ord}_xf+\mathrm{ord}_xg$.

	Suppose that $H^{(a)}_\Lambda(u)=H^{(a)}_{\Lambda'}(u)$ for all $a$. Then we have
	\begin{equation}
		1=\frac{H^{(a)}_\Lambda(u)}{H^{(a)}_{\Lambda'}(u)}=\frac{\prod\limits_{\mathfrak{b}\in\Lambda\backslash\Lambda'}\phi^{ab}(u-\mathfrak{b})\phi^{ab}(-u-\mathfrak{b})}{\prod\limits_{\mathfrak{b}\in\Lambda'\backslash\Lambda}\phi^{ab}(u-\mathfrak{b})\phi^{ab}(-u-\mathfrak{b})}\;,
	\end{equation}
	where $\Lambda\backslash\Lambda'$ and $\Lambda'\backslash\Lambda$ are finite. Take any $x\in w+\mathfrak{L}$, and consider $\mathrm{ord}_x$ for this equation:
	\begin{itemize}
		\item The zeros and poles of $\phi^{ab}(-u-\mathfrak{b})$ lie in $-\mathfrak{b}+\mathfrak{L}+\Gamma\subseteq-w+\mathfrak{L}+\Gamma$. Therefore, these factors do not contribute to $\mathrm{ord}_x$.
		\item For $\phi^{ab}(u-\mathfrak{b})$, we have $\mathrm{ord}_x\phi^{ab}(u-\mathfrak{b})=\mathtt{I}_+-\mathtt{I}_-$, where $\mathtt{I}_+$ (resp.~$\mathtt{I}_-$) is the number of arrows $I\in\{a\rightarrow b\}$ with $\mathfrak{b}=x+\epsilon_I$ (resp.~the number of arrows $I\in\{b\rightarrow a\}$ with $\mathfrak{b}=x-\epsilon_I$).
	\end{itemize}
	Every node $a$ is the target of exactly three arrows with weights $\epsilon_{1,2,3}$, as well as the source of exactly three arrows with weights $\epsilon_{1,2,3}$. Therefore, summing over $a\in Q_0$, we get
	\begin{equation}
		\sum_{i=1}^3\left(\updelta(x+\epsilon_i)-\updelta(x-\epsilon_i)\right)=0\;,\quad\forall \;x\in w+\mathfrak{L}\;.
	\end{equation}
	Here, $\updelta(y)$ is the number of boxes of weight $y$ in $\Lambda\backslash\Lambda'$ minus that in $\Lambda'\backslash\Lambda$. This is equivalent to
	\begin{equation}
		\left(t_1^{-1}-t_1+t_2^{-1}-t_2+t_1t_2-t_1^{-1}t_2^{-1}\right)M=0\;,
	\end{equation}
	where
	\begin{equation}
		M=\sum_{y\in w+\mathfrak{L}}\updelta(y)t^{y-w}\in\mathbb{Z}\left[t_1^{\pm1},t_2^{\pm1}\right]
	\end{equation}
	and $t^{m_1\epsilon_1+m_2\epsilon_2}=t_1^{m_1}t_2^{m_2}$. As the first factor is non-zero and $\mathbb{Z}\left[t_1^{\pm1},t_2^{\pm1}\right]$ is an integral domain, we have $M=0$. In other words, $\Lambda$ and $\Lambda'$ contain the same number of boxes for each weight. For 1d and 2d Young diagrams, different boxes have different weights. For 3d Young diagrams, the boxes of the same weight lie on the same diagonal, which also uniquely fixes the configuration since it needs to be a valid partition. Therefore, $\Lambda=\Lambda'$, and this proves the thinness.
	
	To prove irreducibility, it is completely the same as the proof for $\mathtt{A}$. Let us also recall it here. Consider a submodule generated by $|w\rangle=\sum\limits_{i=1}^kw_i|\Lambda_i\rangle$ with $w_i\in\mathbb{C}^*$. Since $|\Lambda_i\rangle$ have distinct charge functions/$\ell$-weights, one can apply different $h^{(a)}_n$ to get $k$ linearly independent vectors $|w'\rangle=\sum\limits_{i=1}^kw'_i|\Lambda_i\rangle$. Therefore, $|\Lambda_1\rangle$ is contained in the submodule. Then by applying a sequence of the $b$ generators, one can reach any other $|\Lambda\rangle$. As a result, the submodule is actually the whole module $V$.
\end{proof}

We can also write down the restricted representations that come from the quiver BPS algebras, such as the plane partitions with pits in \cite{bershtein2018plane,Harada:2018bkb,Harada:2021xnm}, the PT modules in \cite{Gaiotto:2020dsq,Kimura:2025lfo}, the strange modules in \cite{Litvinov:2021phc}, etc\footnote{Besides these crystal representations, one may also consider the vertex operator representations which are closely related to $\mathcal{W}$-algebras. See, for example, \cite{Noshita:2022dxv} and the references therein.}. For instance, when the plane partitions truncate, we can also have some valid representations in the limit $\epsilon_1=\epsilon_2$, and hence get some restricted representations for $n\geq3$.

As $\epsilon_1=\epsilon_2$, the boxes along the same anti-diagonal (on each $(\epsilon_1,\epsilon_2)$-plane) would give rise to the same pole that appears in the actions in \cref{crystalrep}:
\begin{equation}
	\includegraphics[width=3cm]{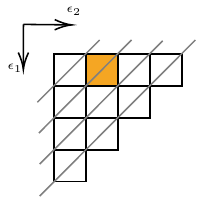}\;.
\end{equation}
The actions would then lose efficacy due to the higher order poles. Since different boxes have different colours with different bond factors, the higher order poles would not appear immediately after one adds boxes to the empty room configuration. However, higher order poles would develop when the (maximal) length of the anti-diagonal is sufficiently large.

Therefore, we may choose the framing factors ${}^\#\phi^{(a)}(u)$ so that the corresponding one has the factor $\zeta(u-w-\epsilon_2)$ to cancel the pole for the orange box above. Then it would be a representation with basis labelled by the 2-dimensional Young diagram on the $(\epsilon_1,\epsilon_3)$-plane. This gives a representation following \cref{iotaquiveralgYoungdiagramrep}.

\subsection{Strict/Symmetric Partitions}\label{strictpartitions}
It was found in \cite{Kac:1999av} that the Witten indices for 1d $\mathcal{N}=4$ supersymmetric quantum mechanics with a single $\USp(2N)$ gauge group and three adjoints are given by the strict partitions of $N$. Moreover, for a single $\SO(N)$ gauge group, the Witten indices are given by the strict odd partitions of $N$. These would give the BPS counting for the corresponding 4d theories on $\mathbb{C}^3$ with orientifolds. The generating functions are
\begin{equation}
	\mathcal{Z}_\USp(t)=\prod_{i=1}^\infty\left(1+t^i\right)=\prod_{i=1}^\infty\frac{1}{1-t^{2i-1}}\;,\quad\mathcal{Z}_\SO(t)=\prod_{i=1}^\infty\left(1+t^{2i-1}\right)\;,
\end{equation}
ignoring the signs from the BPS states. For the strict odd partitions, they are equivalent to the symmetric partitions as can be seen from the following way:
\begin{equation}
	\includegraphics[width=10cm]{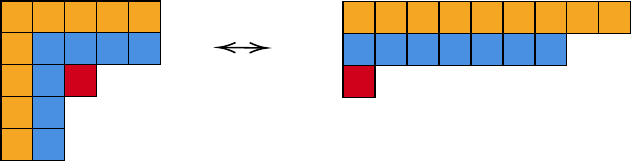}\;.
\end{equation}
Therefore, we shall refer to these as symmetric partitions below.

Although it is not clear whether the $\iota$quiver algebras should be called BPS algebras, we shall now discuss the representations of ${}^\iota\mathtt{A}_1$ that can be labelled by strict/symmetric partitions. In fact, since there is a one-to-one correspondence between strict partitions and symmetric partitions\footnote{By one-to-one correspondence, we do not mean that their partition functions would coincide. What this correspondence means would be explained below.}, they would serve as the same representation.

To see this, let us start with a strict partition. In other words, it is a partition $\lambda$ satisfying $\lambda_1>\lambda_2>\dots>\lambda_L>0$. A box in the Young diagram has two coordinates specified by its row $i$ and column $j$. It would actually be convenient to shift the boxes in row $i$ to the right by $i-1$ units so that the vertical $\epsilon_1$-axis would become diagonal:
\begin{equation}
	\includegraphics[width=8cm]{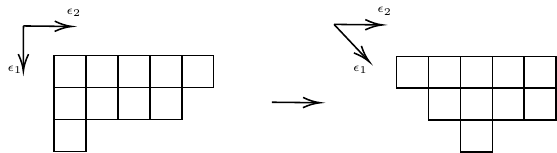}\;.
\end{equation}
As a result, we have $1\leq i\leq L$ and $i\leq j\leq i+\lambda_i-1$. We may also think of this as all the grey boxes already being filled:
\begin{equation}
	\includegraphics[width=2.5cm]{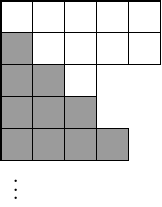}\;.
\end{equation}
We shall take the convention that $\epsilon\left(\square_{(i,j)}\right)=w+(i-1)\epsilon_1+(j-1)\epsilon_2$ such that the origin has weight $w\in\mathbb{C}$. It is obvious that adding or removing a box should keep the resulting configuration a valid strict partition.

One of the benefits of shifting the strict partitions is that we can see the equivalence between strict partitions and symmetric partitions. In the shifted diagram, it is not hard to see that this fills the upper half of the usual partitions. Therefore, filling it with the mirror under the diagonal gives the corresponding symmetric partition:
\begin{equation}
	\includegraphics[width=8cm]{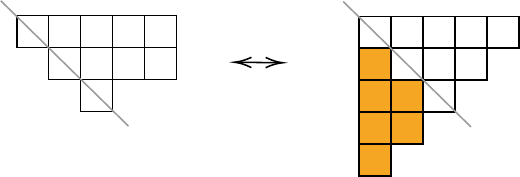}\;.
\end{equation}
The rule of adding and removing boxes remains the same. Therefore, it suffices to consider the shifted diagrams.

\paragraph{Trigonometric and elliptic cases} Let us first consider the trigonometric and the elliptic cases. We shall take $\epsilon_3=1/2$, that is, $\epsilon_1=-\epsilon_2-1/2$. Moreover, for any non-zero integer $m$, we require that $2m\epsilon_2\notin\mathbb{Z}(+\tau\mathbb{Z})$. In particular, $\epsilon(\square)=w+(j-i)\epsilon_2-\frac{i-1}{2}$.

\begin{theorem}
	For $n=1$ and generic $w$, the trigonometric/elliptic $\iota$quiver algebra ${}^\iota\mathtt{A}_n$ has a representation with the basis labelled by strict/symmetric partitions. The actions are given by
	\begin{subequations}
		\begin{align}
			&h(u)|\lambda\rangle=H_\lambda(u)|\lambda\rangle\;,\\
			&b(u)|\lambda\rangle=\sum_{\square\in\mathrm{Add}(\lambda)}\delta(u-\square)E_{\lambda}(\square)|\lambda+\square\rangle+\sum_{\square\in\mathrm{Rem}(\lambda)}\delta(u+\square)F_{\lambda}(\square)|\lambda-\square\rangle\;,
		\end{align}
	\end{subequations}
	where
	\begin{subequations}
		\begin{align}
			&H_\lambda(u)=\frac{\zeta(u-w-\epsilon_1)\zeta(u+w+\epsilon_1)}{\zeta(u-w)\zeta(u+w)}\prod_{\square\in\lambda}\frac{\phi\left(u-\square\right)}{\phi\left(u+\square\right)}\nonumber\\
			&\qquad~~=\frac{\zeta(u-w+\epsilon_2+1/2)\zeta(u+w-\epsilon_2-1/2)}{\zeta(u-w)\zeta(u+w)}\prod_{i=1}^L\prod_{j=i}^{i+\lambda_i-1}\frac{\phi\left(u-\square_{(i,j)}\right)}{\phi\left(u+\square_{(i,j)}\right)}\;,\\
			&E_\lambda\left(\square_{(i,j)}\right)=\prod_{\square\in\mathtt{L}_\lambda(i,j)}\phi\left(\square_{(i,j)}-\square\right)=\prod_{k=1}^{i-1}\prod_{l=j+1}^{k+\lambda_k-1}\phi\left(\square_{(i,j)}-\square_{(k,l)}\right)\;,\\
			&F_\lambda(\square)=\frac{c}{E_{\lambda-\square}(\square)}\left(1-\frac{\zeta(2\square-\epsilon_2)\zeta(2\square+\epsilon_2+1/2)}{\zeta(2\square+\epsilon_2)\zeta(2\square-\epsilon_2-1/2)}\right)\left(\lim_{x\rightarrow\square}\zeta(x-\square)H_{\lambda-\square}(x)\right)\;.
		\end{align}
	\end{subequations}
\end{theorem}

Before we give the proof, let us make a few comments which might help us understand the actions more intuitively:
\begin{itemize}
	\item There is a factor of $\zeta(u-w-\epsilon_1)$ (together with its counterpart $\zeta(u+w+\epsilon_1)$) in the numerator of $H_\lambda(u)$. This can be thought of as an analogue of the truncation introduced by the framing factor as mentioned above, where the truncation happens as the first grey box $\epsilon(\square)=w+\epsilon_1$ here.
	\item In $E_\lambda\left(\square_{(i,j)}\right)$, we have introduced $\mathtt{L}_\lambda(i,j)$ whose definition should be clear from the above expression. This is the collection of boxes circled out by the hook of $\square_{(i,j)}$ flipped along the $\epsilon_1$-axis:
	\begin{equation}
		\includegraphics[width=3cm]{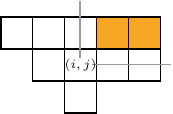}\;.
	\end{equation}
	Since it only appears for the addable and removable boxes, this gives the boxes in the upper right region of the addable or removable box, i.e., those satisfying $k<i$ and $l>j$.
	\item The first bracket in $F_\lambda(\square)$ is actually $1+\phi(-2\square)$. Since $\zeta(u+\epsilon_3)/\zeta(u-\epsilon_3)=\zeta(u+1/2)/\zeta(u-1/2)=-1$, we have the above factor.
	\item Overall, the coefficients resemble the form of the embedding in \cref{embeddingtrigell}. The charge function $H_\lambda(u)$ is like the double copy of the charge function in the quiver BPS algebra case. For the function $E_\lambda(\square)$, it might be very roughly thought of as the ``relic'' after the factors from $\psi$ and $e$ cancel each other, leaving some contributions near certain boundaries. For the function $F_\lambda(\square)$, we can also see the factor $1+\phi(-2\square)$, as well as the second bracket that is similar to the action of $f$.
\end{itemize}

It would also be helpful to show that $H_\lambda(u)$ gives the right poles.
\begin{lemma}
	The poles of the charge function $H_\lambda(u)$ are in one-to-one correspondence with the weights and negative weights of the addable and removable boxes of $\lambda$. In particular, they are always simple poles in the sense that each $\zeta(u-p_i)=0$ in the denominator has multiplicity one.
\end{lemma}
\begin{proof}
	We can write the bond factor as
	\begin{equation}
		\phi(u)=-\frac{\zeta(u+\epsilon_2)\zeta(u-\epsilon_2-1/2)}{\zeta(u-\epsilon_2)\zeta(u+\epsilon_2+1/2)}=\frac{\zeta(u+\epsilon_2)\zeta(u-\epsilon_2+1/2)}{\zeta(u-\epsilon_2)\zeta(u+\epsilon_2+1/2)}\;,
	\end{equation}
	where we have used $\zeta(u+1)=-\zeta(u)$ in the second equality. This also indicates the choice of $\epsilon_3=1/2$.

	Given the $i^\text{th}$ row with $\lambda_i=m$ boxes, consider the product of the bond factors. The factors in the middle all get cancelled, and we have
	\begin{equation}
		\prod_{r=0}^{m-1}\phi(u'-r\epsilon_2)=\frac{\zeta(u'+\epsilon_2)\zeta(u')\zeta(u'-(m-1)\epsilon_2+1/2)\zeta(u'-m\epsilon_2+1/2)}{\zeta(u'+\epsilon_2+1/2)\zeta(u'+1/2)\zeta(u'-(m-1)\epsilon_2)\zeta(u'-m\epsilon_2)}\;,\label{prodphi}
	\end{equation}
	where $u'=u-w-(i-1)\epsilon_1-(i-1)\epsilon_2=u-w+(i-1)/2$ (recall that $j$ starts from $i$ in each row). The factor $\frac{\zeta(u'+\epsilon_2)\zeta(u')}{\zeta(u'+\epsilon_2+1/2)\zeta(u'+1/2)}$ may be thought of as the contribution from the initial part while the remaining $m$-dependent factor gives the contribution from the endpoint. As we can see, to avoid the coincidence of the factors, we need to keep $2m\epsilon_2\notin\mathbb{Z}(+\tau\mathbb{Z})$.
	
	It would also be useful to notice that
	\begin{equation}
		\phi(u+1/2)=\phi(u)^{-1}\;.
	\end{equation}
	As a result, the contributions of successive rows would alternate between \eqref{prodphi} and its inverse.

	By virtue of these properties, after multiplying the rows, the contribution of the factors $\frac{\zeta(u'+\epsilon_2)\zeta(u')}{\zeta(u'+\epsilon_2+1/2)\zeta(u'+1/2)}$, including $\frac{\zeta(u-w+\epsilon_2+1/2)}{\zeta(u-w)}$ from the framing factor, would be $\frac{\zeta(u-w+\epsilon_2+1/2)}{\zeta(u-w)}$ for even $L$ and $\frac{\zeta(u-w+\epsilon_2)}{\zeta(u-w+1/2)}$ for odd $L$. Together with the $m$-dependent endpoint contributions, they would give poles at $w+(\lambda_i-1)\epsilon_2-\frac{i-1}{2}$ and $w+\lambda_i\epsilon_2-\frac{i-1}{2}$ corresponding to the (possibly) removable and addable boxes respectively. The zeros are at $w+(\lambda_i-1)\epsilon_2-\frac{i}{2}$ and $w+\lambda_i\epsilon_2-\frac{i}{2}$.

	If $\lambda_i=\lambda_{i+1}+1$, then the would-be addable box in the $(i+1)^\text{th}$ row would have its pole cancelled by the zero in the $i^\text{th}$ row. Therefore, these factors would give the addable and removable boxes of $\lambda$. As the weights of the boxes are $w+(j-i)\epsilon_2-\frac{i-1}{2}$, they would not coincide for different addable and removable boxes. The remaining factors in $H_\lambda(u)$ are simply the reflected factors with the factors $\zeta(u-\dots)$ becoming $\zeta(u+\dots)$, giving the negative weights of the addable and removable boxes. As a result, we have this one-to-one correspondence, and all the poles are simple, i.e., of multiplicity one.
\end{proof}

Now, let us prove that the strict/symmetric partitions form a representation.
\begin{proof}
	We need to show that the actions are consistent with the relations of the currents. The $hh$ relations are straightforward. For the $hb$ relation, this follows from
	\begin{equation}
		\frac{H_{\lambda+\square}(u)}{H_\lambda(u)}=\frac{\phi(u-\square)}{\phi(u+\square)}\;.
	\end{equation}

	For the $bb$ relation, suppose that two boxes can be added in any order. The box, say $\square$, must live in $\mathtt{L}_\lambda(\square')$ of the other box $\square'$. Then
	\begin{equation}
		E_{\lambda+\square}(\square')=\phi(\square'-\square)E_\lambda(\square')\;,\quad E_{\lambda+\square'}(\square)=E_\lambda(\square)\;.
	\end{equation}
	As a result,
	\begin{equation}
		\frac{E_\lambda(\square)E_{\lambda+\square}(\square')}{E_\lambda(\square')E_{\lambda+\square'}(\square)}=\phi(\square'-\square)\;.
	\end{equation}
	Likewise, when removing two boxes at generic positions, the ratio of the $F_\lambda(\square)$ factors (where $E_{\lambda-\square}(\square)$ and $\lim\limits_{x\rightarrow\square}\zeta(x-\square)H_{\lambda-\square}(x)$ would give uncancelled contributions) is $\phi(\square-\square')$. Notice that the formal $\delta$-functions are $\delta(u-\square)$ and $\delta(u+\square)$ respectively when adding and removing boxes. Therefore, the contributions from adding/removing two boxes at generic positions would cancel in $b(u)b(v)$ and $b(v)b(u)$.

	If one of the boxes can only be added/removed after the other has been added/removed, then they are either in the same row or in two consecutive rows whose lengths differ by one before adding/removing them. As a result, $\square-\square'=\epsilon_2$ or $\square-\square'=\epsilon_1$, and the prefactor in $b(u)b(v)$ would vanish due to the formal $\delta$-function.

	For the crossed terms with both $E$ and $F$, if the added and the removed boxes are different, then $EF$ and $FE$ contributions would cancel each other since
	\begin{equation}
		\frac{E_{\lambda}(\square)F_{\lambda+\square}(\square')}{F_{\lambda}(\square')E_{\lambda-\square'}(\square)}=\phi(-\square-\square')\;.
	\end{equation}
	The remaining part would be the same box that is added and removed. If it is an addable box of $\lambda$, then we have
	\begin{equation}
		E_\lambda(\square)F_{\lambda+\square}(\square)(\delta(u+\square)\delta(v-\square)-\phi(2\square)\delta(u-\square)\delta(v+\square))|\lambda\rangle
	\end{equation}
	in $(b(u)b(v)-\phi(u-v)b(v)b(u))|\lambda\rangle$. Since $\phi(2\square)\phi(-2\square)=1$ and
	\begin{equation}
		E_\lambda(\square)F_{\lambda+\square}(\square)=c(1+\phi(-2\square))\left(\lim_{x\rightarrow\square}\zeta(x-\square)H_\lambda(x)\right)\;,
	\end{equation}
	we get
	\begin{equation}
		-c(1+\phi(2u))\delta(u+v)(\delta(u-\square)-\delta(u+\square))\left(\lim_{x\rightarrow\square}\zeta(x-\square)H_\lambda(x)\right)|\lambda\rangle\;.
	\end{equation}
	As $H_\lambda(u)$ is an even function, we have
	\begin{equation}
		\lim_{x\rightarrow-\square}\zeta(x+\square)H_\lambda(x)=-\lim_{x\rightarrow\square}\zeta(x-\square)H_\lambda(x)\;.
	\end{equation}
	Therefore, this contains the poles for all addable boxes and their negative weights. Likewise, the removable boxes of $\lambda$ give the remaining poles in $H_\lambda(u)$. As $H_\lambda(u)$ only has simple poles, we have
	\begin{equation}
		\sum_{p\in\text{poles}(H_\lambda)}\lim_{x\rightarrow p}\zeta(x-p)H_\lambda(x)\delta(u-p)\delta(-v-p)=\delta(u+v)([H_\lambda(u)]_+-[H_\lambda(u)]_-)\;.
	\end{equation}
	This proves the $bb$ relation.

	For the Serre relation, let us recall that it reads
	\begin{equation}
		\mathrm{Sym}_{u_1,u_2,u_3}\zeta_0(2(u_2-u_3))\mathfrak{D}([b(u_1),[b(u_2),b(u_3)]])=0\;,
	\end{equation}
	where
	\begin{align}
		\mathfrak{D}(b(u_1)b(u_2)b(u_3))=&\prod\limits_{i=1}^{3}\frac{\zeta_0(u_1-u_2+\epsilon_i)\zeta_0(u_1-u_3+\epsilon_i)\zeta_0(u_2-u_3+\epsilon_i)}{\zeta(u_1-u_2+\epsilon_i)\zeta(u_1-u_3+\epsilon_i)\zeta(u_2-u_3+\epsilon_i)}\nonumber\\
		&\bigg(b(u_1)b(u_2)b(u_3)-c(1+\phi(-2u_2))\delta(u_1+u_2)h_+(u_2)b(u_3)\nonumber\\
		&-c\frac{1+\phi(-2u_3)}{\phi(u_3-u_2)}\delta(u_1+u_3)h_+(u_3)b(u_2)\nonumber\\
		&-c(1+\phi(-2u_3))\delta(u_2+u_3)b(u_1)h_+(u_3)\bigg)\;.
	\end{align}
	We are going to show that the Serre relation can actually be derived from the other relations when $\epsilon_1=-\epsilon_2-1/2$, and is therefore not independent. Take
	\begin{align}
		&\zeta_0(2(u_2-u_3))\mathfrak{D}([b(u_1),[b(u_2),b(u_3)]])\nonumber\\
		=&\zeta_0(2(u_2-u_3))(\mathfrak{D}(b(u_1)b(u_2)b(u_3))-\mathfrak{D}(b(u_1)b(u_3)b(u_2))-\mathfrak{D}(b(u_2)b(u_3)b(u_1))+\mathfrak{D}(b(u_3)b(u_2)b(u_1)))
	\end{align}
	in the symmetric sum. Consider the difference involving the exchange of $b(u_1)$ and $b(u_2)$:
	\begin{align}
		&\prod\limits_{i=1}^{3}\frac{\zeta_0(u_1-u_2+\epsilon_i)\zeta_0(u_1-u_3+\epsilon_i)\zeta_0(u_2-u_3+\epsilon_i)}{\zeta(u_1-u_2+\epsilon_i)\zeta(u_1-u_3+\epsilon_i)\zeta(u_2-u_3+\epsilon_i)}\nonumber\\
		&\left(b(u_1)b(u_2)b(u_3)-\left(\prod_{i=1}^{3}\frac{\zeta_0(u_1-u_2+\epsilon_i)}{\zeta_0(u_1-u_2-\epsilon_i)}\right)b(u_2)b(u_1)b(u_3)\right)\;.
	\end{align}
	Together with the prefactor in the first line, the coefficients in the second line satisfy
	\begin{equation}
		\phi(u_1-u_2)[\text{first term}]=[\text{second term}]\;.
	\end{equation}
	Therefore, the second line in the above difference has the following terms remained:
	\begin{align}
		&(b(u_1)b(u_2)-\phi(u_1-u_2)b(u_2)b(u_1))b(u_3)\nonumber\\
		&-c\delta(u_1+u_2)((1+\phi(-2u_2))h_+(u_2)-\phi(u_1-u_2)(1+\phi(-2u_1))h_+(u_1))b(u_3)\nonumber\\
		&+c(1+\phi(-2u_3))\delta(u_1+u_3)\left(\phi(u_1-u_2)b(u_2)h_+(u_3)-\frac{h_+(u_3)b(u_2)}{\phi(u_3-u_2)}\right)\nonumber\\
		&+c(1+\phi(-2u_3))\delta(u_2+u_3)\left(\frac{\phi(u_1-u_2)}{\phi(u_3-u_1)}h_+(u_3)b(u_1)-b(u_1)h_+(u_3)\right)\;.
	\end{align}
	The first two lines cancel each other due to the $bb$ relation. The third line vanishes due to the $hb$ relation, and likewise for the fourth line. Similarly, the difference involving the exchange of $b(u_2)$ and $b(u_3)$,
	\begin{equation}
		\mathfrak{D}(b(u_1)b(u_2)b(u_3))-\left(\prod_{i=1}^{3}\frac{\zeta_0(u_2-u_3+\epsilon_i)}{\zeta_0(u_2-u_3-\epsilon_i)}\right)\mathfrak{D}(b(u_1)b(u_3)b(u_2))\;,
	\end{equation}
	would also vanish. Now, since $\epsilon_1=-\epsilon_2-1/2$, we have
	\begin{equation}
		\prod_{i=1}^{3}\frac{\zeta_0(u+\epsilon_i)}{\zeta_0(u-\epsilon_i)}=\frac{\zeta_0(2u)+\zeta_0(2\epsilon_2)}{\zeta_0(2u)-\zeta_0(2\epsilon_2)}\;.
	\end{equation}
	As a result,
	\begin{equation}
		\zeta_0(2(u_2-u_3))\mathfrak{D}([b(u_1),[b(u_2),b(u_3)]])=\zeta_0(2\epsilon_2)\mathfrak{D}([b(u_1),\{b(u_2),b(u_3)\}])\;.
	\end{equation}
	Adding these factors in the symmetric sum altogether, it follows that the sum is zero. This completes the proof.
\end{proof}

\paragraph{Rational cases} As we can see, the representation labelled by strict/symmetric partitions in the trigonometric and elliptic cases relies on the periodicity of $\zeta(u)$. Therefore, this would not work for the rational case. Here, let us give another construction for the rational case whose representation still has the strict/symmetric partitions as the basis.

Let us take the degenerate limit $\epsilon_3=0$, and $\epsilon_1=-\epsilon_2\neq0$. Then the bond factor becomes $\phi(u)=1$. The relations would also be simplified.

For the strict partitions, it would be convenient to introduce the polynomial basis
\begin{equation}
	|\lambda\rangle=q_{\lambda_1}\dots q_{\lambda_L}\;,
\end{equation}
satisfying
\begin{equation}
	\mathcal{Q}(z)=\sum_{r\geq0}q_rz^r=\exp\left(2\sum_{j\geq1}\frac{p_{2j-1}}{2j-1}z^{2j-1}\right)\;,
\end{equation}
where $q_0=1$ and $q_{r<0}=0$. Let us recall some facts of this polynomial basis \cite{macdonald1998symmetric}. Since $\mathcal{Q}(z)\mathcal{Q}(-z)=1$, the coefficients of $z^{2r}$ should satisfy
\begin{equation}
	q_r^2=2\sum_{i=0}^{r-1}(-1)^{r+i+1}q_{2r-i}q_i\;.
\end{equation}
In other words, we have
\begin{equation}
	q_{2r}=\sum_{i=1}^{r-1}(-1)^{i-1}q_{2r-i}q_i+\frac{1}{2}(-1)^{r-1}q_r^2\;.
\end{equation}
In other words, all the variables with even subscripts can be expressed using those with odd subscripts. Notice that $q_{2j-1}=\frac{2}{2j-1}p_{2j-1}+[\text{polynomial in }p_{2j-3},\dots,p_1]$. Therefore, the ring $\mathbb{C}[q_1,q_2,q_3,\dots]=\mathbb{C}[q_1,q_3,q_5,\dots]=\mathbb{C}[p_1,p_3,p_5,\dots]$, and $q_1,q_3,q_5,\dots$ are algebraically independent.

As a result, the dimension of the degree $L$ part is the number of partitions of $L$ into odd numbers. Since the odd partitions and the strict partitions are equivalent by a theorem of Euler, i.e., $\prod\limits_{j\geq1}\left(1-t^{2j-1}\right)^{-1}=\prod\limits_{r=1}(1+t^r)$, this is also equal to the number of strict partitions.

In this polynomial basis, we can write the representation of ${}^\iota\mathtt{Y}_1$ labelled by strict partitions as follows.
\begin{theorem}
	Introducing the formal variable $x$ and taking $\mathbb{C}[p_1,p_3,p_5,\dots][\![x]\!]$, we have
	\begin{subequations}
		\begin{align}
			&h(u)=1+\sum_{j\geq1}\frac{x^{2j}}{u^2-j^2}\;,\\
			&b(u)=\sum_{j\geq1}x^j\left(\frac{p_{2j-1}}{u-j}+\frac{c}{j(u+j)}\frac{\partial}{\partial p_{2j-1}}\right)\;,
		\end{align}
	\end{subequations}
	when $\epsilon_1=-\epsilon_2$ for ${}^\iota\mathtt{Y}_1$.
\end{theorem}
\begin{proof}
	The $hh$ relations are straightforward. The $h(u)$ and $b(u)$ currents also commute in this case. Since
	\begin{equation}
		\left[p_{2j-1},\frac{\partial}{\partial p_{2j-1}}\right]=-1\;,
	\end{equation}
	we have
	\begin{align}
		[b(u),b(v)]=&-c\sum_{j\geq1}\frac{x^{2j}}{j}\left(\frac1{(u-j)(v+j)}-\frac{1}{(u+j)(v-j)}\right)\nonumber\\
		=&2c\sum_{j\geq1}\frac{x^{2j}(u-v)}{(u^2-j^2)(v^2-j^2)}=-2c\frac{h(u)-h(v)}{u+v}\;,
	\end{align}
	which gives the $bb$ relation in this case. Moreover, it follows that
	\begin{equation}
		[b(u_1),[b(u_2),b(u_3)]]=0\;.
	\end{equation}
	The right hand side of the Serre relation consists of commutators of $h$ and $b$, which also vanishes. This completes the proof.
\end{proof}

Let us illustrate the action of $b(u)$ with some examples. From the expressions of $q_r$ in terms of $p_{2j-1}$, we have
\begin{subequations}
	\begin{align}
		&p_d|\lambda\rangle=\left(\frac{d}{2}q_d+\frac{1}{2}\sum_{r=1}^{(d-1)/2}(-1)^r(d-2r)q_{d-r}q_r\right)\prod_{i=1}^Lq_{\lambda_i}\;,\\
		&\frac{\partial}{\partial p_d}|\lambda\rangle=\frac{2}{d}\sum_{i=1}^Lq_{\lambda_i-d}\prod_{\substack{1\leq j\leq L\\j\neq i}}q_{\lambda_j}\;.
	\end{align}
\end{subequations}
Therefore, at the first two levels, we have
\begin{subequations}
	\begin{align}
		&b(u)|\varnothing\rangle=\frac{x}{2(u-1)}|(1)\rangle+\frac{x^2}{u-2}\left(\frac{3}{2}|(3)\rangle-\frac{1}{2}|(2,1)\rangle\right)+\dots\;,\\
		&b(u)|(1)\rangle=x\left(\frac{1}{u-1}|(2)\rangle+\frac{2c}{u+1}|\varnothing\rangle\right)+\frac{x^2}{u-2}\left(2|(4)\rangle-\frac{1}{2}|(3,1)\rangle\right)+\dots\;,
	\end{align}
\end{subequations}
where the ellipses indicate terms of order $x^{k\geq3}$.

Let us comment that some $\mathcal{W}_{1+\infty}$ subalgebras of BCD types were found to have representations labelled by symmetric partitions in \cite{Bourgine:2021gnb}. It would be interesting to see if there could be any connections between the $\iota$quiver algebras and these $\mathcal{W}$-algebras, similar to the cases in quiver BPS algebras.

\subsection{Bicoloured Strict/Symmetric Partitions}\label{bicolour}
For $n\geq2$, we find it more difficult to get some representations labelled by strict/symmetric partitions with different colours (in particular, we cannot take $\epsilon_1=-\epsilon_2-1/2$ in the cases with $n\geq3$). Nevertheless, when $n=2$, there is a construction with only the box-adding operation.

Now, the strict/symmetric partitions would have two colours corresponding to $a=0,1$. The pattern of the colouring is the same as the shifted diagram, i.e.,
\begin{equation}
	\includegraphics[width=2.5cm]{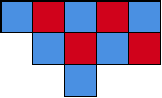}\;.
\end{equation}
Therefore, we may assign the colour of a box to be $j-i\pmod{2}$.

Again, we shall consider the trigonometric and the elliptic cases, and take $\epsilon_1=-\epsilon_2-1/2$. Moreover, for any non-zero integer $m$, we still require that $2m\epsilon_2\notin\mathbb{Z}(+\tau\mathbb{Z})$. In particular, $\epsilon(\square)=w+(j-i)\epsilon_2-\frac{i-1}{2}$, and
\begin{equation}
	\phi^{00}(u)=\phi^{11}(u)=\frac{\zeta(u+\epsilon_3)}{\zeta(u-\epsilon_3)}=-1\;.
\end{equation}

\begin{theorem}
	For $n=2$, the trigonometric/elliptic $\iota$quiver algebra ${}^\iota\mathtt{A}_n$ has a representation with the basis labelled by strict/symmetric partitions. The actions are given by
	\begin{subequations}
		\begin{align}
			&h^{(a)}(u)|\lambda\rangle=H^{(a)}_\lambda(u)|\lambda\rangle\;,\\
			&b^{(a)}(u)|\lambda\rangle=\sum_{\mathfrak{a}\in\mathrm{Add}(\lambda)}\delta(u-\mathfrak{a})E^{(a)}_{\lambda}(\mathfrak{a})|\lambda+\mathfrak{a}\rangle\;,
		\end{align}
	\end{subequations}
	where
	\begin{subequations}
		\begin{align}
			&H^{(a)}_\lambda(u)=\prod_{\mathfrak{b}\in\lambda}\frac{\phi^{ab}(u-\mathfrak{b})}{\phi^{ab}(u+\mathfrak{b})}\;,\\
			&E^{(a)}_\lambda(\mathfrak{a})=\prod_{\mathfrak{b}\in\mathtt{L}_\lambda(\mathfrak{a})}\phi^{ab}(\mathfrak{a}-\mathfrak{b})\;.
		\end{align}
	\end{subequations}
\end{theorem}
Since the proof is similar to the strict/symmetric partition representation for the $n=1$ case, we shall omit it here. In fact, the proof would actually become simpler since, for example, $\left\{b^{(a)}(u),b^{(a)}(v)\right\}=0$ in this case, and therefore the second part that removes the boxes appeared in the $n=1$ case would not appear here.

This representation looks quite different from the above ones as the actions of the currents would only add boxes to $\lambda$ but not removing them. As a result, this representation is reducible. This can be seen by considering any state $|\lambda\rangle$ as a starting state. Then applying the currents would never reach the states that are ``predecessors'' of $|\lambda\rangle$, such as $|\varnothing\rangle$.

Nevertheless, the states with $K$ boxes for a given $K$ may still be separated. Suppose that acting a $b^{(a)}(u)$ current gives a linear combination of these states. Then they can be separated by applying sufficiently many modes $b^{(a)}_i$ in the current expansion to their predecessors.

\section*{Acknowledgement} I am indebted to Meer Ashwinkumar, Duncan Laurie, Kang Lu, Mamoru Ueda, Masahito Yamazaki, Bufan Zheng, and Dongao Zhou for inspiring discussions and/or collaborations on related topics.

\paragraph{On the use of AI} Some of the computations were done with the help of ChatGPT-6 Astra and Claude Opus 5.5. The author has verified all the results and takes full responsibility for the content of this paper.

\appendix

\section{Quiver BPS Algebras}\label{quiverbpsalg}
Let us briefly recall the quiver BPS algebras for $\mathbb{C}\times\mathbb{C}^2/\mathbb{Z}_n$. Here, we shall not include the extra central extension in the trigonometric and the elliptic cases, and we shall only consider the unshifted algebras. The relations are
\begin{subequations}
	\begin{align}
	&\psi^{(a)}_{\varsigma}(z)\psi^{(b)}_{\varsigma'}(w)=\psi^{(b)}_{\varsigma'}(w)\psi^{(a)}_{\varsigma}(z)\;,\quad\varsigma,\varsigma'\in\{\pm\}\;,\\
	&\psi^{(a)}_{\pm}(z)e^{(b)}(w)=\phi^{ab}(z-w)e^{(b)}(w)\psi^{(a)}_{\pm}(z)\;,\\
	&\psi^{(a)}_{\pm}(z)f^{(b)}(w)=\phi^{ab}(z-w)^{-1}f^{(b)}(w)\psi^{(a)}_{\pm}(z)\;,\\
	&e^{(a)}(z)e^{(b)}(w)=\phi^{ab}(z-w)e^{(b)}(w)e^{(a)}(z)\;,\\
	&f^{(a)}(z)f^{(b)}(w)=\phi^{ab}(z-w)^{-1}f^{(b)}(w)f^{(a)}(z)\;,\\
	&\left[e^{(a)}(z),f^{(b)}(w)\right]=-\delta_{ab}\left(\delta(z-w)\psi^{(a)}_+(z)-\delta(z-w)\psi^{(a)}_-(w)\right)\;,\\
	&\mathrm{Sym}_{u_1,u_2,u_3}\zeta_0\left(2(u_2-u_3)\right)\mathscr{D}_\nu\left([x_\nu(u_1),[x_\nu(u_2),x_\nu(u_3)]]\right)=0\;,\quad n=1\;,\\
	&\mathrm{Sym}_{u_1,u_2,u_3}\mathscr{D}_\nu\left(\left[x_\nu^{(a)}(u_1),\left[x_\nu^{(a)}(u_2),\left[x_\nu^{(a)}(u_3),x_\nu^{(a+1)}(v)\right]_{\kappa^2}\right]\right]_{\kappa^{-2}}\right)=0\;,\quad n=2\;,\\
	&\mathrm{Sym}_{u_1,u_2}\mathscr{D}_\nu\left(\left[x_\nu^{(a)}(u_1),\left[x_\nu^{(a)}(u_2),x_\nu^{(b)}(v)\right]_{\kappa^{-1}}\right]_{\kappa}\right)=0\;,\quad b=a\pm1\;,\quad n\geq3\;.
    \end{align}
\end{subequations}
The current expansions of $\psi_\pm^{(a)}(u)$ (resp.~$e^{(a)}(u)$ and $f^{(a)}(u)$) have the same range as the one of $h_\pm^{(a)}(u)$ (resp.~$b^{(a)}(u)$)\footnote{For chiral quivers, the $\psi$ currents may have negative modes. Therefore, if there is any twisted version for the chiral quivers, we would expect that (the analogues of) the $h$ currents would also have negative modes.}. In the Serre relations, we have introduced $x_\nu^{(a)}(u)$ with $\nu=+$ (resp.~$\nu=-$) being $e^{(a)}(u)$ ($f^{(a)}(u)$) for simplicity. Moreover,
\begin{equation}
	\mathscr{D}_\nu\left(x_\nu^{(a_1)}(u_1)\cdots x_\nu^{(a_m)}(u_m)\right)=\left(\prod\limits_{1\leq r<s\leq m}K_\nu^{a_ra_s}(u_r-u_s)\right)x_\nu^{(a_1)}(u_1)\cdots x_\nu^{(a_m)}(u_m)\;,
\end{equation}
where
\begin{equation}
	K_+^{ab}(u)=\prod_{I\in\{a\rightarrow b\}}\frac{\zeta_0(u+\epsilon_I)}{\zeta(u+\epsilon_I)}\;,\quad K_-^{ab}(u)=\prod_{I\in\{b\rightarrow a\}}\frac{\zeta_0(u-\epsilon_I)}{\zeta(u-\epsilon_I)}=K_+^{ba}(-u)\;.
\end{equation}

For the rational cases, let us write down the relations in terms of the modes. For $n\geq3$, we have
\begin{subequations}
	\begin{align}
		&\left[\psi^{(a)}_i,\psi^{(b)}_j\right]=0\;,\\
		&\left[e^{(a)}_i,f^{(b)}_j\right]=\delta_{ab}\psi^{(a)}_{i+j}\;,\\
		&\left[\psi^{(a)}_{i+1},e^{(b)}_j\right]-\left[\psi^{(a)}_i,e^{(b)}_{j+1}\right]=\sigma^{ba}_1\psi^{(a)}_ie^{(b)}_j+\sigma^{ab}_1e^{(b)}_j\psi^{(a)}_i\;,\\
		&\left[\psi^{(a)}_{i+1},f^{(b)}_j\right]-\left[\psi^{(a)}_i,f^{(b)}_{j+1}\right]=-\sigma^{ab}_1\psi^{(a)}_if^{(b)}_j-\sigma^{ba}_1f^{(b)}_j\psi^{(a)}_i\;,\\
		&\left[e^{(a)}_i,e^{(b)}_j\right]=\left[f^{(a)}_i,f^{(b)}_j\right]=0\qquad(\sigma_1^{ab}=0)\;,\\
		&\left[e^{(a)}_{i+1},e^{(b)}_j\right]-\left[e^{(a)}_i,e^{(b)}_{j+1}\right]=\sigma^{ba}_1e^{(a)}_ie^{(b)}_j+\sigma^{ab}_1e^{(b)}_je^{(a)}_i\qquad(\sigma_1^{ab}\neq0)\;,\\
		&\left[f^{(a)}_{i+1},f^{(b)}_j\right]-\left[f^{(a)}_i,f^{(b)}_{j+1}\right]=-\sigma^{ab}_1f^{(a)}_if^{(b)}_j-\sigma^{ba}_1f^{(b)}_jf^{(a)}_i\qquad(\sigma_1^{ab}\neq0)\;,\\
		&\mathrm{Sym}_{i_1,i_2}\left[e^{(a)}_{i_1},\left[e^{(a)}_{i_2},e^{(a\pm1)}_j\right]\right]=\mathrm{Sym}_{i_1,i_2}\left[f^{(a)}_{i_1},\left[f^{(a)}_{i_2},f^{(a\pm1)}_j\right]\right]=0\;,
	\end{align}
\end{subequations}
where $\sigma^{ab}_1$ is the sum of weights of the arrows from $a$ to $b$. For $n=2$, we have
\begin{subequations}
	\begin{align}
	&\left[\psi^{(a)}_i,\psi^{(b)}_j\right]=0\;,\\
	&\left[e^{(a)}_i,f^{(b)}_j\right]=\delta_{ab}\psi^{(a)}_{i+j}\;,\\
	&\left[\psi^{(a)}_{i+1},e^{(a)}_j\right]-\left[\psi^{(a)}_i,e^{(a)}_{j+1}\right]=\epsilon_3\left\{\psi^{(a)}_i,e^{(a)}_j\right\}\;,\\
	&\left[\psi^{(a)}_{i+1},f^{(a)}_j\right]-\left[\psi^{(a)}_i,f^{(a)}_{j+1}\right]=-\epsilon_3\left\{\psi^{(a)}_i,f^{(a)}_j\right\}\;,\\
	&\left[e^{(a)}_{i+1},e^{(a)}_j\right]-\left[e^{(a)}_i,e^{(a)}_{j+1}\right]=\epsilon_3\left\{e^{(a)}_i,e^{(a)}_j\right\}\;,\\
	&\left[f^{(a)}_{i+1},f^{(a)}_j\right]-\left[f^{(a)}_i,f^{(a)}_{j+1}\right]=-\epsilon_3\left\{f^{(a)}_i,f^{(a)}_j\right\}\;,\\
	&\left[\psi^{(a)}_{i+2},e^{(a+1)}_j\right]-2\left[\psi^{(a)}_{i+1},e^{(a+1)}_{j+1}\right]+\left[\psi^{(a)}_i,e^{(a+1)}_{j+2}\right]\nonumber\\
	=&-\epsilon_1\epsilon_2\left[\psi^{(a)}_i,e^{(a+1)}_j\right]-\epsilon_3\left\{\psi^{(a)}_{i+1},e^{(a+1)}_j\right\}+\epsilon_3\left\{\psi^{(a)}_i,e^{(a+1)}_{j+1}\right\}\;,\\
	&\left[\psi^{(a)}_{i+2},f^{(a+1)}_j\right]-2\left[\psi^{(a)}_{i+1},f^{(a+1)}_{j+1}\right]+\left[\psi^{(a)}_i,f^{(a+1)}_{j+2}\right]\nonumber\\
	=&-\epsilon_1\epsilon_2\left[\psi^{(a)}_i,f^{(a+1)}_j\right]+\epsilon_3\left\{\psi^{(a)}_{i+1},f^{(a+1)}_j\right\}-\epsilon_3\left\{\psi^{(a)}_i,f^{(a+1)}_{j+1}\right\}\;,\\
	&\left[e^{(a)}_{i+2},e^{(a+1)}_j\right]-2\left[e^{(a)}_{i+1},e^{(a+1)}_{j+1}\right]+\left[e^{(a)}_i,e^{(a+1)}_{j+2}\right]\nonumber\\
	=&-\epsilon_1\epsilon_2\left[e^{(a)}_i,e^{(a+1)}_j\right]-\epsilon_3\left\{e^{(a)}_{i+1},e^{(a+1)}_j\right\}+\epsilon_3\left\{e^{(a)}_i,e^{(a+1)}_{j+1}\right\}\;,\\
	&\left[f^{(a)}_{i+2},f^{(a+1)}_j\right]-2\left[f^{(a)}_{i+1},f^{(a+1)}_{j+1}\right]+\left[f^{(a)}_i,f^{(a+1)}_{j+2}\right]\nonumber\\
	=&-\epsilon_1\epsilon_2\left[f^{(a)}_i,f^{(a+1)}_j\right]+\epsilon_3\left\{f^{(a)}_{i+1},f^{(a+1)}_j\right\}-\epsilon_3\left\{f^{(a)}_i,f^{(a+1)}_{j+1}\right\}\;,\\
	&\mathrm{Sym}_{i_1,i_2,i_3}\left[e^{(a)}_{i_1},\left[e^{(a)}_{i_2},\left[e^{(a)}_{i_3},e^{(a+1)}_j\right]\right]\right]=\mathrm{Sym}_{i_1,i_2,i_3}\left[f^{(a)}_{i_1},\left[f^{(a)}_{i_2},\left[f^{(a)}_{i_3},f^{(a+1)}_j\right]\right]\right]=0\;.
\end{align}
\end{subequations}
For $n=1$, we have
\begin{subequations}
	\begin{align}
		&[\psi_i,\psi_j]=0\;,\\
		&[e_i,f_j]=\psi_{i+j}\;,\\
		&[\psi_{i+3},e_j]-3[\psi_{i+2},e_{j+1}]+3[\psi_{i+1},e_{j+2}]-[\psi_i,e_{j+3}]+\sigma_2([\psi_{i+1},e_j]-[\psi_i,e_{j+1}])=\sigma_3\{\psi_i,e_j\}\;,\\
		&[e_{i+3},e_j]-3[e_{i+2},e_{j+1}]+3[e_{i+1},e_{j+2}]-[e_i,e_{j+3}]+\sigma_2([e_{i+1},e_j]-[e_i,e_{j+1}])=\sigma_3\{e_i,e_j\}\;,\\
		&[\psi_{i+3},f_j]-3[\psi_{i+2},f_{j+1}]+3[\psi_{i+1},f_{j+2}]-[\psi_i,f_{j+3}]+\sigma_2([\psi_{i+1},f_j]-[\psi_i,f_{j+1}])=-\sigma_3\{\psi_i,f_j\}\;,\\
		&[f_{i+3},f_j]-3[f_{i+2},f_{j+1}]+3[f_{i+1},f_{j+2}]-[f_i,f_{j+3}]+\sigma_2([f_{i+1},f_j]-[f_i,f_{j+1}])=-\sigma_3\{f_i,f_j\}\;,\\
		&\mathrm{Sym}_{i_1,i_2,i_3}[e_{i_1},[e_{i_2},e_{i_3+1}]]=\mathrm{Sym}_{i_1,i_2,i_3}[f_{i_1},[f_{i_2},f_{i_3+1}]]=0\;,
	\end{align}
\end{subequations}
together with the initial conditions
\begin{subequations}
	\begin{align}
		&[\psi_0,e_i]=0\;,\quad[\psi_1,e_i]=0\;,\quad[\psi_2,e_i]=2e_i\;,\\
		&[\psi_0,f_i]=0\;,\quad[\psi_1,f_i]=0\;,\quad[\psi_2,f_i]=-2f_i\;.
	\end{align}
\end{subequations}
Moreover, we have used the convention $\psi(u)=1+\sigma_3\sum\limits_{i=0}^\infty\frac{\psi_i}{u^{i+1}}$ in the current expansion (while the $e$ and $f$ currents have the same convention as the others).

\addcontentsline{toc}{section}{References}
\bibliographystyle{ytamsalpha}
\bibliography{references}

\end{document}